\documentclass[aps,prd,10pt,notitlepage,nofootinbib,superscriptaddress]{revtex4-1}

\usepackage[utf8]{inputenc}
\usepackage{amsmath}
\usepackage{amssymb}
\usepackage{amsfonts}
\usepackage{newtxtext,newtxmath}
\usepackage{bbold}
\usepackage{bm}
\usepackage{graphicx}
\usepackage[usenames,dvipsnames]{xcolor}
\usepackage{color}
\usepackage[colorlinks=true,linkcolor=blue,urlcolor=blue,citecolor=blue]{hyperref}

\usepackage{slashed}
\usepackage[english]{babel}
\usepackage{dcolumn}
\usepackage{blindtext}
\usepackage{epsfig}
\usepackage{pifont}
\usepackage{dsfont,mathrsfs}
\usepackage{cancel}
 \usepackage{bigints}
 \usepackage{accents}
 \usepackage{soul}
 \usepackage{multirow}
 \usepackage{simpler-wick}

\newcommand{\nn}{\nonumber}

\newcommand{\ensembleaverage}[1]{\left\langle#1\right\rangle}

\newcommand{\FB}[1]{\left(#1\right)}
\newcommand{\fb}[1]{(#1)}
\newcommand{\SB}[1]{\left\{#1\right\}}
\newcommand{\TB}[1]{\left[#1\right]}

\newcommand{\mcT}{\mathcal{T}}
\newcommand{\mcTc}{\mathcal{T}_C}
\newcommand{\mcN}{\mathcal{N}}
\newcommand{\scrL}{\mathscr{L}}

\newcommand{\scrD}{\mathscr{D}}

\newcommand{\munu}{{\mu\nu}}

\newcommand{\alphabeta}{{\alpha\beta}}
\newcommand{\mnab}{{\mu\nu\alpha\beta}}
\newcommand{\IM}{\text{Im}}

\newcommand{\Tr}{\text{Tr}}

\newcommand{\kpll}{k_\parallel}
\newcommand{\qpll}{q_\parallel}
\newcommand{\ppll}{p_\parallel}

\newcommand{\kper}{k_\perp}
\newcommand{\pper}{p_\perp}

\newcommand{\gpll}{g_\parallel}
\newcommand{\gper}{g_\perp}

\newcommand{\del}{\partial}

\newcommand{\psibar}{\overline{\psi}}

\newcommand{\xitil}{\tilde{\xi}}
\newcommand{\ftil}{\tilde{f}}
\newcommand{\Dtil}{\tilde{D}}

\newcommand{\wkl}{\omega_{kl}}

\newcommand{\vq}{{\vec q}}
\newcommand{\vk}{{\vec k}}
\def\be {\begin{equation}}
\def\ee {\end{equation}}
\def\nn {\nonumber}
\def\bea {\begin{eqnarray}}
\def\eea {\end{eqnarray}}
\newcommand{\om}{\omega}

\begin{document}
\title{One-loop Kubo estimations of the shear and bulk viscous coefficients for hot and magnetized Bosonic and Fermionic systems}

\author{Snigdha Ghosh}
\email{snigdha.physics@gmail.com}
\affiliation{Government General Degree College Kharagpur-II, Madpur, Paschim Medinipur - 721149, West Bengal, India}

\author{Sabyasachi Ghosh}
\email{sabyaphy@gmail.com}
\affiliation{Indian Institute of Technology Bhilai, GEC Campus, Sejbahar, Raipur - 492015, Chhattisgarh, India}

\begin{abstract}
The expressions of the shear viscosity and the bulk viscosity components in the presence of an arbitrary external magnetic field for a system of hot charged scalar Bosons (spin-0) as well as for a system of hot charged Dirac Fermions (spin-$\frac{1}{2}$) have been derived by employing the one-loop Kubo Formalism. This is done by explicitly evaluating the thermo-magnetic spectral functions of the energy momentum tensors using the real time formalism of finite temperature field theory and the Schwinger proper time formalism. In the present work, a rich quantum field theoretical structure in the expressions of the viscous coefficients in non-zero magnetic field are found, which are different from their respective expressions obtained earlier via kinetic theory based calculations; though, in absence of magnetic field, the one-loop Kubo and the kinetic theory based expressions for the viscosities are known to be identical. We have identified that Kubo and kinetic theory based results of viscosity components follow similar kind of temperature and magnetic field dependency. The relaxation time and the synchrotron frequency in the kinetic theory formalism are realized to be connected respectively with the thermal width of propagator and the transitions among the Landau levels of the charged particles in the Kubo formalism. We believe that, the connection of latter quantities are quite new and probably the present work is the first time addressing this interpretation along with the new expressions of viscosity components, not seen in existing works.
\end{abstract}

\maketitle
%
\section{INTRODUCTION}\label{sec.intro}
The heavy ion collision (HIC) experiment at relativistic energy can produce a super hot quark gluon plasma (QGP), which may be exposed under a strong magnetic field $B$ if the nucleus-nucleus collision is non-central. This magnetic field could be of the order of $\sim 10^{18}$ Gauss and comparable to the quantum chromodynamics (QCD) scale ($eB\sim m_\pi^2$ for RHIC-LHC energies)~\cite{Tuchin}, for which many interesting QCD-linked phenomena~\cite{Rev1,Rev2,Rev3,Rev4} can be observed. Among a long list of interesting quantities, transport coefficients like shear viscosity and bulk viscosity is our aimed quantities in the present work. Owing to this fact, a long list of Refs.~\cite{Tuchin_s,Li_s,Asutosh_s,NJLB_s,JD_s,JD_QP_s,Arpan_s,Manu_s1,Denicol_s,Greiner_s,DQCD,G_HRGB,Hattori_b,Agasian_b1,Agasian_b2,Kadam_b,Hattori_e1,Hattori_e2,Manu_e1,Sedrakian_e1, Nam_e,Arpan_e1,Arpan_e2,Somnath_e,G_NJLB_e,Lata_e,G_HRGBQM_e,Patra_e,SS_PR_e} have focused on the microscopic calculation of the transport coefficients, like the shear viscosity in Refs.~\cite{Tuchin_s,Li_s,Asutosh_s,NJLB_s,JD_s,JD_QP_s,Arpan_s,Manu_s1,Denicol_s,Greiner_s,DQCD,G_HRGB}, bulk viscosity in Refs.~\cite{Hattori_b,Agasian_b1,Agasian_b2,Kadam_b} and electrical conductivity in Refs.~\cite{Hattori_e1,Hattori_e2,Manu_e1,Sedrakian_e1,Nam_e,Arpan_e1,Arpan_e2,Somnath_e,G_NJLB_e,Lata_e,G_HRGBQM_e,Patra_e,SS_PR_e} for the hot and/or dense QCD matter in presence of magnetic field. If we analyze the frameworks of those microscopic calculations, they are mostly in the kinetic theory based approaches.

For the shear viscosity in presence of external magnetic field, one can get five independent trace-less tensors, with which five shear viscosity components will be linked, while a single trace-less tensor with isotropic shear viscosity is found in $B=0$ case. There are two possible sets of these five independent trace-less tensors as proposed in  Ref.~\cite{Landau_10} and Ref.~\cite{XGH1} respectively. Using former set, proposed in Ref.~\cite{Landau_10}, the authors of Refs.~\cite{Tuchin_s,Asutosh_s,NJLB_s,JD_s,JD_QP_s,Arpan_s,G_HRGB} have obtained five shear viscosity components ${\tilde\eta}_n$ ($n=0,1,2,3,4$); while, using latter set proposed in Ref.~\cite{XGH1}, the authors of Refs.~\cite{Denicol_s,Greiner_s} have obtained their five shear viscosity components $\eta_n$. However, the ${\tilde\eta}_n$'s and $\eta_n$'s are inter-connected and ultimately can be expressed in terms of the parallel, perpendicular and Hall components~\cite{JD_s}. The general expressions of ${\tilde \eta}_n$ are obtained in the relaxation time approximation (RTA) of kinetic theory approach in Refs.~\cite{Asutosh_s,JD_s,JD_QP_s,Arpan_s,G_HRGB} and the same using the strong magnetic field approximation in Refs.~\cite{Tuchin_s,NJLB_s}. In Ref.~\cite{Denicol_s}, the author have obtained $\eta_n$ in RTA based moment methods but its RTA based kinetic theory calculation can be seen in Ref.~\cite{JD_s}.

However, a quantum field theoretical treatment at finite temperature and magnetic field via the Kubo relations have never been attempted in details which could reveal rich quantum structure in shear and bulk viscosity components. Though, in Ref.~\cite{Greiner_s}, the Kubo-type correlation structure has been considered  but it is explored through a box simulation, not through a field theoretical calculation. So as far as our best of knowledge, we are going to address for the first time a one-loop Kubo expressions of viscosity components in presence of magnetic field by using real time formalism of finite temperature field theory and the Schwinger proper time framework.

Let us revisit quickly the Kubo expression of shear viscosity of any Bosonic or Fermionic medium in absence of magnetic field~\cite{Kubo,Ghosh:2014yea,G_etaN,Nicola,Lang,Jeon}. Owing to Kubo relation, shear viscosity can be related to the static limit (zero four-momentum limit) of the two point correlation functions of the local viscous stress tensor $\pi^{\mu\nu}$. The simplest Feynman diagrams from the Bosonic and Fermion free Lagrangian (densities) will be a Boson-Boson or a Fermion-Fermion loops, whose propagators must carry finite thermal width $\Gamma$. Without this $\Gamma$, one can not get any non-divergent values; so this imposition of finite $\Gamma$ brings a quasi-particle picture, where the $\Gamma$ can be estimated from interaction Lagrangian of a particular system and the numerical value of the shear viscosity of that system is mainly controlled by the strength of $\Gamma$. Interestingly, this one-loop Kubo expression~\cite{Ghosh:2014yea,G_etaN,Nicola,Lang,Jeon} of shear viscosity becomes exactly identical to the RTA expression~\cite{Gavin,Kapusta} with the relaxation
time $\tau_c=1/\Gamma$ in the absence of magnetic field.

Now, at finite magnetic field, one should not expect the same expressions of viscosity components from the Kubo framework and the RTA framework~\cite{Tuchin_s,Asutosh_s,NJLB_s,JD_s,JD_QP_s,Arpan_s,Denicol_s,G_HRGB}. The present work is going to reveal a difference between the expressions of the viscosities between the two frameworks at finite magnetic field picture. Here, we have found a rich quantum structure in the expressions of viscous coefficients, which might (not) be obtained by appropriate quantum extension of RTA frameworks. Ref.~\cite{Manu_s1} for shear viscosity and Refs.~\cite{G_HRGBQM_e,Hattori_e1,Hattori_e2,Manu_e1,G_HRGBQM_e,Patra_e} for electrical conductivity calculations have followed the Landau quantization version of RTA. However, our Kubo expressions will have additional structures which are not the mere Landau quantized version of the corresponding RTA expressions of the viscosities. 
In the present work, we have considered two different systems: (i) system of charged scalar Bosons (spin-0) and (ii) system of charged Dirac Fermions (spin-$\frac{1}{2}$), and have calculated the corresponding thermo-magnetic spectral functions of the energy-momentum tensors (EMTs). The spectral function is the imaginary part of the Fourier transform of the local EMT-EMT two-point correlator which is obtained using the real time formalism of finite temperature field theory and the Schwinger proper time formalism to incorporate the effects of finite temperature and external magnetic field respectively. Then viscous coefficients are estimated from the thermo-magnetic spectral functions using the Kubo relations in the covariant tensor basis of Ref.~\cite{XGH1}.

The article is organized as follows. We first start with the calculation of the in-medium spectral function of the energy momentum tensor at zero magnetic field in Sec.~\ref{sec.spectral} and at non-zero magnetic field in Sec.~\ref{sec.spectral.B}. Next, Sec.~\ref{sec.viscosity} has demonstrated how to obtain the shear and bulk viscosity from those spectral functions using the Kubo relations. After getting the new expressions viscosity components, their numerical outcomes have been sketched and been tried to interpret in Sec.~\ref{sec.results}. This is followed by Sec.~\ref{sec.summary} where we have summarized the investigation. To compensate the calculation gaps, we have provided detailed Appendices at the end.

\section{THE SPECTRAL FUNCTION OF THE ENERGY-MOMENTUM TENSOR} \label{sec.spectral}
\begin{figure}[h]
	\begin{center}
		\includegraphics[angle=0,scale=0.5]{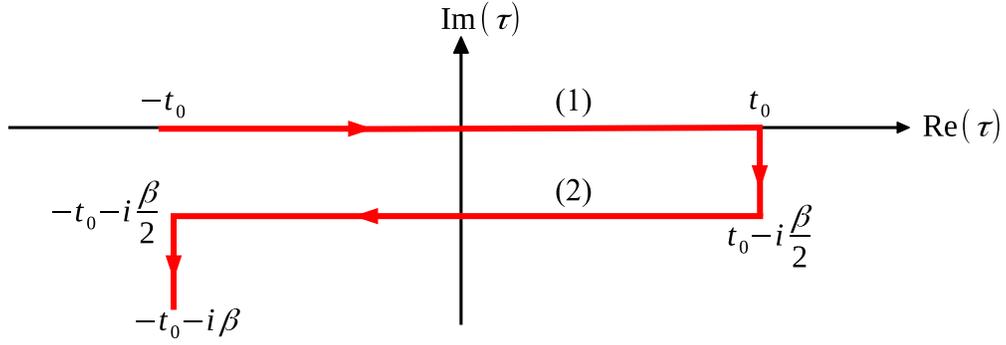}
	\end{center}
	\caption{The symmetric Schwinger-Keldysh contour $C$ in the complex time plane used in the RTF with $t_0\to\infty$ and $\beta = 1/T$. The two horizontal segments of the contour are 
		referred as labels `(1)' and `(2)' respectively.}
	\label{fig.contour}
\end{figure}

The key microscopic quantity that is required to calculate the viscous coefficients of a thermal medium 
using the Kubo formalism~\cite{Kubo} is the in-medium spectral function $\rho^\mnab(q)$, given by
\begin{eqnarray}
\rho^\mnab(q) = \IM~i\int d^4x  e^{iq\cdot x}\ensembleaverage{T^\munu(x)T^\alphabeta(0)}_R~,
\end{eqnarray}
where $T^\munu(x)$ is the local EMT and $\ensembleaverage{\cdots}_R$ 
represents the ensemble average of the retarded two point correlation function. We will be using 
metric tensor with signature $g^\munu=\text{diag}(1,-1,-1,-1)$. 
In order to use field theoretic methods, it is more convenient to express the spectral function in 
terms of time-ordered correlator as~\cite{Bellac:2011kqa,Mallik:2016anp}
\begin{eqnarray}
\rho^\mnab(q) = \tanh\FB{\frac{q^0}{2T}}\IM~ i\int d^4x e^{iq\cdot x} \ensembleaverage{\mcTc T^\munu(x)T^\alphabeta(0)}_{11}
\label{eq.spectral.1}
\end{eqnarray}
where, $\mcTc$ is the time-ordering with respect to the symmetric Schwinger-Keldysh contour $C$ in complex time plane shown in Fig.~\ref{fig.contour} as used in the real time formalism (RTF) of finite temperature field theory . The subscript $11$ in the above equation implies that the two points are on the real horizontal segment `(1)' of the contour $C$.

The form of the local EMT $T^\munu(x)$ appearing in Eq.~\eqref{eq.spectral.1} depends on the particular system considered. 
In this work, we will be mainly considering two systems: (i) system of charged scalar Bosons $B^\pm$ (spin-0) and 
(ii) system of charged Dirac Fermions $F^\pm$ (spin-$\frac{1}{2}$). 
They are respectively described by the fields $\phi(x)$ and $\psi(x)$ which constructs the following free Lagrangian (densities)~\cite{Greiner:1996zu}
\begin{eqnarray}
\scrL_\text{Scalar} &=& \del^\mu\phi^\dagger\del_\mu\phi - m^2\phi^\dagger\phi~, \label{eq.lag.scalar}\\
\scrL_\text{Dirac} &=& \frac{i}{2}\FB{\psibar\gamma^\mu\del_\mu\psi-\del_\mu\psibar\gamma^\mu\psi}-m\psibar\psi
\label{eq.lag.dirac}
\end{eqnarray} 
in which $m$ is the mass of the particles. Symmetric EMTs can be constructed out of the above Lagrangians as~\cite{Peskin:1995ev,Lahiri:2005sm,Greiner:1996zu}
\begin{eqnarray}
T^\munu_\text{Scalar} &=& \del^\mu\phi^\dagger\del^\nu\phi - \frac{1}{2}g^\munu \scrL_\text{Scalar} + (\mu \leftrightarrow \nu)~, \label{eq.emt.scalar}\\
T^\munu_\text{Dirac} &=& \frac{i}{4}\FB{\psibar\gamma^\mu\del^\nu\psi-\del^\nu\psibar\gamma^\mu\psi} - \frac{1}{2}g^\munu \scrL_\text{Dirac}  + (\mu \leftrightarrow \nu). \label{eq.emt.dirac}
\end{eqnarray} 
Using Eqs.~\eqref{eq.emt.scalar} and \eqref{eq.emt.dirac}, we now calculate the EMT correlation function $\ensembleaverage{\mcTc T^\munu(x)T^\alphabeta(0)}_{11}$ appearing in Eq.~\eqref{eq.spectral.1} for both the scalar and Dirac fields. The calculation have been provided in Appendix~\ref{app.emt.corr} and we get from Eqs.~\eqref{eq.corr.scalar.3} and \eqref{eq.corr.dirac.4}
\begin{eqnarray}
\ensembleaverage{\mcTc T_\text{Scalar}^\munu(x)T_\text{Scalar}^\alphabeta(0)}_{11} &=& 
-\int\!\!\!\!\int\!\!\frac{d^4p}{(2\pi)^4}\frac{d^4k}{(2\pi)^4} e^{-ix\cdot(p-k)} D_{11}(p;m)D_{11}(k;m)
\mcN_\text{Scalar}^\mnab(k,p)~, \label{eq.corr.scalar}\\
\ensembleaverage{\mcTc T_\text{Dirac}^\munu(x)T_\text{Dirac}^\alphabeta(0)}_{11} &=& 
-\int\!\!\!\!\int\!\!\frac{d^4p}{(2\pi)^4}\frac{d^4k}{(2\pi)^4} e^{-ix\cdot(p-k)} \Dtil_{11}(p;m)\Dtil_{11}(k;m)
\mcN_\text{Dirac}^\mnab(k,p) \label{eq.corr.dirac}
\end{eqnarray}
where, $D_{11}$, $\Dtil_{11}$ and $\mcN_\text{Scalar,Dirac}^\mnab$ can be respectively read off from Eqs.~\eqref{eq.D11}, \eqref{eq.D11til}, \eqref{eq.N.scalar} and \eqref{eq.N.dirac}. 
We now substitute the EMT correlators of Eqs.~\eqref{eq.corr.scalar} and \eqref{eq.corr.dirac} into Eq.~\eqref{eq.spectral.1} and perform the $d^4x$ integral which yields the Dirac delta 
function $\delta^4(q-p+k)$. The Dirac delta function is in turn used to perform the $d^4p$ integral and the spectral function of EMT becomes,
\begin{eqnarray}
\rho_\text{Scalar}^\mnab(q) &=& -\tanh\FB{\frac{q^0}{2T}}\IM~ i\int\!\!\!\frac{d^4k}{(2\pi)^4} D_{11}(k;m)D_{11}(p=q+k;m)\mcN_\text{Scalar}^\mnab(k,p=q+k)~, \\
\rho_\text{Dirac}^\mnab(q) &=& -\tanh\FB{\frac{q^0}{2T}}\IM~ i\int\!\!\!\frac{d^4k}{(2\pi)^4} \Dtil_{11}(k;m)\Dtil_{11}(p=q+k;m)\mcN_\text{Dirac}^\mnab(k,p=q+k).
\end{eqnarray}
Substituting $D_{11}$ and $\Dtil_{11}$ from Eqs.~\eqref{eq.D11} and \eqref{eq.D11til} into the above two equations followed by performing the $dk^0$ integral, we get after some simplifications 
\begin{eqnarray}
\rho^\mnab(q) = \tanh\FB{\frac{q^0}{2T}}\pi\int\!\!\!\frac{d^3k}{(2\pi)^3} \frac{1}{4\omega_k\omega_p} \Big[
\SB{1+af_a(\omega_k)+af_a(\omega_p)+2f_a(\omega_k)f_a(\omega_p)} \nn \\
\times ~\big\{N^\mnab(k^0=\omega_k)\delta(q^0-\omega_k-\omega_p) 
+ N^\mnab(k^0=-\omega_k)\delta(q^0+\omega_k+\omega_p)\big\} \nn \\
+ \SB{af_a(\omega_k)+af_a(\omega_p)+2f_a(\omega_k)f_a(\omega_p)} 
 \big\{N^\mnab(k^0=-\omega_k)\delta(q^0-\omega_k+\omega_p) \nn \\
+~ N^\mnab(k^0=\omega_k)\delta(q^0+\omega_k-\omega_p)\big\}
\Big]
\label{branch}
\end{eqnarray}
where $\omega_k=\sqrt{\vec{k}^2+m^2}$, $\omega_p=\sqrt{(\vec{k}+\vec{p})^2+m^2}$ 
and $f_a(x) = \dfrac{1}{e^{x/T}-a}$ with 
 $a = \begin{cases}
1 ~~\text{for}~~ \text{scalar} \\
-1 ~~\text{for}~~ \text{Dirac}
\end{cases} $.
Above Eq.~(\ref{branch}) carries four Dirac delta functions which will give rise to the branch cuts of the spectral function in the complex $q^0$ plane. The kinematic regions where these Dirac delta functions are non-zero are respectively: (i) $\sqrt{\vq^2+4m^2}<q_0<\infty$, (ii) $-\infty<q_0<-\sqrt{\vq^2+4m^2}$, (iii) and (iv) $|q^0| < |\vq|$ as they appear in Eq.~\eqref{branch}. Regions (i) and (ii) are respectively called the unitary-I and unitary-II cuts whereas regions (iii) and (iv) are called the Landau-II and Landau-I cuts respectively~\cite{Mallik:2016anp,Ghosh:2014yea}. 
For the evaluation of the viscous coefficients, we need to take the static limit $\vec{q}=\vec{0}, q^0\to 0$ of the EMT spectral function for which the unitary cuts do not contribute. Considering the Landau cuts only, we are left with
\begin{eqnarray}
\rho^\mnab(q^0,\vec{q}=\vec{0}) = \tanh\FB{\frac{q^0}{2T}}\pi\int\!\!\!\frac{d^3k}{(2\pi)^3} \frac{1}{2\omega_k^2} \delta(q^0)
f_a(\omega_k)\SB{a+f_a(\omega_k)} \big\{N^\mnab(k^0=-\omega_k)+ N^\mnab(k^0=\omega_k)\big\} \\
= \lim\limits_{\Gamma\to0}\tanh\FB{\frac{q^0}{2T}}\int\!\!\!\frac{d^3k}{(2\pi)^3} \frac{1}{2\omega_k^2} \frac{\Gamma}{q_0^2+\Gamma^2}
f_a(\omega_k)\SB{a+f_a(\omega_k)} \big\{N^\mnab(k^0=-\omega_k)+ N^\mnab(k^0=\omega_k)\big\} 
\end{eqnarray}
where a Breit-Wigner representation of the Dirac-delta function has been used. According to the definition of Kubo relation, the dissipation coefficients are related to the zero four-momentum limit of $\rho^\mnab/q_0$ or $\mathcal{S}^\mnab = \frac{\del\rho^\mnab}{\del q^0}$, owing to the L'Hospital's rule. Differentiating the spectral function with respect to $q^0$ and taking limit $q^0\to0$, we finally obtain
\begin{eqnarray}
\mathcal{S}^\mnab = \frac{\del\rho^\mnab}{\del q^0}\Bigg|_{\vec{q}=\vec{0},q^0\to 0} \!\!\!=  \lim\limits_{\Gamma\to0}\frac{1}{T}
\int\!\!\!\frac{d^3k}{(2\pi)^3} \frac{1}{4\omega_k^2\Gamma} 
f_a(\omega_k)\SB{a+f_a(\omega_k)} \Big[N^\mnab(k,k)\Big|_{k^0=\omega_k}+ N^\mnab(k,k)\Big|_{k^0=-\omega_k}\Big]
\label{eq.spectral}
\end{eqnarray}
where, the simplified expressions of $N_\text{Scalar,Dirac}^\mnab(k,k)$ can be obtained from Eqs.~\eqref{eq.Nkk.scalar} and \eqref{eq.Nkk.dirac} as
\begin{eqnarray}
\mcN_\text{Scalar}^\mnab(k,k) &=& 4k^\mu k^\nu k^\alpha k^\beta -2\fb{k^2 -m^2}\fb{g^\munu k^\alpha k^\beta + g^\alphabeta k^\mu k^\nu } +\fb{k^2 -m^2}^2 g^\munu g^\alphabeta, \label{eq.Nkk.scalar.final} \\
\mcN_\text{Dirac}^\mnab(k,k) &=& 
-8 k^\mu k^\nu k^\alpha k^\beta 
+  (k^2-m^2) \big\{ g^{\mu\alpha} k^\nu k^\beta + g^{\nu\alpha} k^\mu k^\beta + g^{\mu\beta} k^\nu k^\alpha + g^{\nu\beta} k^\mu k^\alpha \nn \\ &&
 +  4 g^\munu k^\alpha k^\beta 
+  4 g^\alphabeta k^\mu k^\nu  \big\} 
 - 4(k^2-m^2)^2 g^\munu g^\alphabeta. \label{eq.Nkk.dirac.final}
\end{eqnarray}

To get a non-divergent contribution of EMT spectral function in the zero momentum limit, further calculation will have to be continued with finite value of $\Gamma$. This is the place where the interaction picture is introduced, which should be entered for a dissipative system. This $\Gamma$ can be identified as the thermal width or collision rate of the constituent particles, and it reciprocally measures the dissipative coefficients, like the shear viscosity and the bulk viscosity.

It may be noted that, in the present method, we have introduced the interaction information $\Gamma$ via the transformation of Dirac delta functions to Breit-Wigner functions, from which the non-interacting picture is realized as $\Gamma\rightarrow 0$ limit. This similar kind of transformation from non-interacting ($\Gamma=0$) to interacting ($\Gamma\neq 0$) picture can also be done by introducing $i\Gamma/2$ in the propagators, located in our one-loop diagramatic respresentation of the transport coefficients.

\section{THE SPECTRAL FUNCTION OF THE EMT IN PRESENCE OF EXTERNAL MAGNETIC FIELD} \label{sec.spectral.B}
In the presence of an external electromagnetic field described by the 
four-potential $A^\mu_\text{ext}(x)$, the Lagrangians of Eqs.~\eqref{eq.lag.scalar} and \eqref{eq.lag.dirac} are modified to~\cite{Schwartz} 
\begin{eqnarray}
\scrL_\text{Scalar} &=& D^{*\mu}\phi^\dagger D_\mu\phi - m^2\phi^\dagger\phi~, \label{eq.lag.scalar.B}\\
\scrL_\text{Dirac} &=& \frac{i}{2}\FB{\psibar\gamma^\mu D_\mu\psi-D^{*\mu}\psibar\gamma^\mu\psi}-m\psibar\psi
\label{eq.lag.dirac.B}
\end{eqnarray} 
where, $D^\mu=\del^\mu+ieA^\mu_\text{ext}(x)$ and $D^{*\mu}=\del^\mu-ieA^\mu_\text{ext}(x)$ are the covariant derivatives incorporating the minimal coupling between the charged particle with charge $e$ (we consider $e>0$) and the electromagnetic field. The symmetric EMT in presence of electromagnetic field now becomes~\cite{Schwartz,Greiner:1996zu} 
\begin{eqnarray}
T^\munu_\text{Scalar} &=& D^{*\mu}\phi^\dagger D^\nu\phi - \frac{1}{2}g^\munu \scrL_\text{Scalar} + (\mu \leftrightarrow \nu)~, \label{eq.emt.scalar.B}\\
T^\munu_\text{Dirac} &=& \frac{i}{4}\FB{\psibar\gamma^\mu D^\nu\psi-D^{*\nu}\psibar\gamma^\mu\psi} - \frac{1}{2}g^\munu \scrL_\text{Dirac}  + (\mu \leftrightarrow \nu). \label{eq.emt.dirac.B}
\end{eqnarray} 

Let us now consider a constant magnetic field $\vec{B}=B\hat{z}$ along the positive $\hat{z}$-direction. 
Using Eqs.~\eqref{eq.emt.scalar.B} and \eqref{eq.emt.dirac.B}, we now calculate the EMT correlation function $\ensembleaverage{\mcTc T^\munu(x)T^\alphabeta(0)}^B_{11}$ in presence of external magnetic field  for both the scalar and Dirac field as sketched in Appendix~\ref{app.emt.corr.b}. 
We obtain from Eqs.~\eqref{eq.corr.scalar.B.3} and \eqref{eq.corr.dirac.B.2} 
\begin{eqnarray}
\ensembleaverage{\mcTc T_\text{Scalar}^\munu(x)T_\text{Scalar}^\alphabeta(0)}^B_{11} &=& 
-\int\!\!\!\!\int\!\!\frac{d^4p}{(2\pi)^4}\frac{d^4k}{(2\pi)^4} e^{-ix\cdot(p-k)}
\sum_{l=0}^{\infty} \sum_{n=0}^{\infty} D_{11}(\ppll;m_n) D_{11}(\kpll;m_l) 
\mcN_{ln;\text{Scalar}}^\mnab(k,p)~, \label{eq.corr.scalar.B} \\
\ensembleaverage{\mcTc T_\text{Dirac}^\munu(x)T_\text{Dirac}^\alphabeta(0)}^B_{11} &=& 
-\int\!\!\!\!\int\!\!\frac{d^4p}{(2\pi)^4}\frac{d^4k}{(2\pi)^4} e^{-ix\cdot(p-k)}
\sum_{l=0}^{\infty} \sum_{n=0}^{\infty} \Dtil_{11}(\ppll;m_n)\Dtil_{11}(\kpll;m_l) 
\mcN_{ln;\text{Dirac}}^\mnab(k,p)
\label{eq.corr.dirac.B}
\end{eqnarray}
where, $D_{11}$, $\Dtil_{11}$ and $\mcN_{ln;\text{Scalar,Dirac}}^\mnab$ can be respectively read off from Eqs.~\eqref{eq.D11}, \eqref{eq.D11til}, \eqref{eq.N.scalar.B} and \eqref{eq.N.dirac.B} and 
$m_l=\sqrt{m^2+(2l+1-2s)eB}$ with $s$ being the spin of the particle ($s=0$ for the scalar and $s=1/2$ for the Dirac).

Similar to the zero magnetic field case, we now substitute the EMT correlators of Eqs.~\eqref{eq.corr.scalar.B} and \eqref{eq.corr.dirac.B} into Eq.~\eqref{eq.spectral.1} and perform the $d^4x$ integral which yields the Dirac delta 
function $\delta^4(q-p+k)$. The Dirac delta function is in turn used to perform the $d^4p$ integral and the thermo-magnetic spectral function of EMT becomes,
\begin{eqnarray}
\rho_\text{Scalar}^\mnab(q) &=& -\tanh\FB{\frac{q^0}{2T}}\IM~ i
\int\!\!\!\frac{d^4k}{(2\pi)^4} \sum_{l=0}^{\infty} \sum_{n=0}^{\infty} 
D_{11}(\kpll;m_l)D_{11}(\ppll=\qpll+\kpll;m_n)\mcN_{ln;\text{Scalar}}^\mnab(k,p=q+k)~, \label{eq.rho.scalar}\\
\rho_\text{Dirac}^\mnab(q) &=& -\tanh\FB{\frac{q^0}{2T}}\IM~ i\int\!\!\!\frac{d^4k}{(2\pi)^4}
\sum_{l=0}^{\infty} \sum_{n=0}^{\infty}
 \Dtil_{11}(\kpll;m_l)\Dtil_{11}(\ppll=\qpll+\kpll;m_n)\mcN_{ln;\text{Dirac}}^\mnab(k,p=q+k)~. \label{eq.rho.dirac}
\end{eqnarray}
Substituting $D_{11}$ and $\Dtil_{11}$ from Eqs.~\eqref{eq.D11} and \eqref{eq.D11til} into the above two equations followed by performing the $dk^0$ integral, we get after some simplifications 
\begin{eqnarray}
\rho^\mnab(q) = \tanh\FB{\frac{q^0}{2T}}\pi
\sum_{l=0}^{\infty} \sum_{n=0}^{\infty} 
\int\!\!\!\frac{d^3k}{(2\pi)^3} \frac{1}{4\omega_{kl}\omega_{pn}} \Big[
\SB{1+af_a(\omega_{kl})+af_a(\omega_{pn})+2f_a(\omega_{kl})f_a(\omega_{pn})} \nn \\  
\times~ \big\{N_{ln}^\mnab(k^0=\omega_{kl})\delta(q^0-\omega_{kl}-\omega_{pn}) 
+ N_{ln}^\mnab(k^0=-\omega_{kl})\delta(q^0+\omega_{kl}+\omega_{pn})\big\} \nn \\ 
+ \SB{af_a(\omega_{kl})+af_a(\omega_{pn})+2f_a(\omega_{kl})f_a(\omega_{pn})} 
  \big\{N_{ln}^\mnab(k^0=-\omega_{kl})\delta(q^0-\omega_{kl}+\omega_{pn}) \nn \\ 
+~ N_{ln}^\mnab(k^0=\omega_{kl})\delta(q^0+\omega_{kl}-\omega_{pn})\big\}
\Big] \label{eq.rho.B}
\end{eqnarray}
where, $\omega_{kl}=\sqrt{k_z^2+m_l^2}$, $\omega_{pn}=\sqrt{p_z^2+m_n^2}=\sqrt{(k_z+q_z)^2+m_n^2}$.
Similar to the zero field case, the spectral function in presence of external magnetic field contains four Dirac delta functions giving rise to branch cuts of the spectral function in the complex energy plane. The kinematic regions where these Dirac delta functions are non-zero now depends on the Landau levels of the charged particles as well. Thus, when summed over an infinite number of Landau levels, the kinematic regions for the unitary-I and unitary-II cuts comes out to be $\sqrt{q_z^2+4(m^2+eB)} < q^0 < \infty$ and $-\infty < q^0 < -\sqrt{q_z^2+4(m^2+eB)} $ respectively whereas the kinematic domain for both the Landau cuts becomes $|q^0| < \sqrt{q_z^2+(\sqrt{m^2+eB}-\sqrt{m^2+3eB})^2}$ for the scalar case. 
On the other hand, for the Dirac case, the correspond kinematic domains for unitary-I and unitary-II cuts are $\sqrt{q_z^2+4m^2} < q^0 < \infty$ and $-\infty < q^0 < -\sqrt{q_z^2+4m^2} $ respectively whereas the same for both the Landau cuts is $|q^0| < \sqrt{q_z^2+(m-\sqrt{m^2+2eB})^2}$~\cite{,Ghosh:2019fet,Ghosh:2017rjo,Ghosh:2018xhh}.

As already discussed in Sec.~\ref{sec.spectral}, for the evaluation of the viscous coefficients, we only need the Landau cuts and the spectral function becomes
\begin{eqnarray}
	\rho^\mnab(q^0,\vec{q}=\vec{0}) &=& \tanh\FB{\frac{q^0}{2T}}\pi
	\sum_{l=0}^{\infty} \sum_{n=0}^{\infty} 
	\int\!\!\!\frac{d^3k}{(2\pi)^3} \frac{1}{4\omega_{kl}\omega_{kn}} 
	\SB{af_a(\omega_{kl}) + af_a(\omega_{kn})+ 2f_a(\omega_{kl})f_a(\omega_{kn})} \nn \\ && \hspace{2.0cm} \times \big\{N_{ln}^\mnab(k^0=-\omega_{kl})\delta(q^0-\omega_{kl}+\omega_{kn})+ N_{ln}^\mnab(k^0=\omega_{kl})\delta(q^0+\omega_{kl}-\omega_{kn})\big\} \\
	&=& \lim\limits_{\Gamma\to0}\tanh\FB{\frac{q^0}{2T}}
	\sum_{l=0}^{\infty} \sum_{n=0}^{\infty} 
	\int\!\!\!\frac{d^3k}{(2\pi)^3} \frac{1}{4\omega_{kl}\omega_{kn}} 
	\SB{af_a(\omega_{kl}) + af_a(\omega_{kn})+ 2f_a(\omega_{kl})f_a(\omega_{kn})} \nn \\ && 
	\hspace{0.5cm}\times \bigg\{
	N_{ln}^\mnab(k^0=-\omega_{kl})\frac{\Gamma}{(q^0-\omega_{kl}+\omega_{kn})^2+\Gamma^2}+ N_{ln}^\mnab(k^0=\omega_{kl})\frac{\Gamma}{(q^0+\omega_{kl}-\omega_{kn})^2+\Gamma^2}\bigg\}
\end{eqnarray}
where a Breit-Wigner representation of the Dirac-delta function has again been used. Differentiating the above equation with respect to $q^0$ and taking limit $q^0\to0$, we finally obtain
\begin{eqnarray}
	\mathcal{S}^\mnab = \lim\limits_{\Gamma\to0} 
	\sum_{l=0}^{\infty} \sum_{n=0}^{\infty}\frac{1}{2T} 
	\int\!\!\!\frac{d^3k}{(2\pi)^3} \frac{1}{4\omega_{kl}\omega_{kn}} 
	\frac{\Gamma}{(\omega_{kl}-\omega_{kn})^2+\Gamma^2}
	\SB{af_a(\omega_{kl}) + af_a(\omega_{kn})+ 2f_a(\omega_{kl})f_a(\omega_{kn})} \nn \\ 
	\hspace{1.3cm}
	\times \Big[N_{ln}^\mnab(k,k)\Big|_{k^0=\omega_{kl}}+ N_{ln}^\mnab(k,k)\Big|_{k^0=-\omega_{kl}}\Big]
	\label{eq.S.B}
\end{eqnarray}
where, the simplified expressions of $N_{ln;\text{Scalar,Dirac}}^\mnab(k,k)$ can be obtained from Eqs.~\eqref{eq.Nkk.scalar.B} and \eqref{eq.Nkk.dirac.B} as
\begin{eqnarray}
\mcN_{ln;\text{Scalar}}^\mnab(k,k)
&=& 4 \mathcal{A}_{ln}(\kper^2)	\Big\{
4k^\mu k^\nu k^\alpha k^\beta -2\fb{k^2 -m^2}\fb{g^\munu k^\alpha k^\beta + g^\alphabeta k^\mu k^\nu } +\fb{k^2 -m^2}^2 g^\munu g^\alphabeta \Big\} 
\label{eq.Nkk.scalar.B.final}
\end{eqnarray}
and, 
\begin{eqnarray}
\mcN_{ln;\text{Dirac}}^\mnab(k,k) &=& 
-16 \mathcal{B}_{ln}(\kper^2) \Big[ 
k^\nu k^\beta (2\kper^\mu\kper^\alpha-\kper^2g^{\mu\alpha}) 
-g^\munu k^\beta \kper^2(\kper^\alpha-\kpll^\alpha) 
- g^\alphabeta k^\nu \kper^2(\kper^\mu-\kpll^\mu) \nn \\ &&
+ g^\munu g^\alphabeta \kper^2(\kper^2-\kpll^2+m^2) \Big]
-2 \mathcal{C}_{ln}(\kper^2) \Big[
k^\nu k^\beta \big\{2\kpll^\mu\kpll^\alpha-(\kpll^2-m^2)\gpll^{\mu\alpha}\big\}
- (\kpll^2-m^2) g^\munu k^\beta\kpll^\alpha \nn \\ &&
- (\kpll^2-m^2) g^\alphabeta k^\nu \kpll^\mu
+ g^\munu g^\alphabeta (\kpll^2-m^2)^2 \Big]
-2 \mathcal{D}_{ln}(\kper^2) (\kpll^2-m^2) \Big[
k^\nu k^\beta \gper^{\mu\alpha} 
-  g^\munu k^\beta\kper^\alpha  \nn \\ &&
-  g^\alphabeta k^\nu \kper^\mu
+ g^\munu g^\alphabeta \kper^2 \Big]
-4 \mathcal{E}_{ln}(\kper^2) \Big[
k^\nu k^\beta (\kpll^\mu\kper^\alpha+\kper^\mu\kpll^\alpha) 
- g^\munu k^\beta \big\{(\kpll^2-m^2)\kper^\alpha + \kper^2\kpll^\alpha \big\}  \nn \\ &&
- g^\alphabeta k^\nu \big\{(\kpll^2-m^2)\kper^\mu + \kper^2\kpll^\mu\big\}
+ 2g^\munu g^\alphabeta \kper^2 (\kpll^2-m^2) \Big]
+ (\mu \leftrightarrow \nu) +  (\alpha \leftrightarrow \beta) 
+ (\mu \leftrightarrow \nu, \alpha \leftrightarrow \beta)
\label{eq.Nkk.dirac.B.final}
\end{eqnarray}
in which the functions $\mathcal{A}_{ln}(\kper^2)$, $\mathcal{B}_{ln}(\kper^2)$, $\cdots$,  $\mathcal{E}_{ln}(\kper^2)$ are defined in Eqs~\eqref{eq.Anl} and \eqref{eq.Bnl}-\eqref{eq.Enl}.

Few comments on the Lorentz structure of the EMT spectral functions are in order here. In the current work, we have considered only the symmetric part of the EMT spectral function which is proportional to the imaginary part of the retarded correlator. On the other hand, the real part of the correlator contributing to the antisymmetric piece of the spectral function has been discarded. Another way of saying this is that, in Eqs.~\eqref{eq.rho.scalar} and \eqref{eq.rho.dirac}, if the tensor $\mcN^{\mu\nu\alpha\beta}_{ln}$ contains terms like $i \epsilon_\perp^{\mu\alpha} k^\nu k^\beta$, then Eq.~\eqref{eq.rho.B} will in turn have additional terms related to the real part of the Fourier transformed correlator~\cite{Mamo:2015xkw}. In the current work, we did not get such antisymmetric terms contributing to the viscous coefficients at one-loop order.


\section{VISCOUS COEFFICIENTS FROM THE SPECTRAL FUNCTION IN KUBO FORMALISM}\label{sec.viscosity}
Owing to the Kubo relation~\cite{Kubo}, the viscous coefficients (shear and bulk) can be calculated from the spectral functions of the EMT which have already been obtained in Secs.~\ref{sec.spectral} and \ref{sec.spectral.B}. We will first revisit $B=0$ case~\cite{Nicola,Ghosh:2014yea,G_etaN,Lang,Jeon} before discussing the finite $B$ calculations.

In absence of external magnetic field, the shear viscosity ($\eta$) and bulk viscosity ($\zeta$) are obtained from~\cite{Ghosh:2014yea}
\begin{eqnarray}
\upsilon = \mathcal{P}^{(\upsilon)}_\mnab \mathcal{S}^\mnab ~~~;~ \upsilon\in\{ \eta,\zeta \} 
\label{eq.v}
\end{eqnarray}
where, 
\begin{eqnarray}
\mathcal{P}^{(\eta)}_\mnab &=& \frac{1}{10}
\FB{\Delta^\sigma_\mu \Delta^\rho_\nu - \frac{1}{3}\Delta^{\sigma\rho}\Delta_\munu}
\FB{\Delta_{\sigma\alpha} \Delta_{\rho\beta} - \frac{1}{3}\Delta_{\sigma\rho}\Delta_\alphabeta}~, 
\label{eq.P.eta}\\
\mathcal{P}^{(\zeta)}_\mnab &=& \FB{\frac{1}{3}\Delta_\munu + \theta~ u_\mu u_\nu} 
\FB{\frac{1}{3}\Delta_\alphabeta + \theta~ u_\alpha u_\beta} \label{eq.P.zeta}
\end{eqnarray}
in which $\Delta^\munu=(g^\munu-u^\mu u^\nu)$, 
$\theta = \FB{\dfrac{\del P}{\del \varepsilon}}$, $P$ is the pressure and $\varepsilon$ is the energy density of the system being considered. 
Substituting Eq.~\eqref{eq.spectral} into Eq.~\eqref{eq.v}, we obtain the viscous coefficients at zero magnetic field as
\begin{eqnarray}
\upsilon = \frac{1}{T}\int\!\!\!\frac{d^3k}{(2\pi)^3} \frac{1}{2\omega_k^2\Gamma} 
f_a(\omega_k)\SB{a+f_a(\omega_k)} \mcN^{(\upsilon)}(\vec{k}) ~~~;~ \upsilon\in\{ \eta,\zeta \}
\label{eq.v.1}
\end{eqnarray}
where, 
\begin{eqnarray}
\mcN^{(\upsilon)}(\vec{k}) = \frac{1}{2}\mathcal{P}^{(\upsilon)}_\mnab 
\SB{\mathcal{N}^\mnab(k,k)\big|_{k^0=\omega_k} + \mathcal{N}^\mnab(k,k)\big|_{k^0=-\omega_k}}.
\label{eq.N.v}
\end{eqnarray}

Substituting Eqs.~\eqref{eq.Nkk.scalar.final}, \eqref{eq.Nkk.dirac.final}, \eqref{eq.P.eta} and \eqref{eq.P.zeta} into Eq.~\eqref{eq.N.v} and simplifying, we obtain,
\begin{eqnarray}
\mcN^{(\eta)}_\text{Scalar} = \frac{4}{15}\vec{k}^4 ~~~&,&~~~~ 
\mcN^{(\zeta)}_\text{Scalar} = \frac{4}{9}\SB{m^2+(3\theta-1)\omega_k^2}^2~, \label{eq.Nv.scalar}\\
\mcN^{(\eta)}_\text{Dirac} = -\frac{8}{15}\vec{k}^4 ~~~&,&~~~~
\mcN^{(\zeta)}_\text{Dirac} = -\frac{8}{9}\SB{m^2+(3\theta-1)\omega_k^2}^2~. \label{eq.Nv.dirac}
\end{eqnarray}

It is now easy to check that, substitution of Eqs.~\eqref{eq.Nv.scalar} and \eqref{eq.Nv.dirac} into Eq.~\eqref{eq.v.1} yields the well known expressions of the shear and bulk viscosities for the system of scalar Bosons and system of Dirac Fermions~\cite{Ghosh:2014yea,G_etaN,Nicola,Lang,Jeon}:
\begin{eqnarray}
\eta_\text{Scalar} &=& \frac{2}{15T}\int\!\!\!\frac{d^3k}{(2\pi)^3} \frac{\vec{k}^4}{\omega_k^2\Gamma} f(\omega_k)\SB{1+f(\omega_k)}, 
\label{eta_B0S} \\
\eta_\text{Dirac} &=& \frac{4}{15T}\int\!\!\!\frac{d^3k}{(2\pi)^3} \frac{\vec{k}^4}{\omega_k^2\Gamma} \ftil(\omega_k)\SB{1-\ftil(\omega_k)}, 
\label{eta_B0D} \\
\zeta_\text{Scalar} &=& \frac{2}{9T}\int\!\!\!\frac{d^3k}{(2\pi)^3} \frac{1}{\omega_k^2\Gamma}
\SB{m^2+(3\theta-1)\omega_k^2}^2 f(\omega_k)\SB{1+f(\omega_k)}, 
\label{zeta_B0S} \\
\zeta_\text{Dirac} &=& \frac{4}{9T}\int\!\!\!\frac{d^3k}{(2\pi)^3} \frac{1}{\omega_k^2\Gamma}
\SB{m^2+(3\theta-1)\omega_k^2}^2 \ftil(\omega_k)\SB{1-\ftil(\omega_k)}.
\label{zeta_B0D}
\end{eqnarray}
Above expressions of shear and bulk viscosity for scalar and Dirac system from Kubo framework~\cite{Ghosh:2014yea,G_etaN,Nicola,Lang,Jeon}
are exactly identical to same obtained using the RTA in kinetic theory formalism~\cite{Gavin,Kapusta}.

Let us now switch on the external magnetic field. The main difference between $B=0$ and $B\neq 0$ pictures of shear and bulk viscosity will start from their macroscopic definition, where the viscosity coefficients have basically appeared as the  proportionality constants between thermodynamical force tensors and gradient tensors. Unlike single gradient tensors for $\eta$ and $\zeta$ at $B=0$, one can get five (trace-less) and two (non-zero trace) independent gradient tensors, for which five shear viscosity coefficients $\eta_n$ ($n=0,1,2,3,4$) and two bulk viscosity coefficients $\zeta_{\perp,\parallel}$ will be introduced in $B\neq 0$ picture. As shown in Ref.~\cite{Landau_10} by Landau, in the presence of external magnetic field, there will be seven viscous coefficients (in Landau's notation, they are: $\eta, \zeta, \eta_1, \eta_2, \eta_3, \eta_4$ and $\zeta_1$). The viscosity coefficients appear as the expansion coefficients of the tensorial decomposition (in a particular basis) of the viscous stress tensor. However, the choice of the tensor basis to decompose the viscous stress tensor is not unique. Apart from Ref.~\cite{Landau_10}, in Refs.~\cite{XGH1,XGH2}, Huang \textit{et. al.} have constructed a different tensor basis and thus obtained the seven viscosity coefficients (denoted as: $\eta_0, \eta_1, \eta_2, \eta_3, \eta_4, \zeta_\parallel$ and $\zeta_\perp$ in their notation). In a more recent work~\cite{Hongo:2020qpv}, the authors have constructed a more general tensor basis and thus obtained seven independent transport coefficients namely the two electrical resistivities ($\rho_\parallel$ and $\rho_\perp$) and five viscosities ($\eta_\parallel, \eta_\perp, \zeta_\parallel, \zeta_\perp$ and $\zeta_\times = \zeta'_\times$ ). From a physical point of view, it is understood that, the viscous coefficients defined in Ref.~\cite{Landau_10}, Refs.~\cite{XGH1,XGH2} and Ref.~\cite{Hongo:2020qpv} must be inter-connected and they can be expressed in terms of one another. In the RTA based kinetic theory or moment methods, Refs.~\cite{Tuchin_s,Asutosh_s,NJLB_s,JD_s,JD_QP_s,Arpan_s,G_HRGB} have used the tensor basis of Ref.~\cite{Landau_10} whereas Refs.~\cite{Denicol_s,JD_s} have used the tensor basis of Refs.~\cite{XGH1,XGH2}. In this work, we have used the tensor decomposition of Refs.~\cite{XGH1,XGH2} and thus obtained the seven viscosity coefficients namely $\eta_0, \eta_1, \eta_2, \eta_3, \eta_4, \zeta_\parallel$ and $\zeta_\perp$ using the Kubo formalism. Exploring the same with the tensor basis of Ref.~\cite{Landau_10} as well as Ref.~\cite{Hongo:2020qpv} might be a very interesting future project.

Let us start with connecting relations between viscous coefficients $\upsilon$ and spectral function~\cite{XGH1,XGH2}:
\begin{eqnarray}
\upsilon = - \xi^{(\upsilon)} \eta_0 + \mathcal{P}^{(\upsilon)}_\mnab \mathcal{S}^\mnab ~~~;~ 
\upsilon\in\{ \eta_0, \eta_1, \eta_2, \eta_3, \eta_4, \zeta_\perp, \zeta_\parallel \}
\label{eq.v.B}
\end{eqnarray}
where $\xi^{(\upsilon)} = \begin{cases}
4/3 ~\text{if}~ \upsilon=\eta_1 \\
1 ~\text{if}~ \upsilon=\eta_2 \\
0 ~\text{otherwise}
\end{cases}$ 
and $\mathcal{P}^{(\upsilon)}_\mnab$ are given by
\begin{eqnarray}
\mathcal{P}^{(\eta_0)}_\mnab &=& \frac{1}{4}
\FB{\Xi^\sigma_\mu \Xi^\rho_\nu - \frac{1}{2}\Xi^{\sigma\rho}\Xi_\munu}
\FB{\Xi_{\sigma\alpha} \Xi_{\rho_\beta} - \frac{1}{2}\Xi_{\sigma\rho}\Xi_\alphabeta} ~,
\label{eq.P.eta0}\\
\mathcal{P}^{(\eta_1)}_\mnab &=& 2\FB{b_\mu b_\nu - \theta~ u_\mu u_\nu} 
\FB{\frac{1}{2}\Xi_\alphabeta + (\theta+\phi)~ u_\alpha u_\beta} \label{eq.P.eta1}~, \\
\mathcal{P}^{(\eta_2)}_\mnab &=& -\frac{1}{2} \Xi^\sigma_\mu b_\nu\Xi_{\sigma\alpha}b_\alpha ~,
\label{eq.P.eta2} \\
\mathcal{P}^{(\eta_3)}_\mnab &=& -\frac{1}{8}
\FB{\Xi^\sigma_\mu \Xi^\rho_\nu - \frac{1}{2}\Xi^{\sigma\rho}\Xi_\munu}b_\sigma^{~\lambda}
\FB{\Xi_{\lambda\alpha} \Xi_{\rho_\beta} - \frac{1}{2}\Xi_{\lambda\rho}\Xi_\alphabeta} 
\label{eq.P.eta3}~, \\
\mathcal{P}^{(\eta_4)}_\mnab &=& \frac{1}{2} b_{\rho\sigma}\Xi^\rho_\mu b_\nu\Xi^\sigma_\alpha b_\beta~, \label{eq.P.eta4}\\
\mathcal{P}^{(\zeta_\perp)}_\mnab &=& \frac{1}{3}\FB{\Delta_\munu + (3\theta+2\phi)~ u_\mu u_\nu } 
\FB{\frac{1}{2}\Xi_\alphabeta + (\theta+\phi)~ u_\alpha u_\beta} \label{eq.P.zeta.per}~, \\
\mathcal{P}^{(\zeta_\parallel)}_\mnab &=& -\frac{1}{3}\FB{\Delta_\munu + (\theta+2\phi)~ u_\mu u_\nu }
\FB{b_\alpha b_\beta - \theta~ u_\alpha u_\beta} \label{eq.P.zeta.pll}
\end{eqnarray}
where, $b^\mu = \frac{1}{2B}\varepsilon^\mnab F_{\nu\alpha}u_\beta$, 
$F^\munu=(\del^\mu A^\nu_\text{ext}-\del^\nu A^\mu_\text{ext})$ is the electromagnetic field strength tensor, $b^\munu = \varepsilon^\mnab b_\alpha u_\beta$, $\Xi^\munu = \Delta^\munu+b^\mu b^\nu$ with the convention of the Levi-Civita tensor $\varepsilon^{0123}=1$. In the LRF, $b^\mu_\text{LRF}\equiv (0,0,0,1)$. In Eqs.~\eqref{eq.P.eta0}-\eqref{eq.P.zeta.pll}, the thermodynamic quantities $\theta = \FB{\dfrac{\del P}{\del\varepsilon}}_B$ and 
$\phi=-B\FB{\dfrac{\del M}{\del\varepsilon}}_B$ in which $M$ is the magnetization of the medium.

Substituting Eq.~\eqref{eq.S.B} into Eq.~\eqref{eq.v.B}, we obtain the viscous coefficients in presence of constant external magnetic field as
\begin{eqnarray}
\upsilon = \xi^{(\upsilon)}\eta_0 + \sum_{l=0}^{\infty} \sum_{n=0}^{\infty}\frac{1}{T} 
\int\!\!\!\frac{d^3k}{(2\pi)^3} \frac{1}{4\omega_{kl}\omega_{kn}} 
\frac{\Gamma}{(\omega_{kl}-\omega_{kn})^2+\Gamma^2}
\SB{af_a(\omega_{kl}) + af_a(\omega_{kn})+ 2f_a(\omega_{kl})f_a(\omega_{kn})} 
\mcN_{ln}^{(\upsilon)} (\vec{k})
\label{eq.v.B.1}
\end{eqnarray}
where, 
\begin{eqnarray}
\mcN_{ln}^{(\upsilon)}(\vec{k}) = \frac{1}{2}\mathcal{P}^{(\upsilon)}_\mnab 
\SB{\mathcal{N}_{ln}^\mnab(k,k)\big|_{k^0=\omega_{kl}} + \mathcal{N}_{ln}^\mnab(k,k)\big|_{k^0=-\omega_{kl}}}.
\label{eq.N.v.B}
\end{eqnarray}

Substitution of Eqs.~\eqref{eq.Nkk.scalar.B.final}, \eqref{eq.Nkk.dirac.B.final}, and \eqref{eq.P.eta0}-\eqref{eq.P.zeta.pll} into Eq.~\eqref{eq.N.v.B} 
yields after bit simplifications the following final expressions of $\mcN_{ln}^{(\upsilon)}(\vec{k})$ as:
\begin{eqnarray}
\mcN_{ln;\text{Scalar}}^{(\eta_0)}(\vec{k}) &=& 2 \mathcal{A}_{ln}(\kper^2) \kper^4 \label{eq.N.scalar.eta0}~,\\
\mcN_{ln;\text{Scalar}}^{(\eta_1)}(\vec{k}) &=& -8 \mathcal{A}_{ln}(\kper^2) \big\{ (1-\theta)\wkl^2 + (1+\theta)\kper^2 + (1-\theta)k_z^2 -(1+\theta)m^2 \big\}
\nn \\ &&
\times \big\{  (1-\theta-\phi)\wkl^2 + (\theta+\phi)\kper^2 - (1+\theta+\phi)(k_z^2+m^2) \big\} ~,\\ 
\mcN_{ln;\text{Scalar}}^{(\eta_2)}(\vec{k}) &=&  -8 \mathcal{A}_{ln}(\kper^2) \kper^2 k_z^2 ~,\\
\mcN_{ln;\text{Scalar}}^{(\eta_3)}(\vec{k}) &=&  \mcN_{ln;\text{Scalar}}^{(\eta_4)}(\vec{k}) = 0~,
\end{eqnarray}
\begin{eqnarray}
\mcN_{ln;\text{Scalar}}^{(\zeta_\perp)}(\vec{k}) &=&  \frac{4}{3} \mathcal{A}_{ln}(\kper^2) \big\{  (1-\theta-\phi)\wkl^2 + (\theta+\phi)\kper^2 - (1+\theta+\phi)(k_z^2+m^2)  \big\}
\nn \\ &&
\times \big\{ (3-3\theta-2\phi)\wkl^2 + (1+3\theta+2\phi)\kper^2 - (1+3\theta+2\phi)k_z^2 - (3+3\theta+2\phi)m^2  \big\}~,\\
\mcN_{ln;\text{Scalar}}^{(\zeta_\parallel)}(\vec{k}) &=&  \frac{4}{3} \mathcal{A}_{ln}(\kper^2) \big\{  (1-\theta)\wkl^2 + (1+\theta)\kper^2 + (1-\theta)k_z^2 - (1+\theta) m^2 \big\}
\nn \\ &&
\times \big\{  (3-3\theta-2\phi)\wkl^2 + (1+3\theta+2\phi)\kper^2 - (1+3\theta+2\phi)k_z^2 - (3+3\theta+2\phi)m^2 \big\}~,
\end{eqnarray}
\begin{eqnarray}
\mcN_{ln;\text{Dirac}}^{(\eta_0)}(\vec{k}) &=& 2 \mathcal{D}_{ln}(\kper^2) \kper^2 ( -\wkl^2 +k_z^2 + m^2) ~,\\
\mcN_{ln;\text{Dirac}}^{(\eta_1)}(\vec{k}) &=& 8 \Big[ 8 \mathcal{B}_{ln}(\kper^2) \kper^2 \big\{ -2 \wkl^2 + (1+\theta)(1+2 \theta +2 \phi)\kper^2 + 2 \theta (1+\theta +\phi)  k_z^2 +2 (1+\theta)(1+\theta +\phi)m^2\big\}
\nn \\ &&
+2 \mathcal{C}_{ln}(\kper^2) \big\{ \wkl^4 + \theta (1+\theta +\phi)k_z^4  +(1+\theta)(1+\theta +\phi)m^4  - (1 +3 \theta +2 \phi - \theta \phi - \theta ^2 ) \wkl^2 k_z^2
\nn \\ &&  \hspace{-1cm}
+ (1+2 \theta) (1+\theta +\phi)k_z^2m^2 - (2 +2\theta+\phi - \theta\phi - \theta ^2) \wkl^2 m^2 \big\}
+ \mathcal{D}_{ln}(\kper^2)\kper^2 (1+\theta) (1+2 \theta +2 \phi) (\wkl^2-k_z^2-m^2) \nn \\ &&  
+2 \mathcal{E}_{ln}(\kper^2) \kper^2 \big\{  ( 3+ 4 \theta +2 \phi) \wkl^2
- k_z^2 ( 2 +5 \theta +2 \phi +4 \theta  \phi + 4 \theta ^2) -(1+\theta)(3+4 \theta +4 \phi) m^2 \big\} \Big] ~,\\
\mcN_{ln;\text{Dirac}}^{(\eta_2)}(\vec{k}) &=&  8 \mathcal{B}_{ln}(\kper^2) \kper^4+\mathcal{C}_{ln}(\kper^2) \kper^2 (\wkl^2+k_z^2-m^2)
+2 \mathcal{D}_{ln}(\kper^2) k_z^2 ( \wkl^2 -k_z^2 - m^2)+4 \mathcal{E}_{ln}(\kper^2) \kper^2 k_z^2 ~,\\
\mcN_{ln;\text{Dirac}}^{(\eta_3)}(\vec{k}) &=&  \mcN_{ln;\text{Dirac}}^{(\eta_4)}(\vec{k}) = 0~,
\end{eqnarray}
\begin{eqnarray}
\mcN_{ln;\text{Dirac}}^{(\zeta_\perp)}(\vec{k}) &=& \frac{4}{3} \Big[ 8 \mathcal{B}_{ln}(\kper^2) \kper^2 \big\{ 6 \wkl^2 - (1+2 \theta +2 \phi) (2+3 \theta +2 \phi)\kper^2 
-2 (1+\theta +\phi)(2+3 \theta +2 \phi) k_z^2 \nn \\ &&  
- 2 (1+\theta +\phi +1) (3+3 \theta +2 \phi)m^2\big\}  
+ 2 \mathcal{C}_{ln}(\kper^2) \big\{ -3 \wkl^4 - (1+\theta +\phi)(2+3 \theta +2 \phi)k_z^4 \nn \\ &&  \hspace{-1cm}
- (1+\theta +\phi)(3+3 \theta +2 \phi)m^4   
+ ( 5 + 7\theta + 6 \phi -3 \theta ^2 - 5 \theta \phi  - 2\phi^2 ) \wkl^2 k_z^2   
- (1+\theta +\phi) (5+6 \theta +4 \phi)k_z^2 m^2  \nn \\ &&  \hspace{-1cm}
+  (6 + 6\theta + 5 \phi -3\theta^2 -5 \theta\phi -2 \phi^2) \wkl^2 m^2 \big\} 
+  \mathcal{D}_{ln}(\kper^2) \kper^2 (1+2 \theta +2 \phi) (2+3 \theta +2 \phi)(-\wkl^2+ k_z^2+m^2)
\nn \\ && \hspace{-1.8cm}
+ 2 \mathcal{E}_{ln}(\kper^2) \kper^2 \big\{ -\wkl^2 (7+12 \theta +10 \phi) + (2+3 \theta +2 \phi)(3+4 \theta +4 \phi)k_z^2 
+  (7 +19\theta +16\phi + 12 \theta ^2+  20\theta \phi +  8\phi^2) m^2 \big\} \Big],
\end{eqnarray}
\begin{eqnarray}
\mcN_{ln;\text{Dirac}}^{(\zeta_\parallel)}(\vec{k}) &=& \frac{8}{3} \Big[ 8 \mathcal{B}_{ln}(\kper^2) \kper^2 \big\{ 3 \wkl^2 - (1+\theta)(2+3 \theta +2 \phi)\kper^2 
-\theta (2+3 \theta +2 \phi)k_z^2 -(1+\theta)(3+3 \theta +2 \phi)m^2 \big\}
\nn \\ &&  
- \mathcal{C}_{ln}(\kper^2) \big\{ 3 \wkl^4 + \theta(2+3 \theta +2 \phi)k_z^4  +(1+\theta)(3+3 \theta +2 \phi)m^4  
- \wkl^2 k_z^2 ( 1 +10\theta + 4 \phi -  3 \theta^2  -2 \theta\phi  )
\nn \\ &&  \hspace{-1cm}
+  (1+ 8\theta +2\phi+ 6\theta^2 + 4\theta\phi)k_z^2 m^2 - (6 + 6\theta + 2\phi - 3\theta^2 - 2\theta\phi )\wkl^2m^2 \big\}
+  \mathcal{D}_{ln}(\kper^2)\kper^2 (1+\theta)(2+3 \theta +2 \phi) \nn \\ && \hspace{-1.5cm}
\times (-\wkl^2+k_z^2+m^2)
+ 2 \mathcal{E}_{ln}(\kper^2) \kper^2 \big\{ -\wkl^2 (5+6 \theta +2 \phi) + (1+2 \theta)(2+3 \theta +2 \phi)k_z^2 + (1+\theta)(5+6 \theta +4 \phi) m^2 \big\} \Big]
. \label{eq.N.dirac.zetapll}
\end{eqnarray}
During our entire calculation, $\Gamma$ is introduced as a parameter, although it can be calculated microscopically from the interaction Lagrangian of a particular system and one can get it as temperature ($T$), magnetic field ($B$) and momentum $\vk$ dependent function. By taking appropriate momentum average one can get momentum independent $\Gamma$ and take outside the $d^2\kper$ integral of Eq.~\eqref{eq.v.B.1}. So, considering $\Gamma$ as constant or independent of $\kper$, the $d^2\kper$ integral of Eq.~\eqref{eq.v.B.1} can be analytically performed and we get, the following simplified expressions of the viscous coefficients in presence of constant external magnetic field:
\begin{eqnarray}
\upsilon = \xi^{(\upsilon)}\eta_0 + \sum_{l=0}^{\infty} \sum_{n=0}^{\infty}\frac{1}{T} 
\int_{-\infty}^{\infty}\!\frac{dk_z}{(2\pi)} \frac{1}{4\omega_{kl}\omega_{kn}} 
\frac{\Gamma}{(\omega_{kl}-\omega_{kn})^2+\Gamma^2}
\SB{af_a(\omega_{kl}) + af_a(\omega_{kn})+ 2f_a(\omega_{kl})f_a(\omega_{kn})} 
\tilde{\mcN}_{ln}^{(\upsilon)}(k_z)  
\label{eq.vtil.B.1}
\end{eqnarray}
where, 
\begin{eqnarray}
\tilde{\mcN}_{ln}^{(\upsilon)}(k_z) = \int\!\! \frac{d^2\kper}{(2\pi)^2} \mcN_{ln}^{(\upsilon)}(\vec{k}).
\label{eq.N.vtil.B}
\end{eqnarray}
Substituting Eqs.~\eqref{eq.N.scalar.eta0}-\eqref{eq.N.dirac.zetapll} into Eq.~\eqref{eq.N.vtil.B}, we obtain 
\begin{eqnarray}
\tilde{\mcN}_{ln;\text{Scalar}}^{(\eta_0)}(\vec{k}) &=& 2 \mathcal{A}^{(4)}_{ln} ~,\\
\tilde{\mcN}_{ln;\text{Scalar}}^{(\eta_1)}(\vec{k}) &=&  8 \Big[  \mathcal{A}^{(0)}_{ln} \big\{ (1-\theta)(\wkl^2+k_z^2) - (1+\theta )m^2 \big\}
\big\{ - (1-\theta -\phi) \wkl^2 + (1+\theta +\phi)(k_z^2+m^2)  \big\}
\nn \\ &&  \hspace{-1.5cm}
+ \mathcal{A}^{(2)}_{ln} \big\{ - ( 1+\theta - 2 \theta ^2 - 2 \theta\phi) \wkl^2 (1 +\theta + 2 \theta ^2 +2 \theta\phi )k_z^2 +(1+\theta +1)(1+2\theta +2 \phi) m^2  \big\}
- \mathcal{A}^{(4)}_{ln} (1+\theta)(\theta +\phi ) \Big]~, \\
\tilde{\mcN}_{ln;\text{Scalar}}^{(\eta_2)}(\vec{k}) &=& -8 \mathcal{A}^{(2)}_{ln} k_z^2  ~,\\
\tilde{\mcN}_{ln;\text{Scalar}}^{(\eta_3)}(\vec{k}) &=&  \tilde{\mcN}_{ln;\text{Scalar}}^{(\eta_4)}(\vec{k}) = 0~,
\end{eqnarray}
\begin{eqnarray}
\tilde{\mcN}_{ln;\text{Scalar}}^{(\zeta_\perp)}(\vec{k}) &=& \frac{1}{3} \Big[ 4 \mathcal{A}^{(0)}_{ln} \big\{  (1-\theta -\phi)\wkl^2 - (1+\theta +\phi)(k_z^2+m^2) \big\} 
\big\{  (3-3 \theta -2 \phi)\wkl^2 - (1+3 \theta +2 \phi)k_z^2 -  (3+3 \theta +2 \phi)m^2  \big\}
\nn \\ &&  
+ \mathcal{A}^{(2)}_{ln} \big\{ 4 ( 1 +10\theta +4\phi -6 \theta ^2 - 10\theta\phi - 4\phi^2 )\wkl^2 -4 (1+2 \theta +2 \phi) (1+3 \theta +2 \phi)k_z^2 
\nn \\ &&  
- 4 ( 1 +7\theta +6 \phi + 6\theta^2 + 10 \theta \phi +4 \phi ^2) m^2  \big\}+ 4 \mathcal{A}^{(4)}_{ln} (\theta +\phi ) (1+3 \theta +2 \phi) \Big]~, \\
\tilde{\mcN}_{ln;\text{Scalar}}^{(\zeta_\parallel)}(\vec{k}) &=& \frac{4}{3} \Big[  \mathcal{A}^{(0)}_{ln} \big\{ (1-\theta)(\wkl^2 + k_z^2) - (1+\theta1) m^2\big\}
\big\{  (3-3 \theta -2 \phi)\wkl^2 - (1+3 \theta +2 \phi)k_z^2 - (3+3 \theta +2 \phi)m^2  \big\}
\nn \\ &&  \hspace{-1cm}
+ 2 \mathcal{A}^{(2)}_{ln} \big\{  (2 +\theta - 3\theta^2 -2\theta\phi )\wkl^2 - \theta (1+3 \theta +2 \phi)k_z^2 - (1+\theta)(2+3 \theta +2 \phi) m^2  \big\}
+  \mathcal{A}^{(4)}_{ln} (1+\theta) (1+3 \theta +2 \phi)  \Big]~,
\end{eqnarray}
\begin{eqnarray}
\tilde{\mcN}_{ln;\text{Dirac}}^{(\eta_0)}(\vec{k}) &=& -2 \mathcal{D}^{(2)}_{ln} (\wkl^2-k_z^2-m^2) ~,\\
\tilde{\mcN}_{ln;\text{Dirac}}^{(\eta_1)}(\vec{k}) &=& 8 \Big[ 16 \mathcal{B}^{(2)}_{ln} \big\{ - \wkl^2 + \theta(1+\theta +\phi) k_z^2 + (1+\theta)(1+\theta +\phi) m^2 \big\}
+8 \mathcal{B}^{(4)}_{ln} (1+\theta) (1+2 \theta +2 \phi)
\nn \\ &&  
+2 \mathcal{C}^{(0)}_{ln} \big\{ \wkl^4 + \theta (1+\theta +\phi)k_z^4 +(1+\theta) (1+\theta +\phi)m^4  - ( 1 +3\theta+2 \phi -\theta ^2 - \theta\phi ) \wkl^2 k_z^2
\nn \\ &&    
+  (1+2 \theta)(1+\theta +\phi) k_z^2 m^2  - (2 +2\theta + \phi - \theta ^2 - \theta\phi ) \wkl^2 m^2 \big\}
+ \mathcal{D}^{(2)}_{ln} (1+\theta) (1+2 \theta +2 \phi) (\wkl^2 - k_z^2 -m^2 )
\nn \\ &&  
+2 \mathcal{E}^{(2)}_{ln} \big\{ (3+4 \theta +2 \phi)\wkl^2 - (2 +5 \theta +2 \phi + 4 \theta ^2+4 \theta  \phi )k_z^2  - (1+\theta)(3+4 \theta +4 \phi)m^2  \big\} \Big] ~,\\
\tilde{\mcN}_{ln;\text{Dirac}}^{(\eta_2)}(\vec{k}) &=&  8 \mathcal{B}^{(4)}_{ln}+\mathcal{C}^{(2)}_{ln} (\wkl^2+k_z^2-m^2)
+2 \mathcal{D}^{(0)}_{ln} k_z^2 (\wkl^2 - k_z^2 - m^2) + 4 \mathcal{E}^{(2)}_{ln} k_z^2 ~,\\
\tilde{\mcN}_{ln;\text{Dirac}}^{(\eta_3)}(\vec{k}) &=&  \tilde{\mcN}_{ln;\text{Dirac}}^{(\eta_4)}(\vec{k}) = 0~,
\end{eqnarray}
\begin{eqnarray}
\tilde{\mcN}_{ln;\text{Dirac}}^{(\zeta_\perp)}(\vec{k}) &=& \frac{4}{3} \Big[ 16 \mathcal{B}^{(2)}_{ln} \big\{ 3 \wkl^2 - ( 2 +5\theta +4\phi + 3 \theta ^2 +  5\theta\phi + 2\phi^2)k_z^2 
- (3 +6 \theta +5 \phi +3 \theta ^2 + 5\theta\phi  +2\phi ^2)m^2 \big\}
\nn \\ &&  
-8 \mathcal{B}^{(4)}_{ln} ( 2+7\theta+6 \phi  + 6 \theta ^2+  10\theta \phi +4 \phi ^2)
-2 \mathcal{C}^{(0)}_{ln} \big\{ 3 \wkl^4 + ( 2+5\theta +4\phi  +3\theta ^2+  5\theta\phi + 2\phi^2 ) k_z^4
\nn \\ &&    
+ (3+6 \theta +5 \phi    +3\theta^2 + 5\theta\phi +2\phi^2 )m^4  - ( 5+7\theta +6 \phi -3\theta^2 -5\theta\phi -2\phi^2) \wkl^2k_z^2
\nn \\ && 
+ ( 5+11 \theta +9 \phi +6 \theta ^2+10 \theta\phi +4 \phi ^2) k_z^2 m^2 - (6+6\theta +5\phi -3\theta^2 - 5\theta\phi -2\phi^2 )\wkl^2m^2 \big\}
\nn \\ && 
- \mathcal{D}^{(2)}_{ln} ( 2+7\theta+6 \phi  + 6 \theta^2+ 10\theta\phi  +4\phi^2) (\wkl^2 - k_z^2 - m^2)
-2 \mathcal{E}^{(2)}_{ln} \big\{ \wkl^2 (7+12 \theta +10 \phi) 
\nn \\ && 
- (6+17\theta+14 \phi  +12 \theta ^2+  20\theta\phi  +8 \phi^2)k_z^2  
- m^2 ( 7+19 \theta +16 \phi +12\theta^2+20\theta\phi +8\phi^2 )\big\} \Big] ~,
\end{eqnarray}
\begin{eqnarray}
\tilde{\mcN}_{ln;\text{Dirac}}^{(\zeta_\parallel)}(\vec{k}) &=& \frac{8}{3} \Big[ 8 \mathcal{B}^{(2)}_{ln} \big\{ 3 \wkl^2 - \theta (2+3 \theta +2 \phi) k_z^2 
- (1+\theta)(3+3 \theta +2 \phi)m^2  \big\}
-8 \mathcal{B}^{(4)}_{ln} (1+\theta) (2+3 \theta +2 \phi)
\nn \\ && 
- \mathcal{C}^{(0)}_{ln} \big\{ 3 \wkl^4 + \theta(2+3 \theta +2 \phi) k_z^4  +(1+\theta)(3+3 \theta +2 \phi)m^4  
-  ( 1 +10\theta +4 \phi -3 \theta ^2 - 2\theta\phi ) \wkl^2k_z^2
\nn \\ &&   
+  (1+8\theta +2\phi +6\theta^2 + 4\theta\phi)k_z^2m^2  - (6+6\theta + 2\phi - 3\theta ^2 - 2\theta\phi) \wkl^2m^2 \big\}
-  \mathcal{D}^{(2)}_{ln} (1+\theta) (2+3 \theta +2 \phi) \nn \\ &&  
\times (\wkl^2 - k_z^2 - m^2 )
-2 \mathcal{E}^{(2)}_{ln} \big\{ (5+6 \theta +2 \phi)\wkl^2 - (1+2 \theta)(2+3 \theta +2 \phi)k_z^2 - (1+\theta)(5+6 \theta +4 \phi) m^2   \big\} \Big]~,
\label{eq.tmp}
\end{eqnarray}
where,
\begin{eqnarray}
\mathcal{A}_{ln}^{(j)} &=& \int\!\! \frac{d^2\kper}{(2\pi)^2}\mathcal{A}_{ln}(\kper^2) \FB{\kper^2}^{j/2}, \label{eq.Aln.j}\\
\mathcal{B}_{ln}^{(j)} &=& \int\!\! \frac{d^2\kper}{(2\pi)^2}\mathcal{B}_{ln}(\kper^2) \FB{\kper^2}^{j/2}, \label{eq.Bln.j} \\
\mathcal{C}_{ln}^{(j)} &=& \int\!\! \frac{d^2\kper}{(2\pi)^2}\mathcal{C}_{ln}(\kper^2) \FB{\kper^2}^{j/2}, \label{eq.Cln.j} \\
\mathcal{D}_{ln}^{(j)} &=& \int\!\! \frac{d^2\kper}{(2\pi)^2}\mathcal{D}_{ln}(\kper^2) \FB{\kper^2}^{j/2}, \label{eq.Dln.j} \\
\mathcal{E}_{ln}^{(j)} &=& \int\!\! \frac{d^2\kper}{(2\pi)^2}\mathcal{E}_{ln}(\kper^2) \FB{\kper^2}^{j/2}. \label{eq.Eln.j}
\end{eqnarray}
Exploiting the orthogonality of the Laguerre polynomials present in the functions $\mathcal{A}_{ln}(\kper^2)$, $\mathcal{B}_{ln}(\kper^2)$, $\cdots$,  $\mathcal{E}_{ln}(\kper^2)$, 
the $d^2\kper$ integrals of Eqs.~\eqref{eq.Aln.j}-\eqref{eq.Eln.j} are now performed and the analytic expressions of the quantities $\mathcal{A}^{(j)}_{ln}$, $\mathcal{B}^{(j)}_{ln}$, $\cdots$,  $\mathcal{E}^{(j)}_{ln}$ are listed in Appendix~\ref{app.integral}.

\section{NUMERICAL RESULTS \& DISCUSSIONS}\label{sec.results}
In this section, we will try to explore the numerical outcomes of shear and bulk viscosity components; mainly, their temperature and magnetic field dependent
curves will be our matter of interest. To show numerical results of the viscosities, we have chosen systems consisting of massless particles for simplicity, although the final expressions for the viscous coefficients in Eqs.~\eqref{eq.vtil.B.1}-\eqref{eq.tmp} are valid for massive particles as well. We have also performed a proper numerical consistency check for ensuring the correctness of the analytical expressions as well as the numerical codes as follows: we have taken the numerical limit $B\to 0$ of the viscous coefficients in presence of magnetic field (Eq.~\eqref{eq.vtil.B.1}) and found that, for sufficiently small values of $B$, $\eta_0 \to \eta $, $\eta_1 \to 0$, $\eta_2 \to 0$, $\zeta_\perp \to \zeta$ and $\zeta_\parallel \to \zeta$ where $\eta$ and $\zeta$ are respectively the shear and bulk viscosities in absence of the external magnetic field obtained from Eqs.~\eqref{eta_B0S}-\eqref{zeta_B0D}. This consistency implies that, for sufficiently small values of $B$, a large number of Landau levels contribute to $\upsilon$ in  Eq.~\eqref{eq.vtil.B.1} (the Landau levels become infinitesimally close to each other reaching the continuum), which in turn numerically reproduce the exact continuum results of Eqs.~\eqref{eta_B0S}-\eqref{zeta_B0D}; though it is non-trivial to take an analytic $B\to0$ limit of Eq.~\eqref{eq.vtil.B.1}. For all the numerical results shown in this section, we have considered upto 10,000 Landau levels.
\begin{figure}[ht]
	\begin{center}
		\includegraphics[angle=-90,scale=0.3]{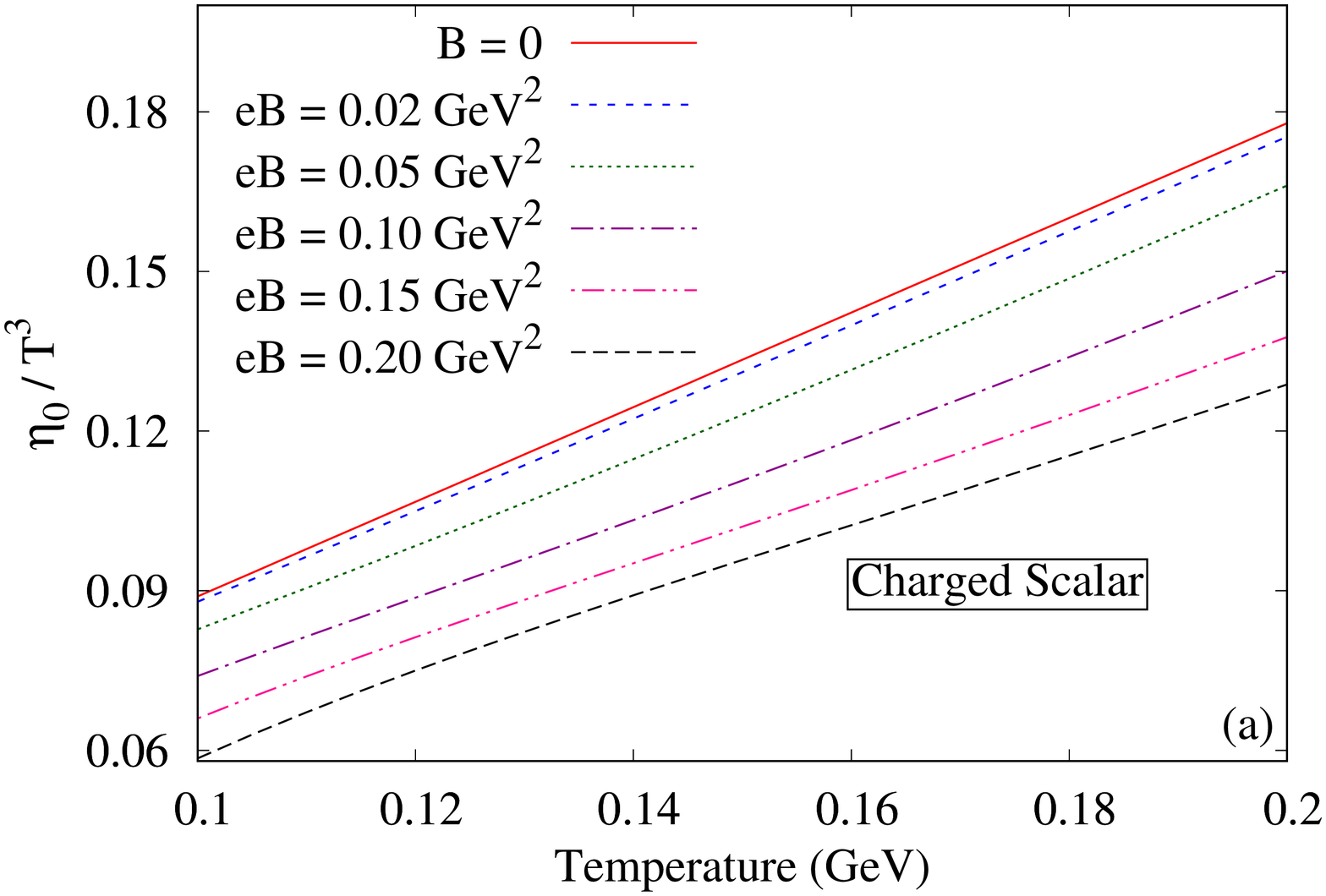}  \includegraphics[angle=-90,scale=0.3]{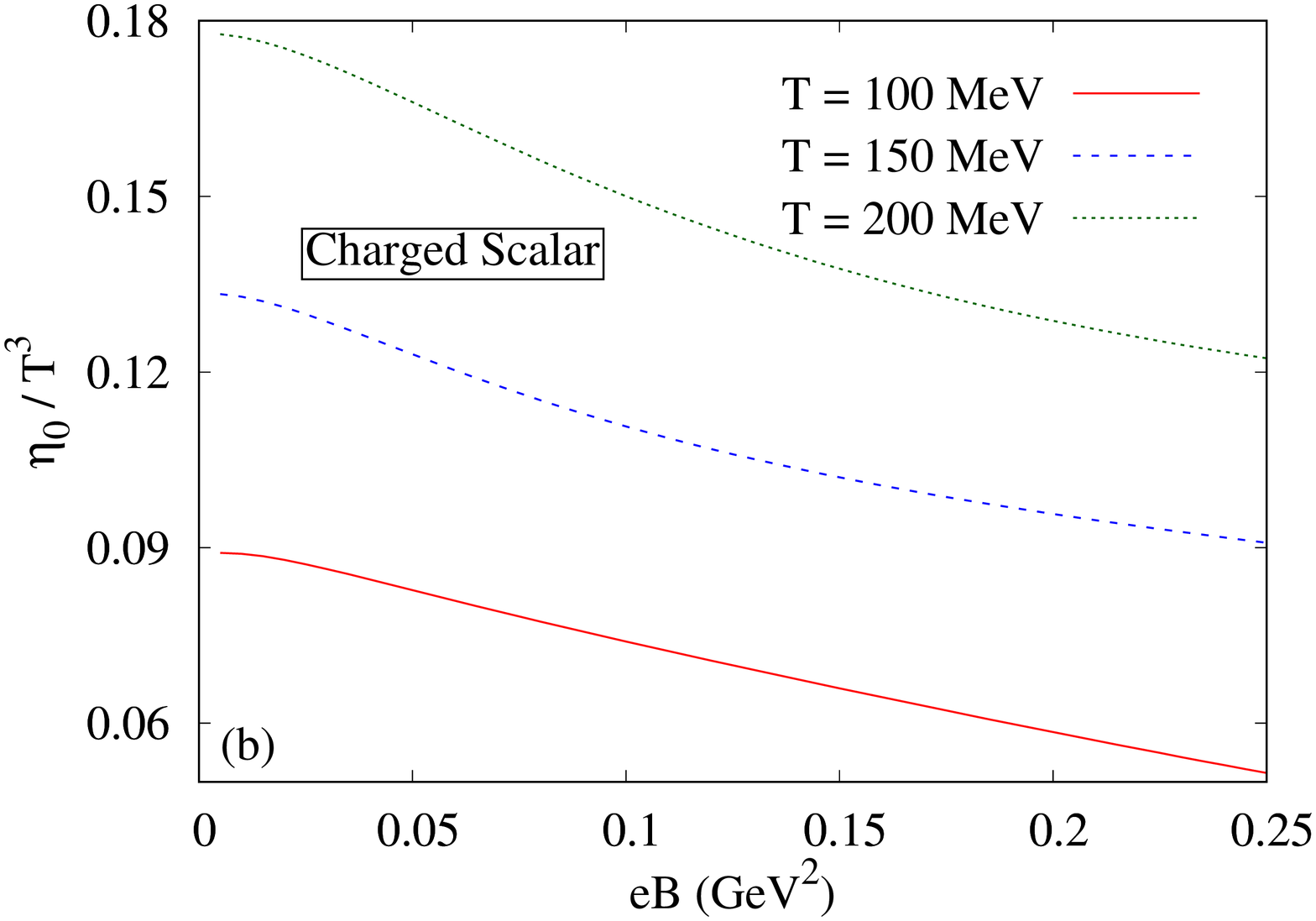}
		\includegraphics[angle=-90,scale=0.3]{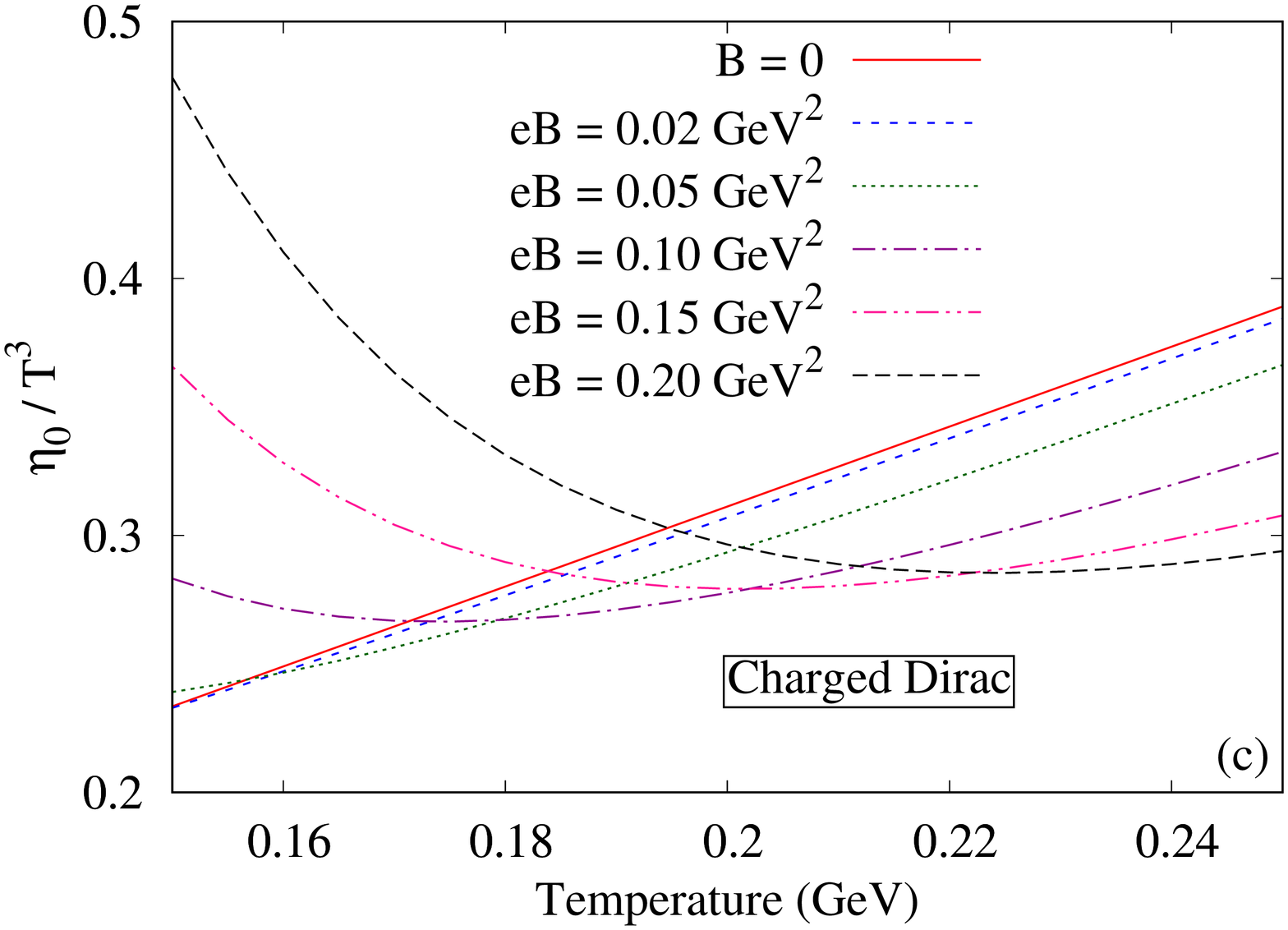}  \includegraphics[angle=-90,scale=0.3]{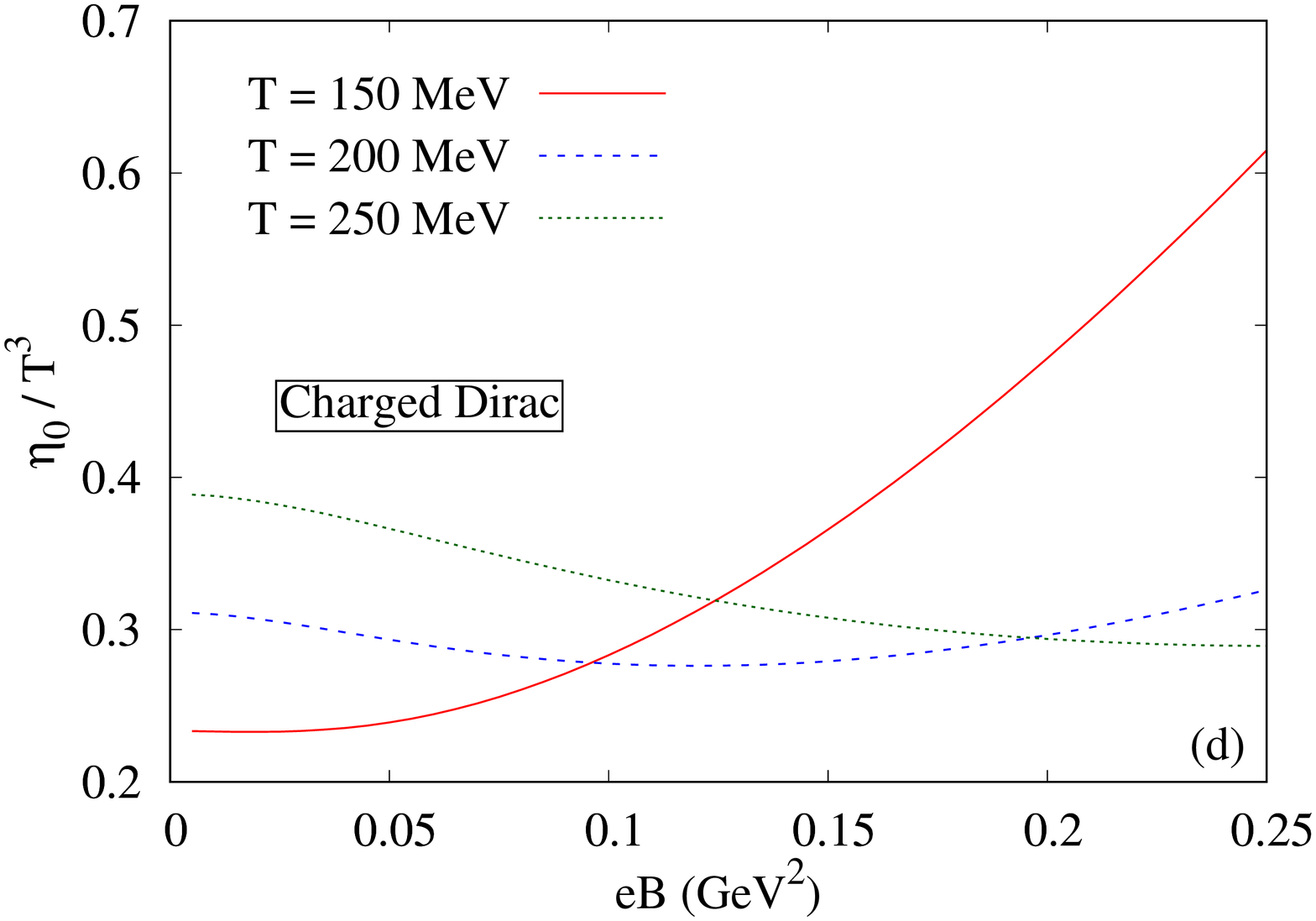}
	\end{center}
	\caption{(Color Online) The variation of $\eta_0/T^3$ as a function of (a) $T$ and (b) $eB$ for system of massless charged scalar Bosons (spin-0) with relaxation time $\tau_c=1/\Gamma=1$ fm. The variation of $\eta_0/T^3$ as a function of (c) $T$ and (d) $eB$ for system of massless charged Dirac Fermions (spin-$\frac{1}{2}$) with relaxation time $\tau_c=1/\Gamma=1$ fm.}
	\label{fig:eta0_TB}
\end{figure}
\begin{figure}[h]
	\begin{center}
		\includegraphics[angle=-90,scale=0.3]{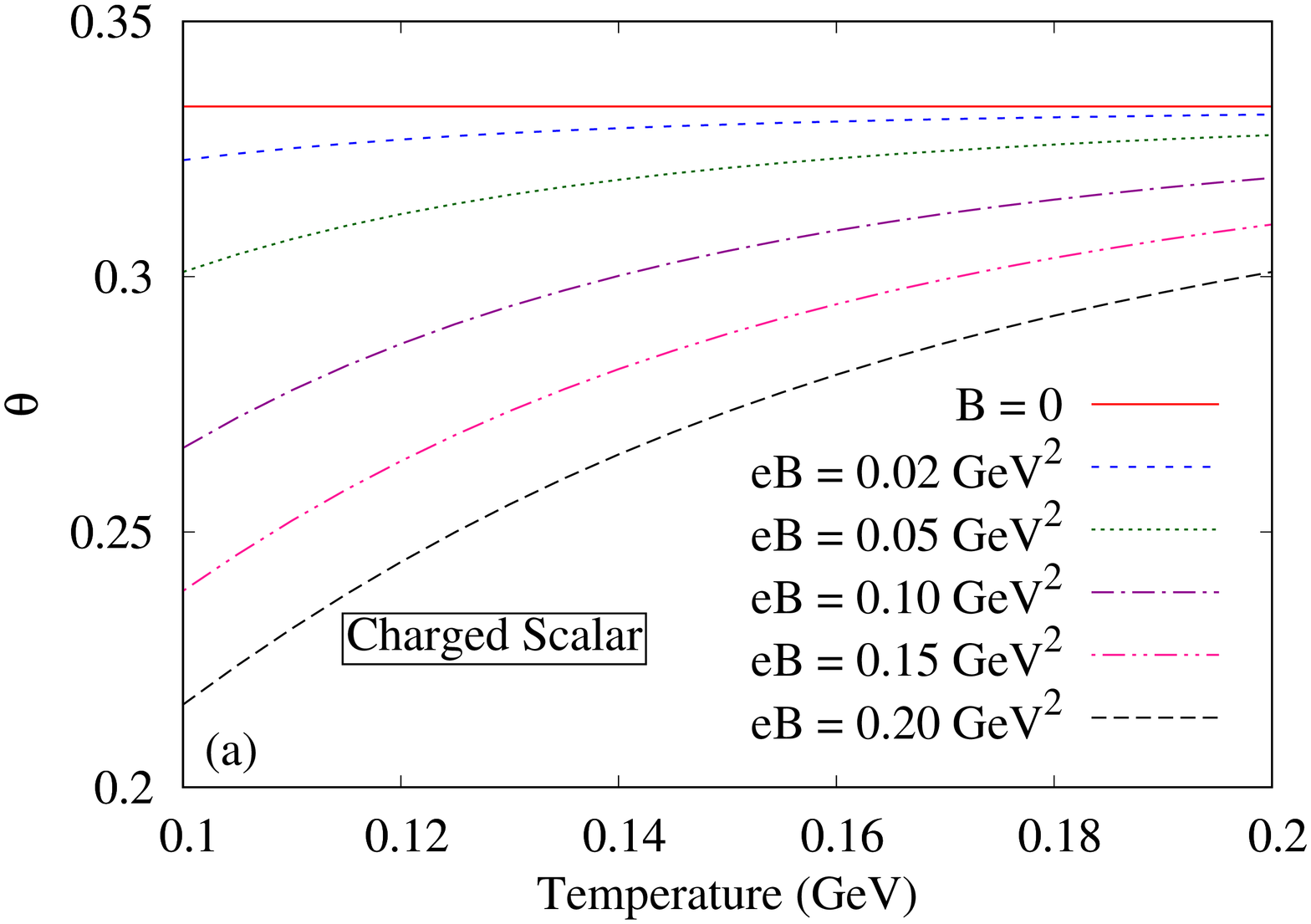}  \includegraphics[angle=-90,scale=0.3]{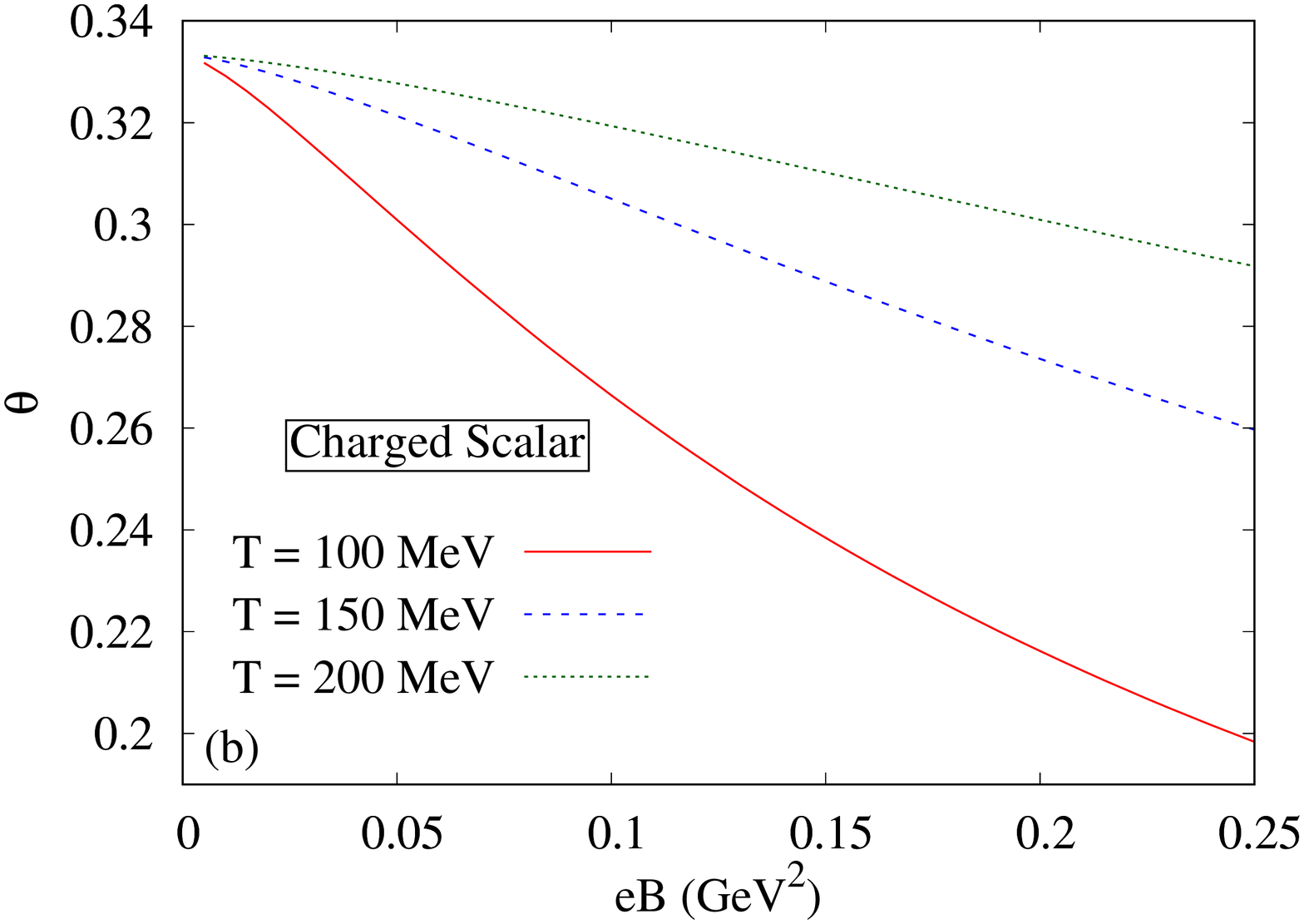}
		\includegraphics[angle=-90,scale=0.3]{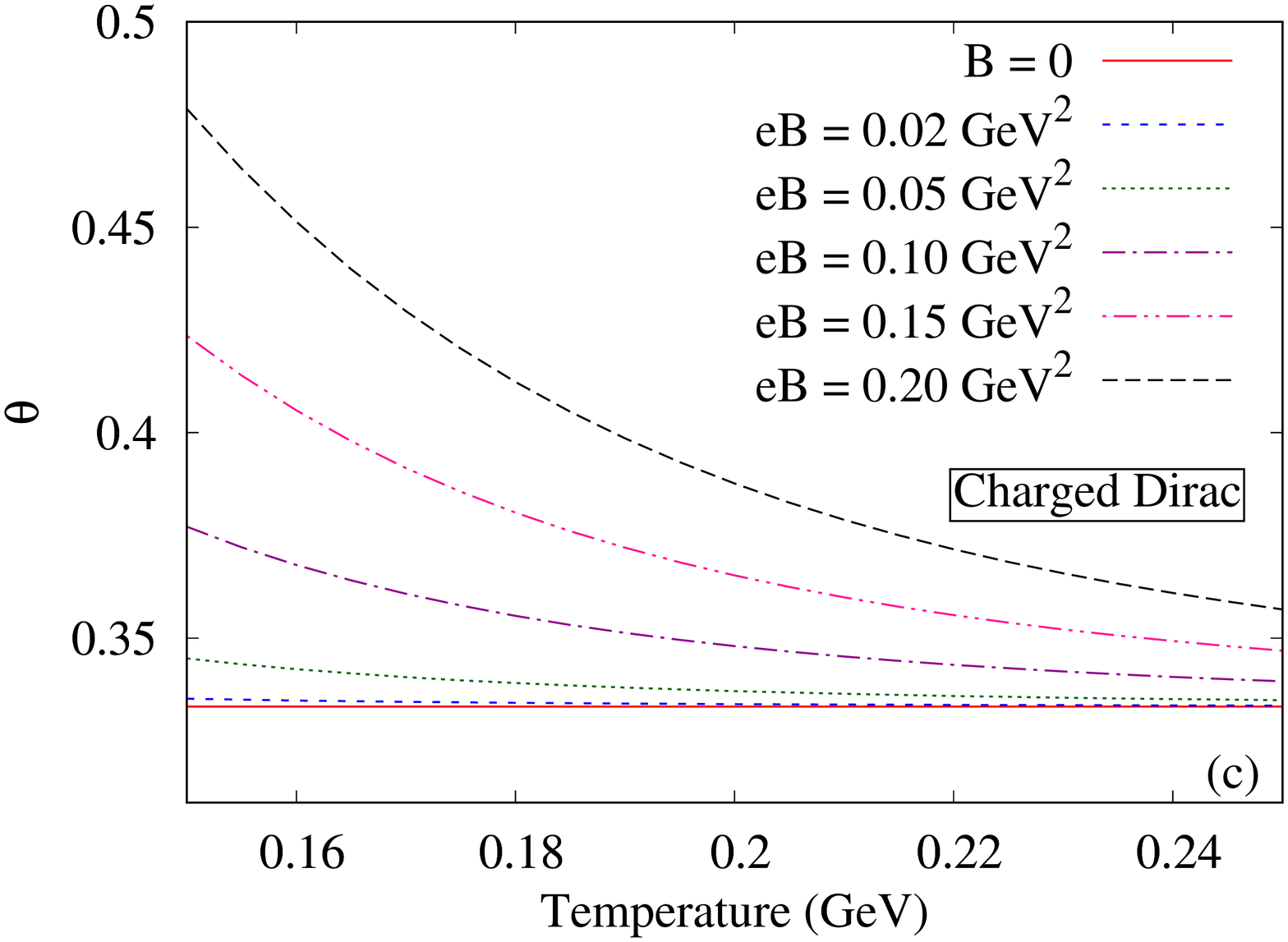}  \includegraphics[angle=-90,scale=0.3]{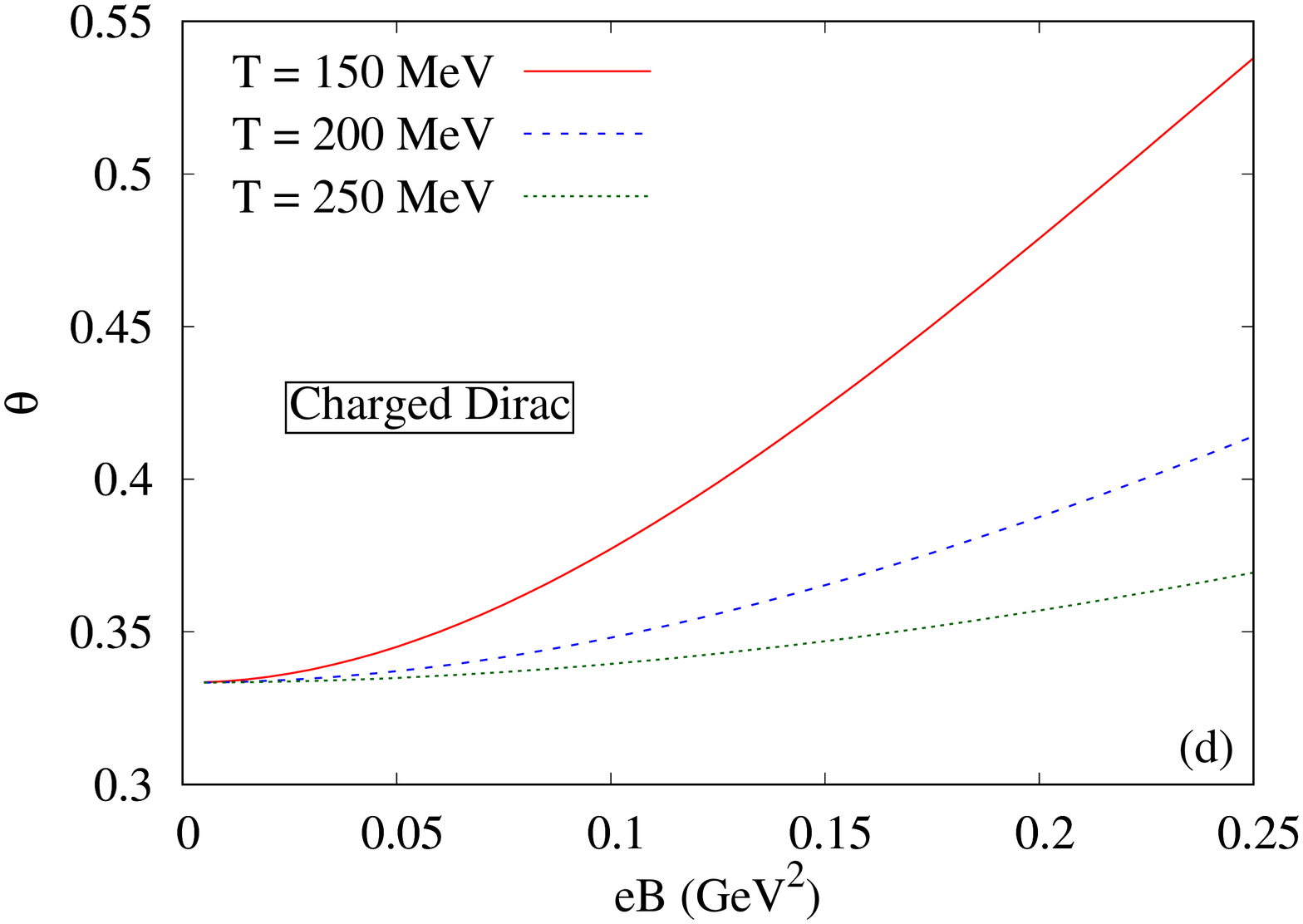}
	\end{center}
	\caption{(Color Online) The variation of $ \theta = \FB{\frac{\del P}{\del \varepsilon}}_B $ as a function of (a) $T$ and (b) $eB$ for system of massless charged scalar Bosons (spin-0). The variation of $ \theta $ as a function of (c) $T$ and (d) $eB$ for system of massless charged Dirac Fermions (spin-$\frac{1}{2}$).}
	\label{fig:theta}
\end{figure}
\begin{figure}[h]
	\begin{center}
		\includegraphics[angle=-90,scale=0.3]{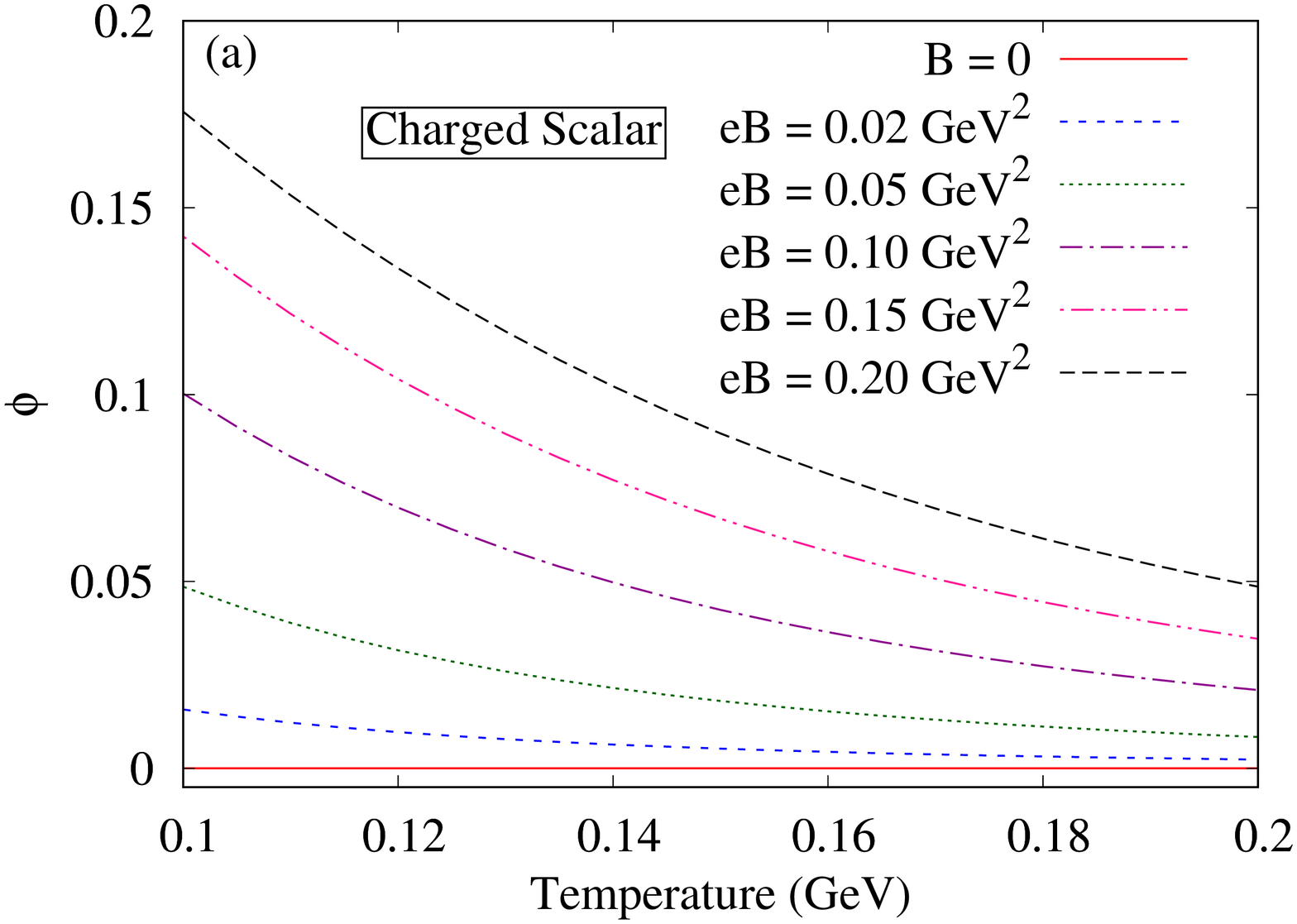}  \includegraphics[angle=-90,scale=0.3]{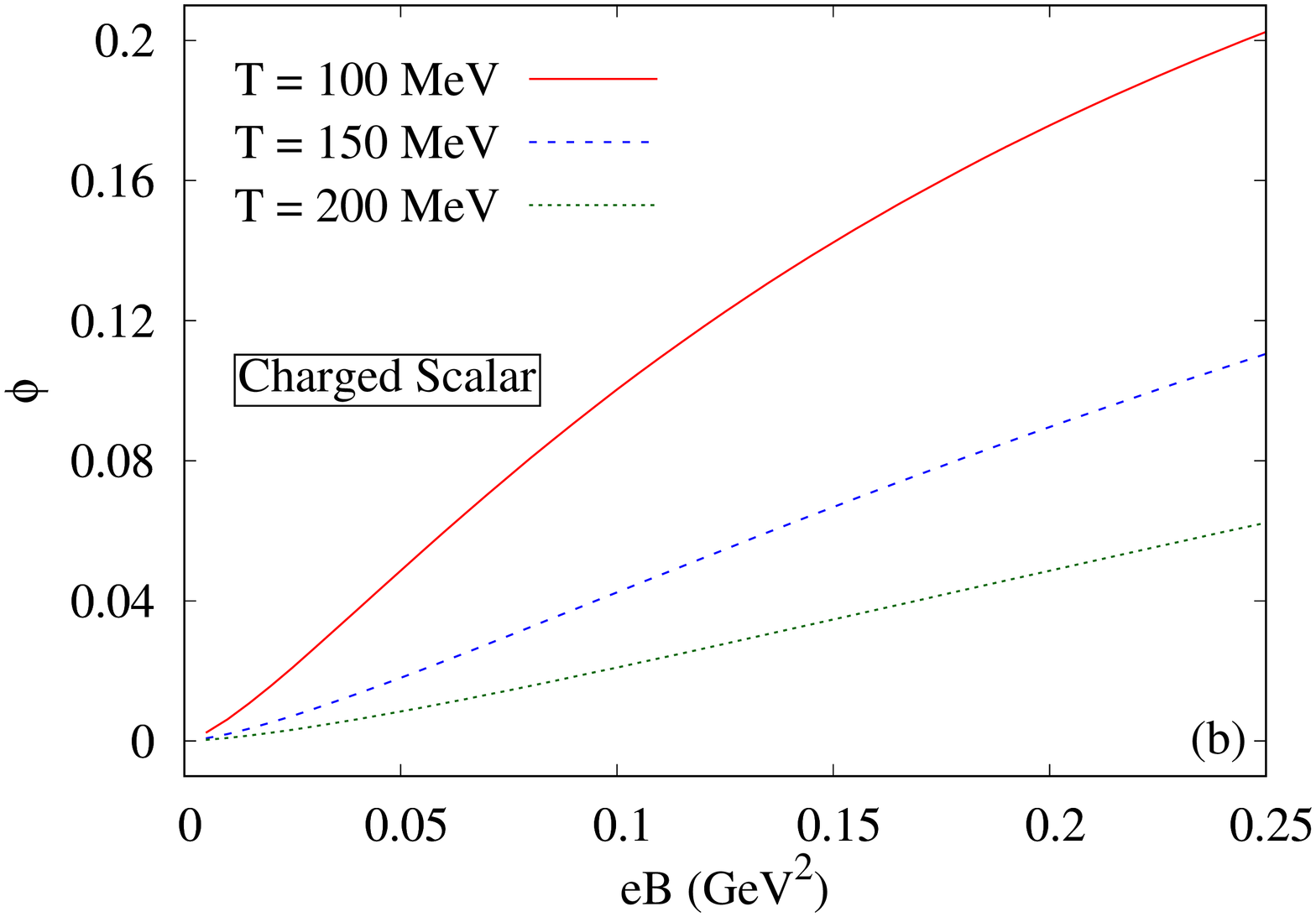}
		\includegraphics[angle=-90,scale=0.3]{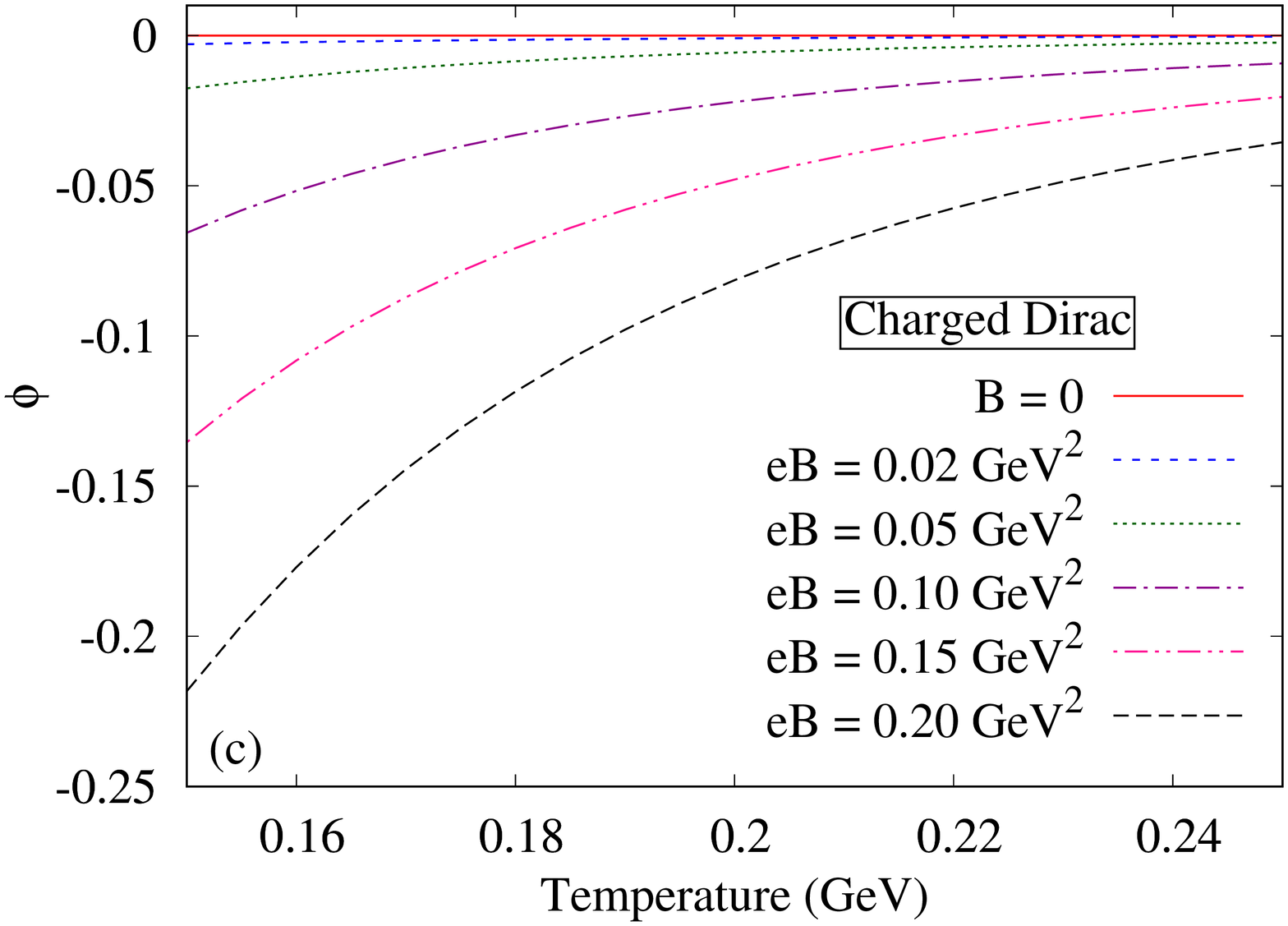}  \includegraphics[angle=-90,scale=0.3]{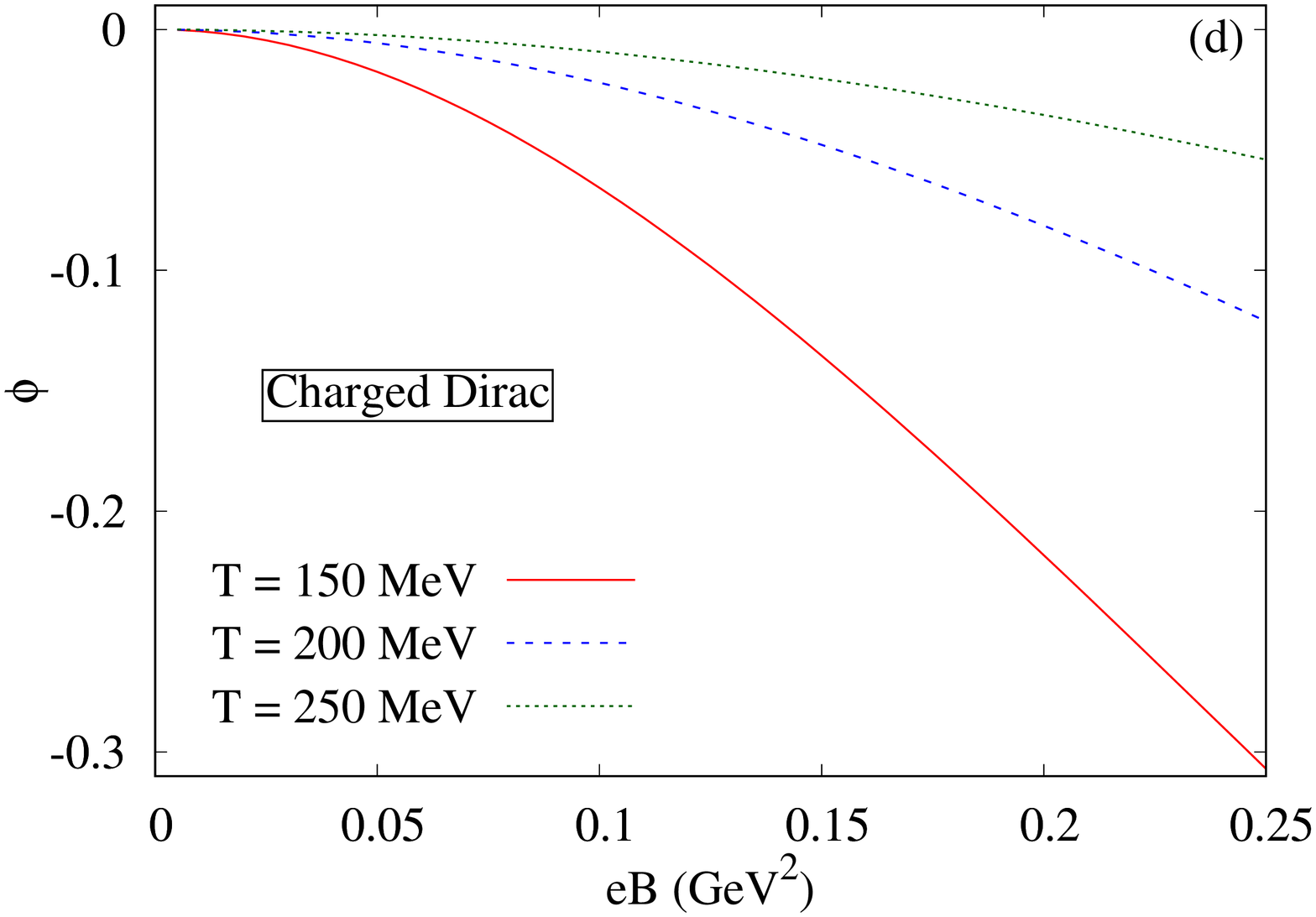}
	\end{center}
	\caption{(Color Online) The variation of $ \phi = -B\FB{\frac{\del M}{\del \varepsilon}}_B $ as a function of (a) $T$ and (b) $eB$ for system of massless charged scalar Bosons (spin-0). The variation of $ \phi $ as a function of (c) $T$ and (d) $eB$ for system of massless charged Dirac Fermions (spin-$\frac{1}{2}$).}
	\label{fig:phi}
\end{figure}
\begin{figure}[h]
	\begin{center}
		\includegraphics[angle=-90,scale=0.3]{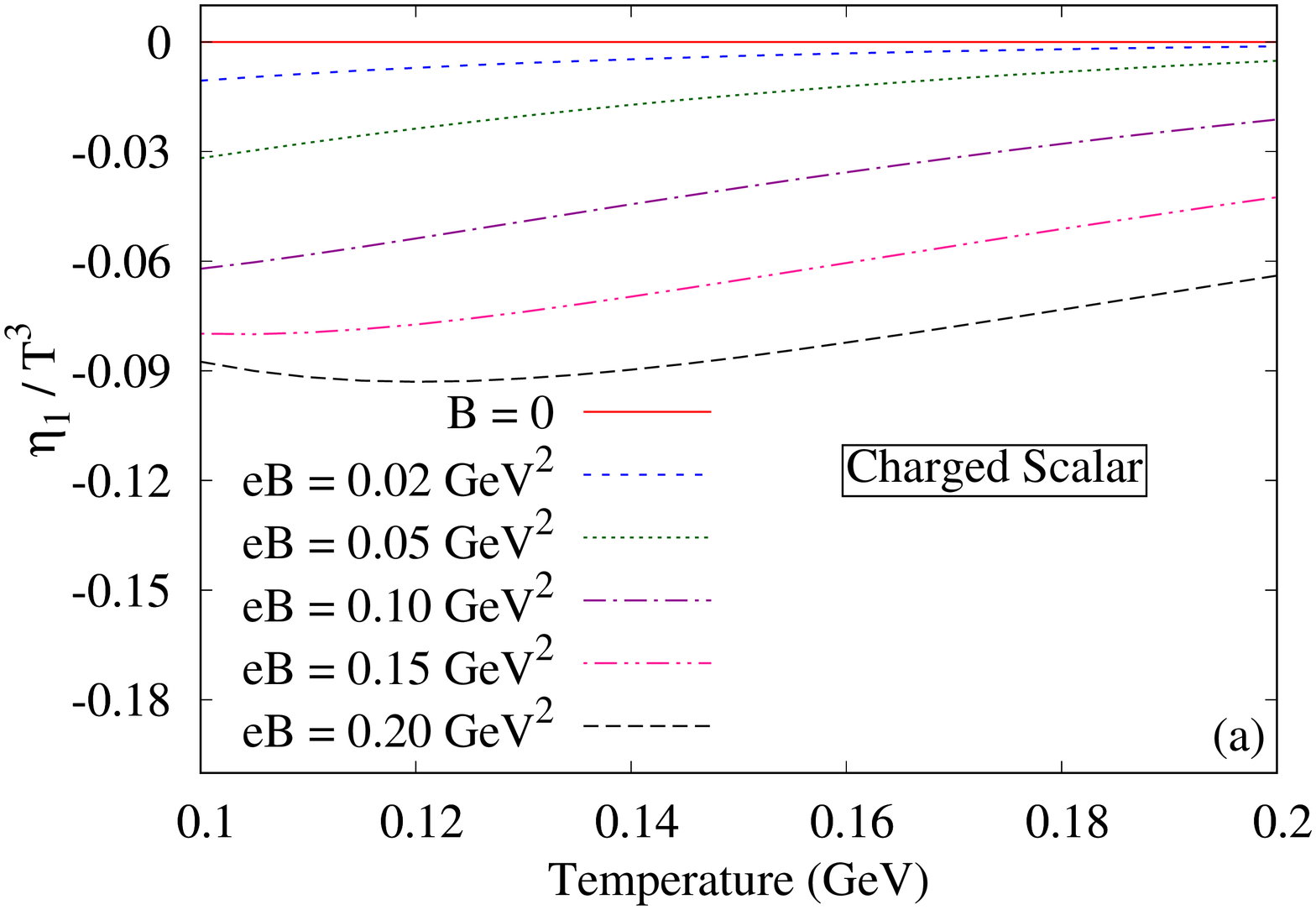}  \includegraphics[angle=-90,scale=0.3]{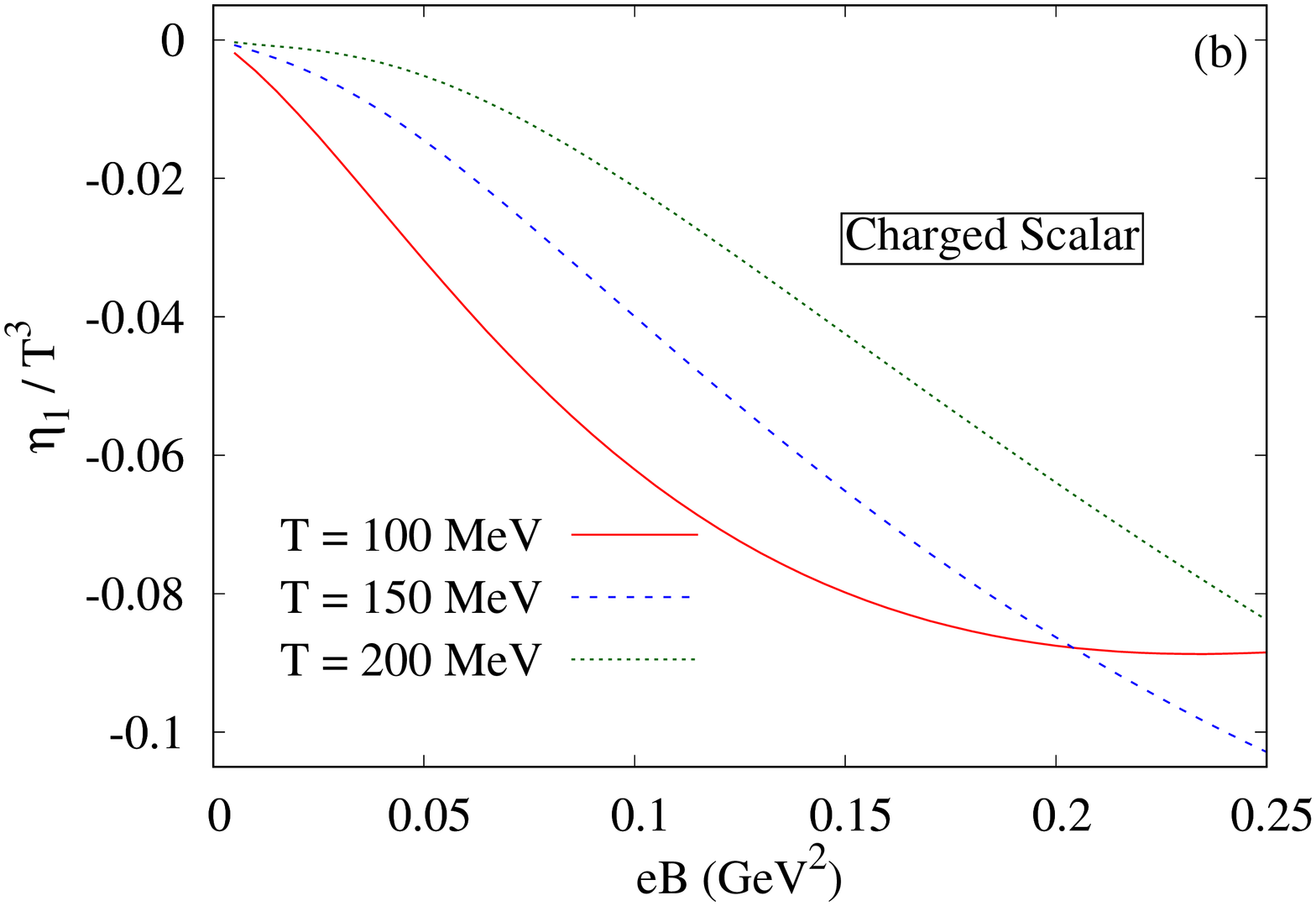}
		\includegraphics[angle=-90,scale=0.3]{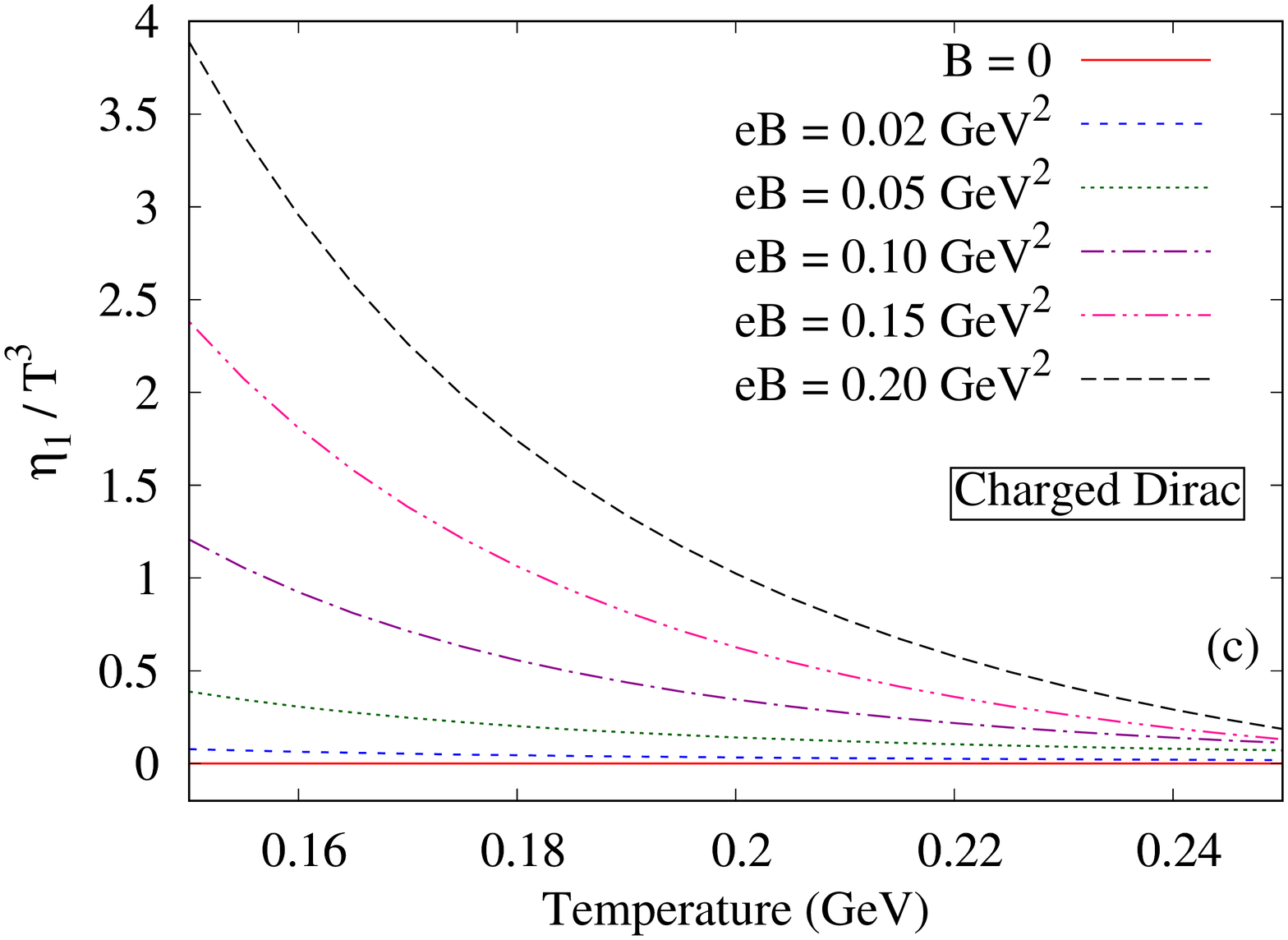}  \includegraphics[angle=-90,scale=0.3]{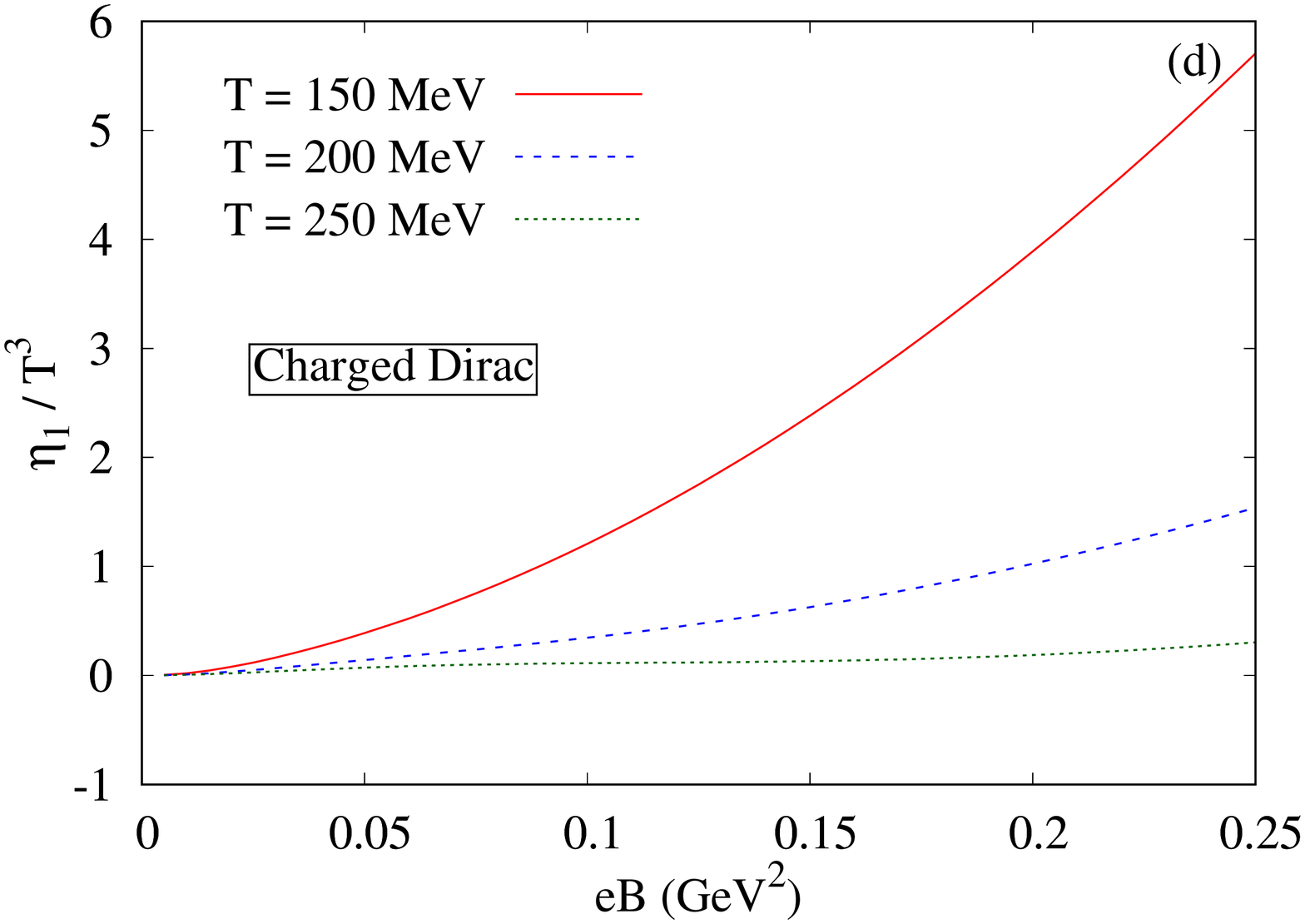}
	\end{center}
	\caption{(Color Online) The variation of $\eta_1/T^3$ as a function of (a) $T$ and (b) $eB$ for system of massless charged scalar Bosons (spin-0) with relaxation time $\tau_c=1/\Gamma=1$ fm. The variation of $\eta_1/T^3$ as a function of (c) $T$ and (d) $eB$ for system of massless charged Dirac Fermions (spin-$\frac{1}{2}$) with relaxation time $\tau_c=1/\Gamma=1$ fm.}
	\label{fig:eta1_TB}
\end{figure}
\begin{figure}[h]
	\begin{center}
		\includegraphics[angle=-90,scale=0.3]{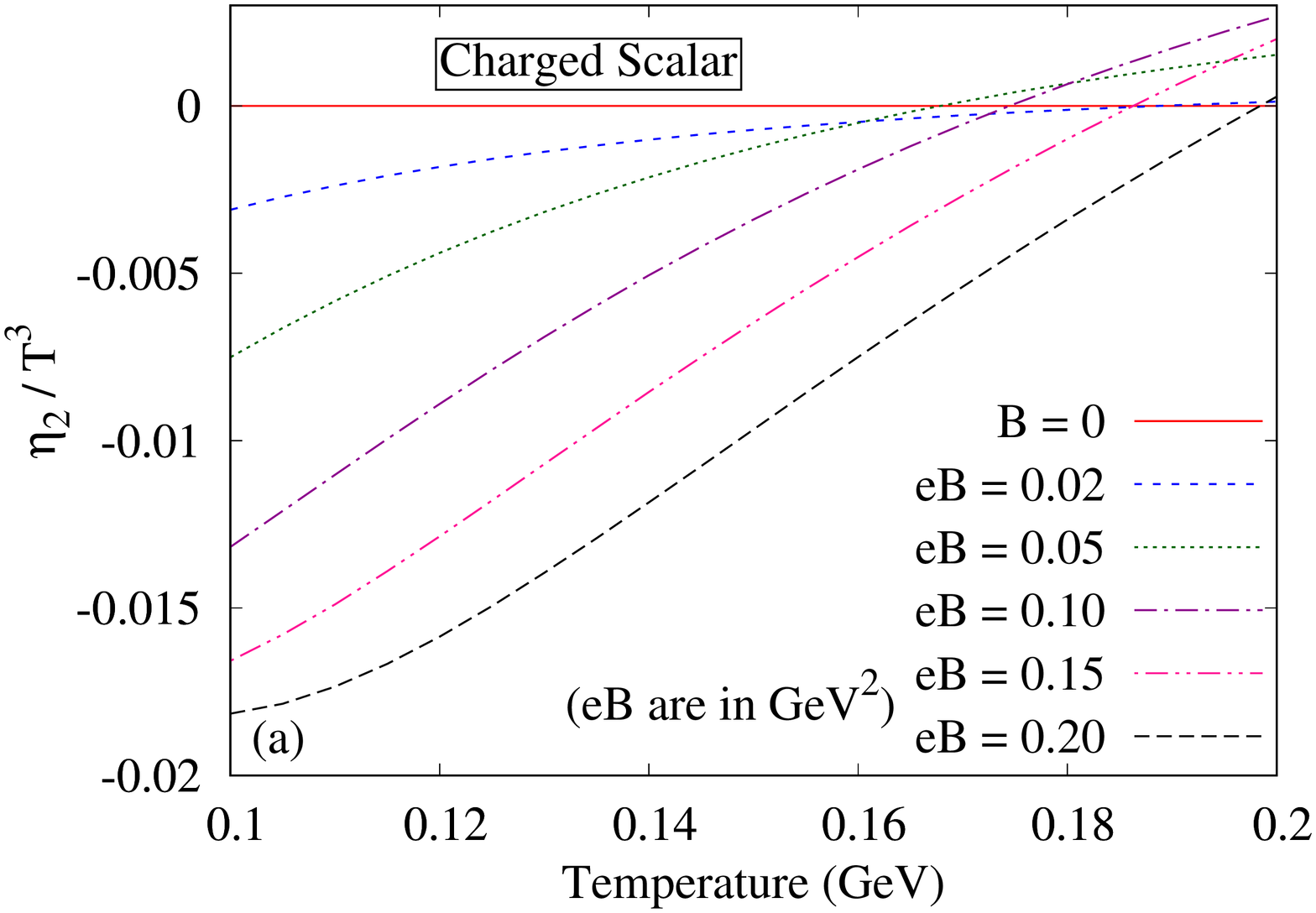}  \includegraphics[angle=-90,scale=0.3]{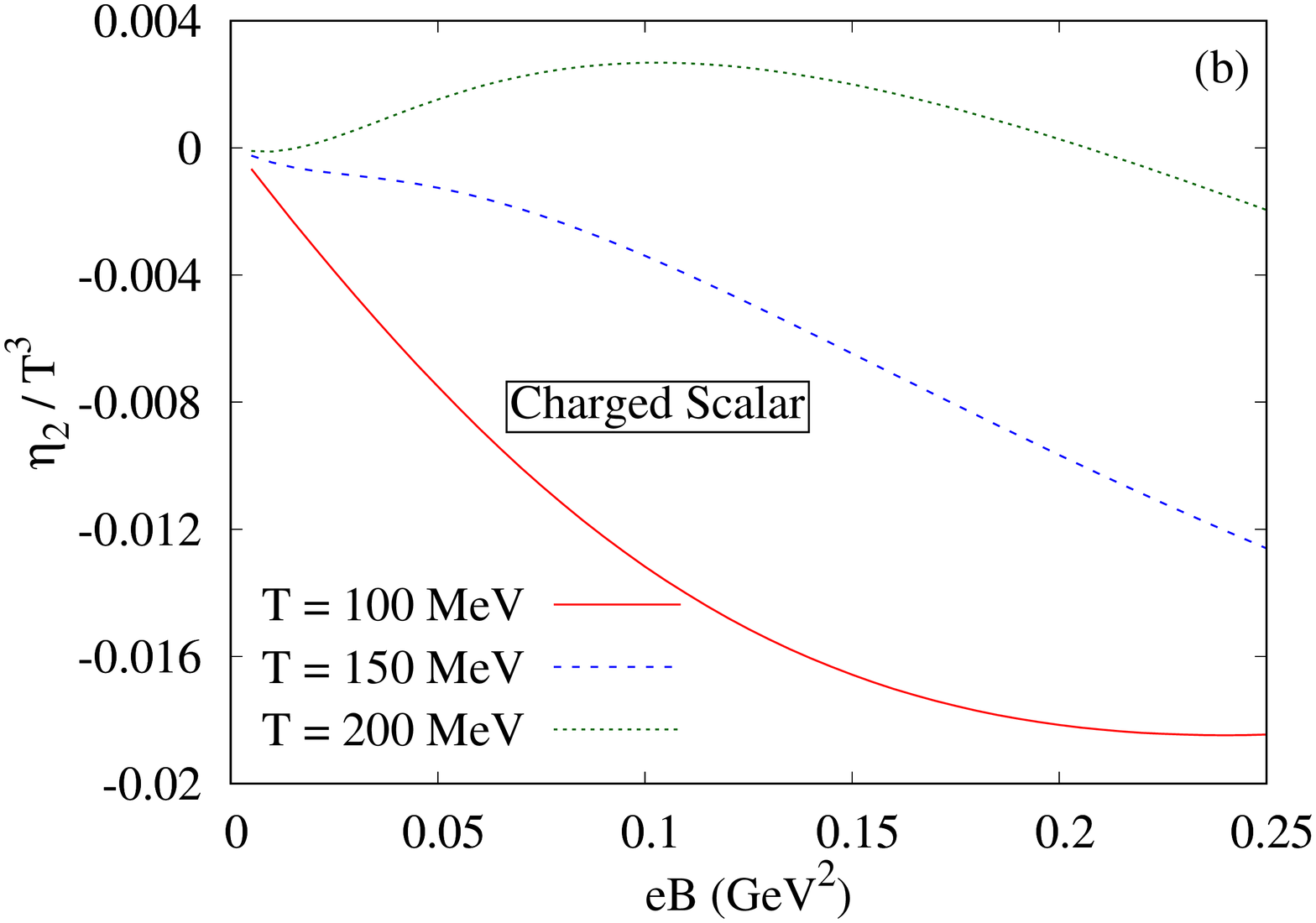}
		\includegraphics[angle=-90,scale=0.3]{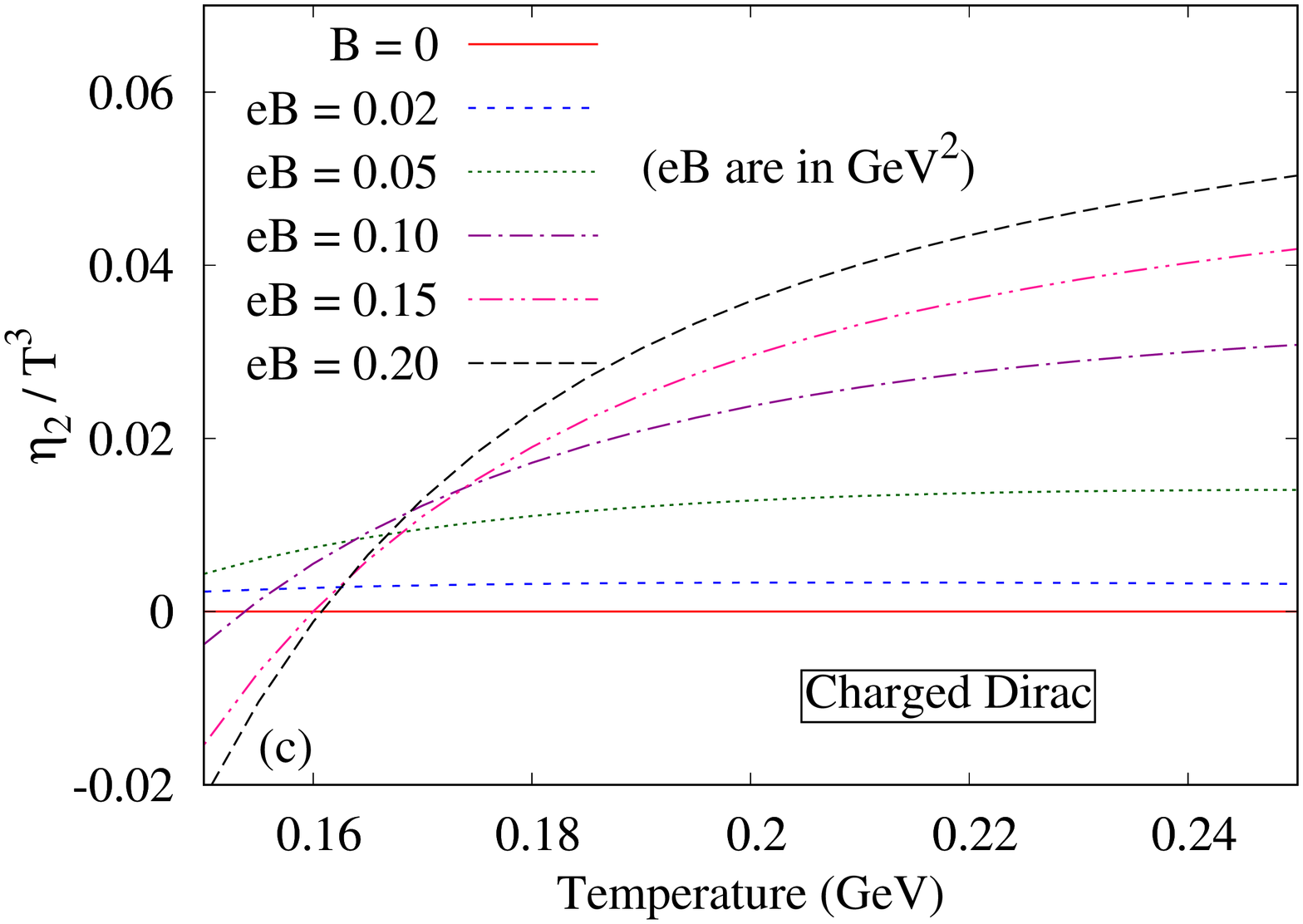}  \includegraphics[angle=-90,scale=0.3]{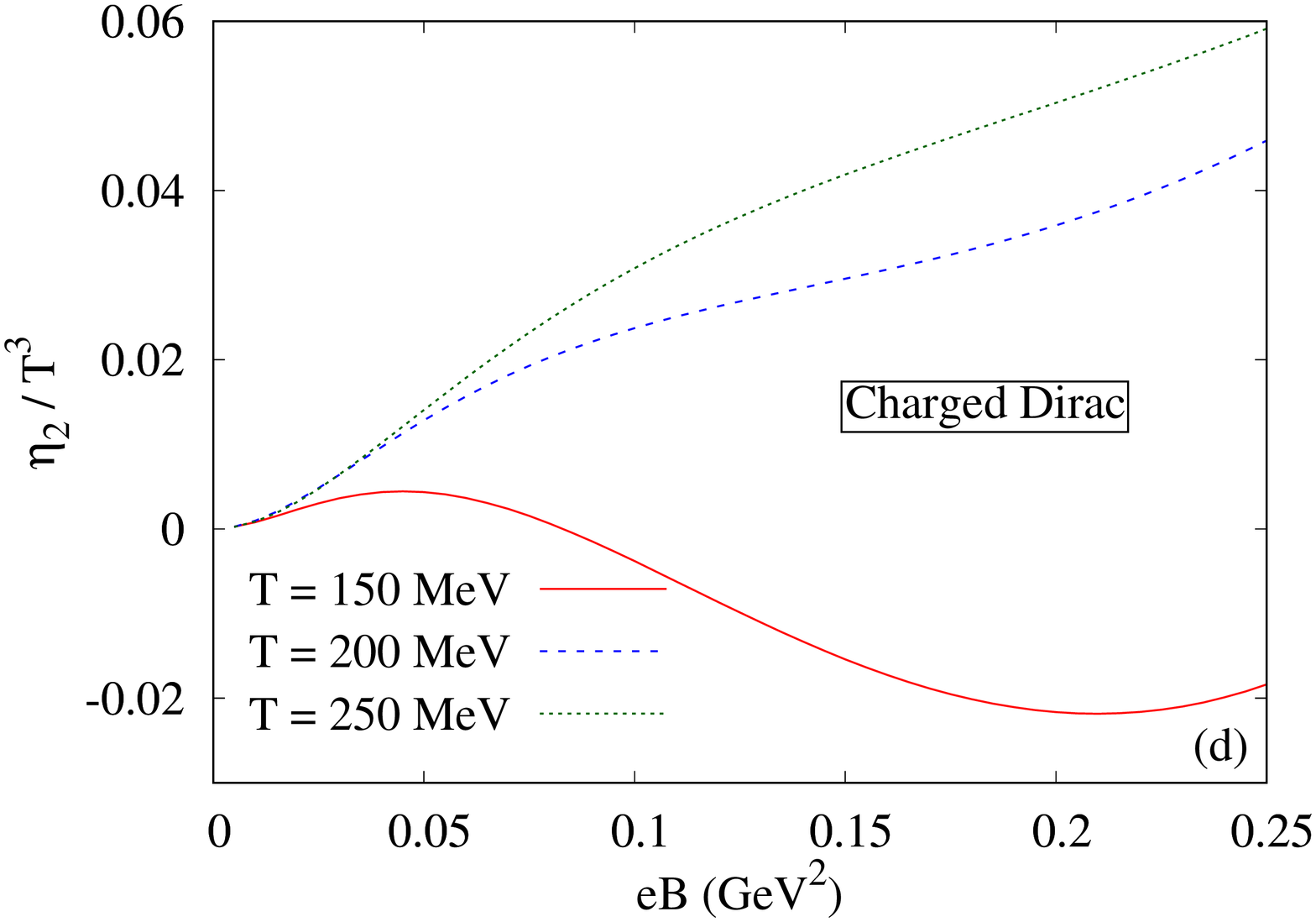}
	\end{center}
	\caption{(Color Online) The variation of $\eta_2/T^3$ as a function of (a) $T$ and (b) $eB$ for system of massless charged scalar Bosons (spin-0) with relaxation time $\tau_c=1/\Gamma=1$ fm. The variation of $\eta_2/T^3$ as a function of (c) $T$ and (d) $eB$ for system of massless charged Dirac Fermions (spin-$\frac{1}{2}$) with relaxation time $\tau_c=1/\Gamma=1$ fm.}
	\label{fig:eta2_TB}
\end{figure}
\begin{figure}[h]
	\begin{center}
		\includegraphics[angle=-90,scale=0.3]{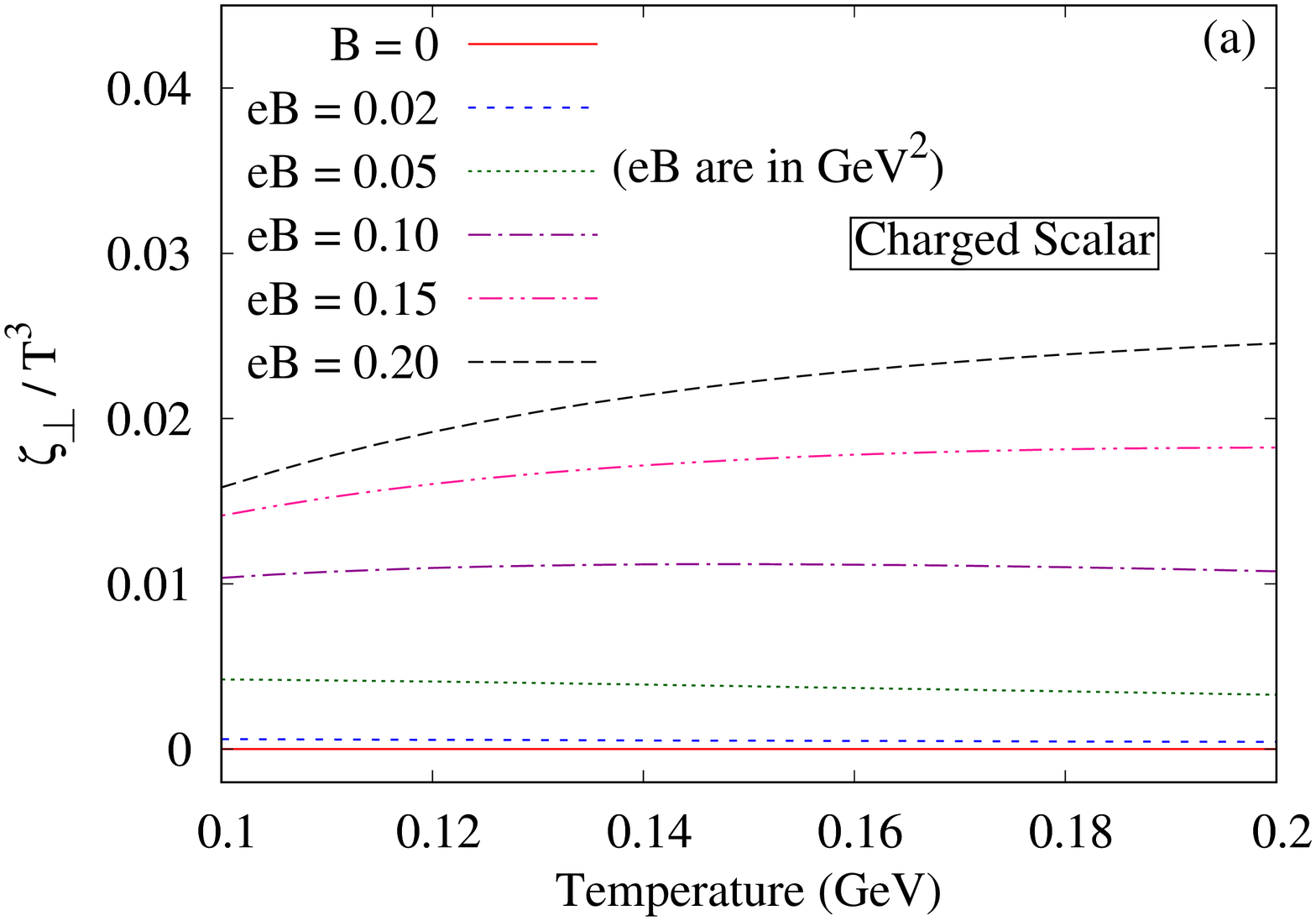}  \includegraphics[angle=-90,scale=0.3]{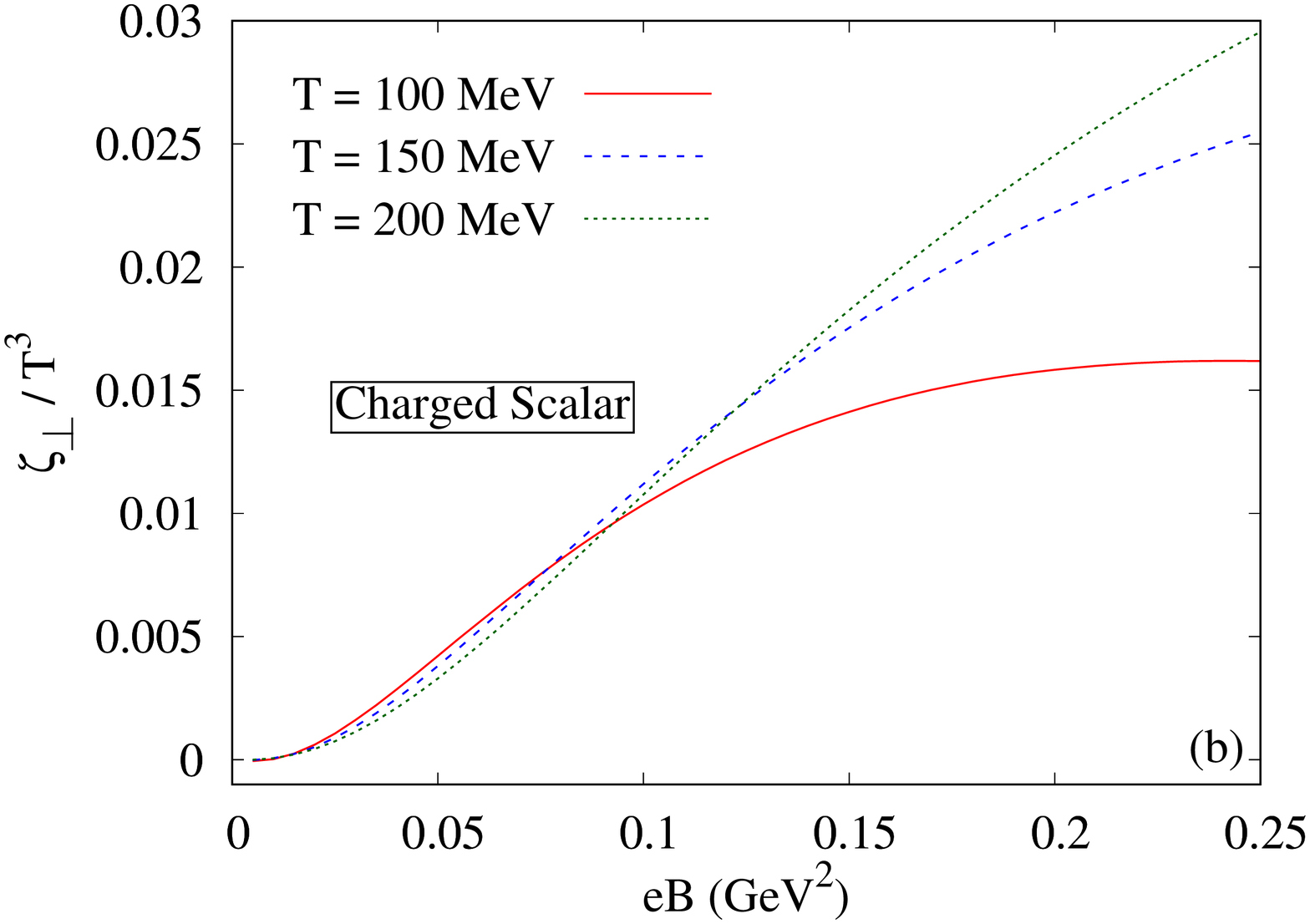}
		\includegraphics[angle=-90,scale=0.3]{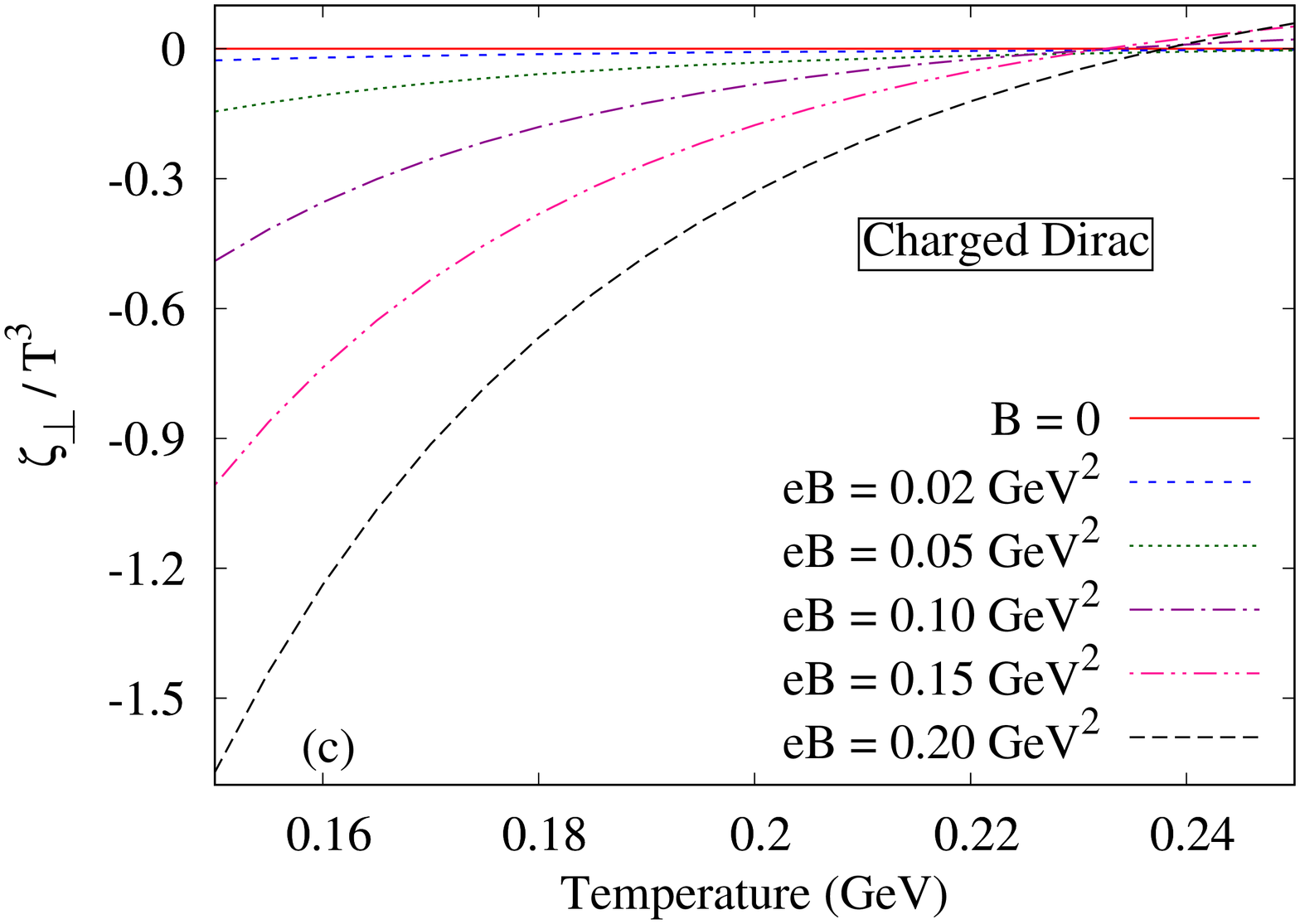}  \includegraphics[angle=-90,scale=0.3]{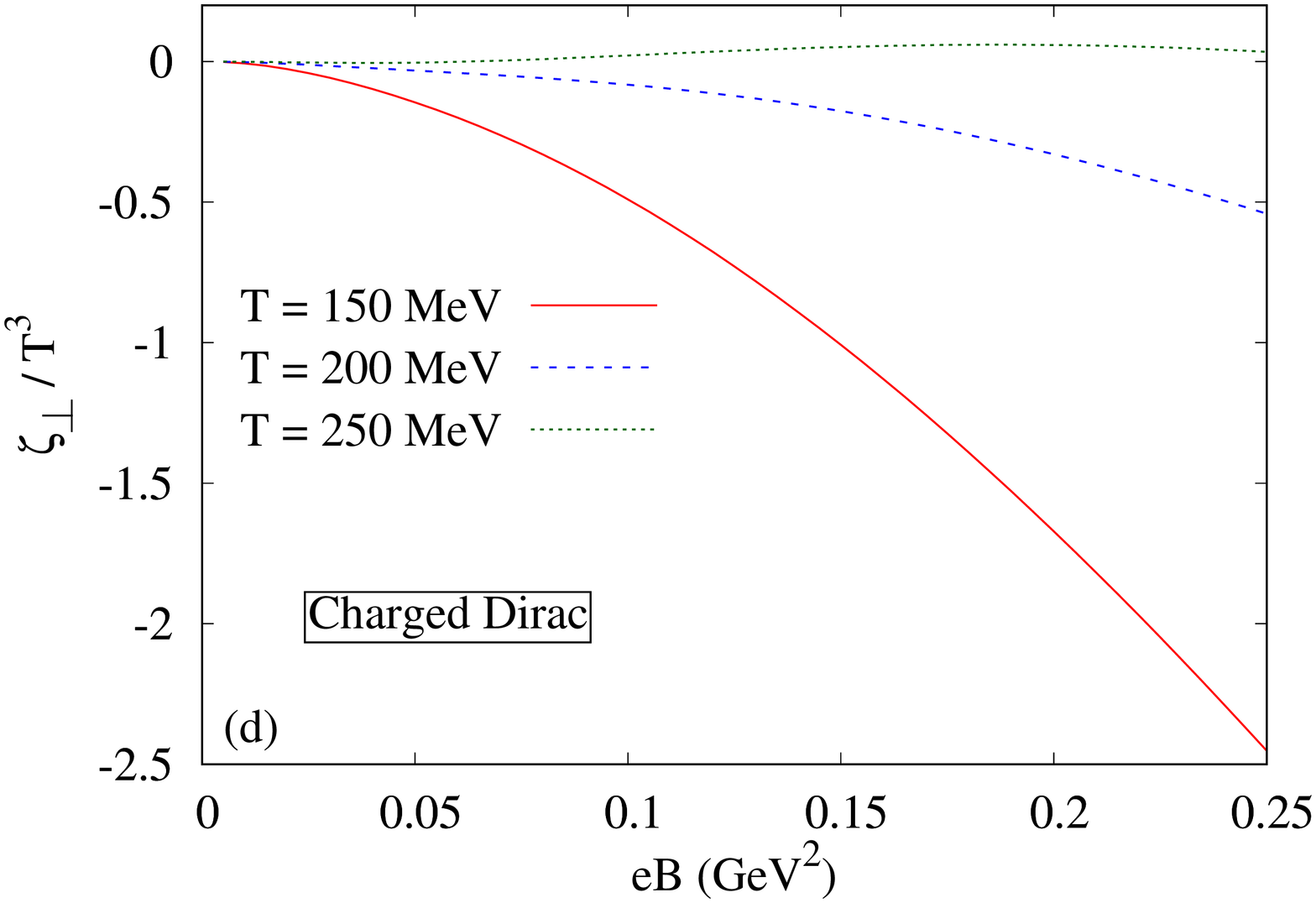}
	\end{center}
	\caption{(Color Online) The variation of $\zeta_\perp/T^3$ as a function of (a) $T$ and (b) $eB$ for system of massless charged scalar Bosons (spin-0) with relaxation time $\tau_c=1/\Gamma=1$ fm. The variation of $\eta_\perp/T^3$ as a function of (c) $T$ and (d) $eB$ for system of massless charged Dirac Fermions (spin-$\frac{1}{2}$) with relaxation time $\tau_c=1/\Gamma=1$ fm.}
	\label{fig:zeta_perp}
\end{figure}
\begin{figure}[h]	
	\begin{center}
		\includegraphics[angle=-90,scale=0.3]{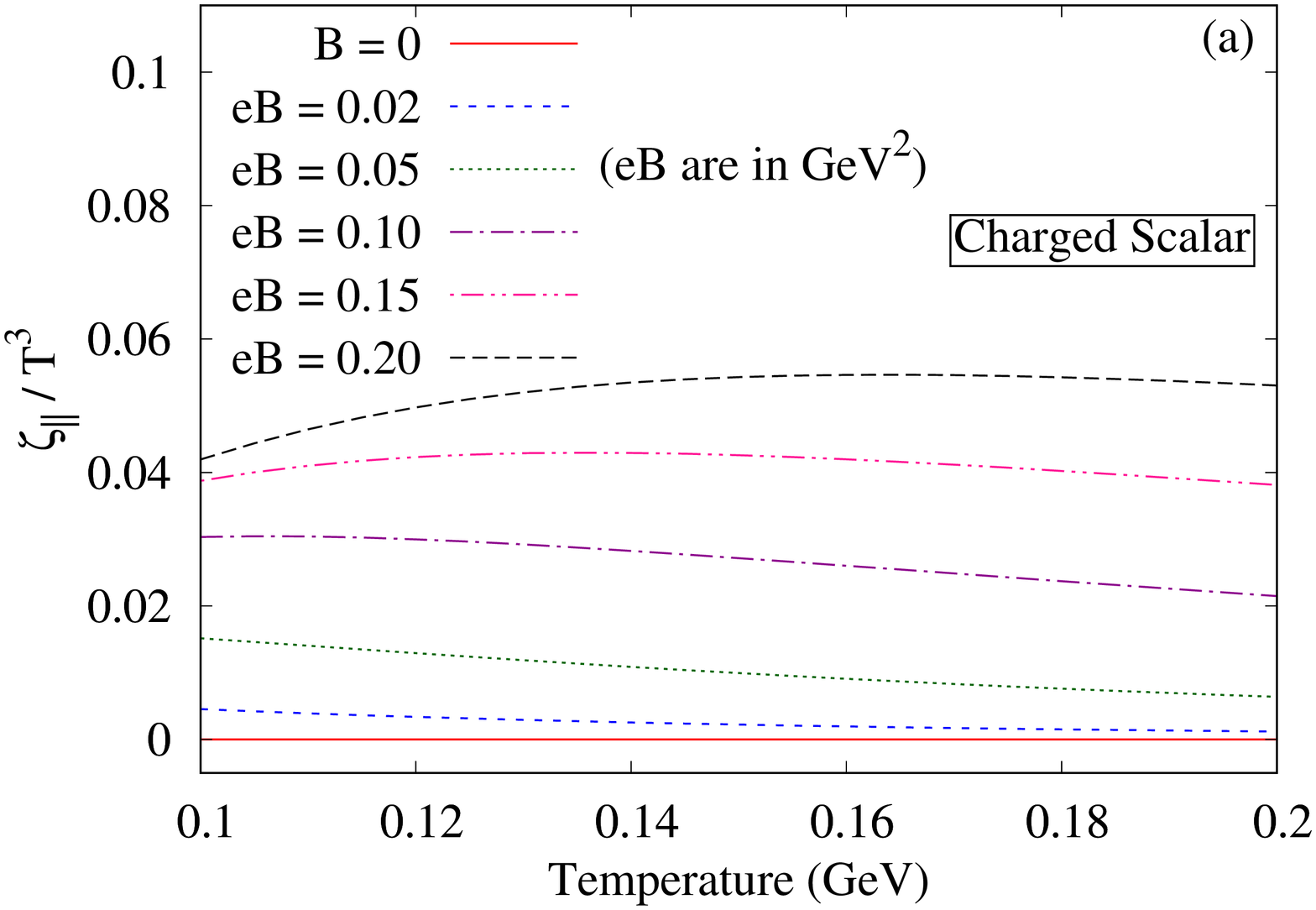}  \includegraphics[angle=-90,scale=0.3]{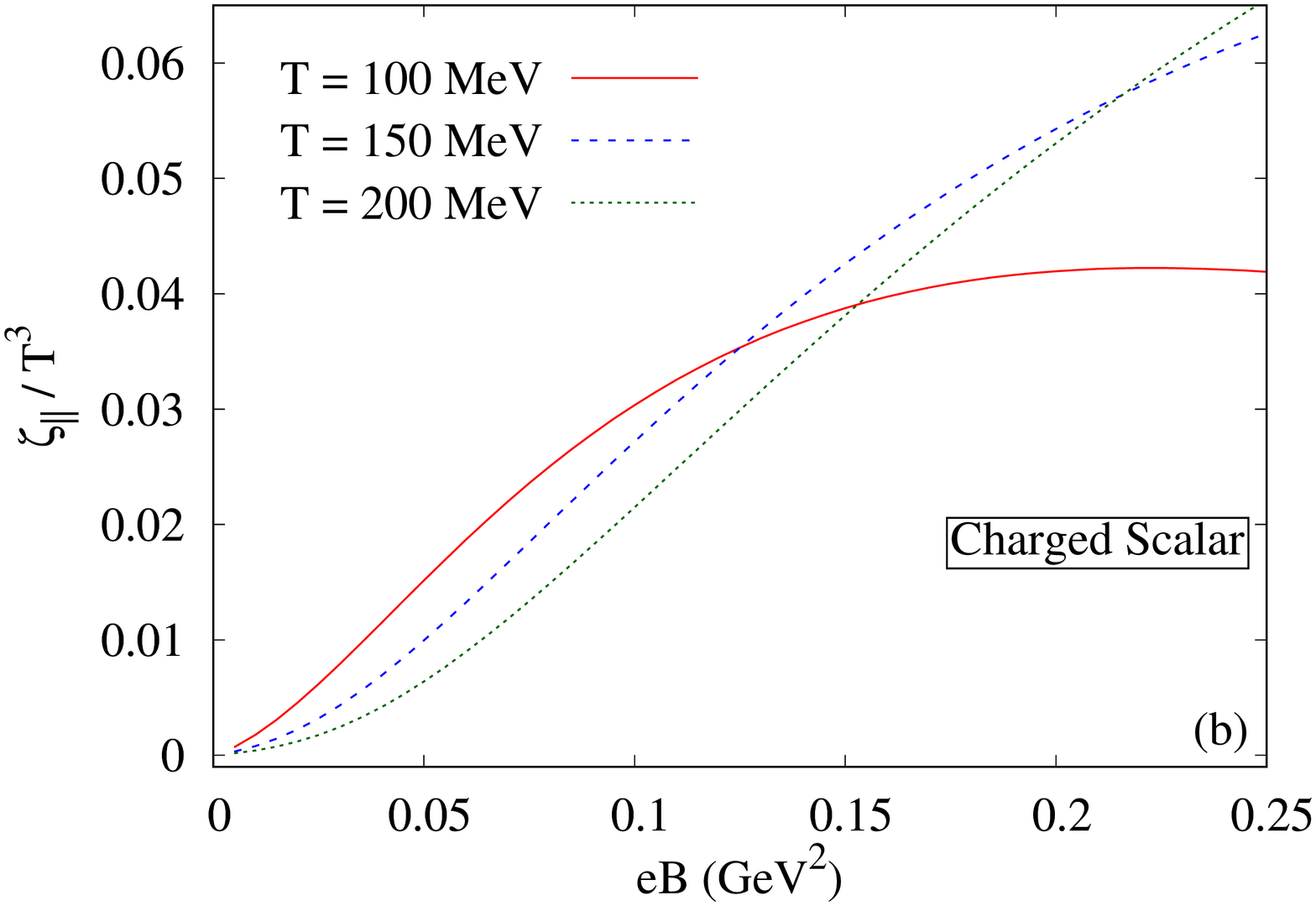}
		\includegraphics[angle=-90,scale=0.3]{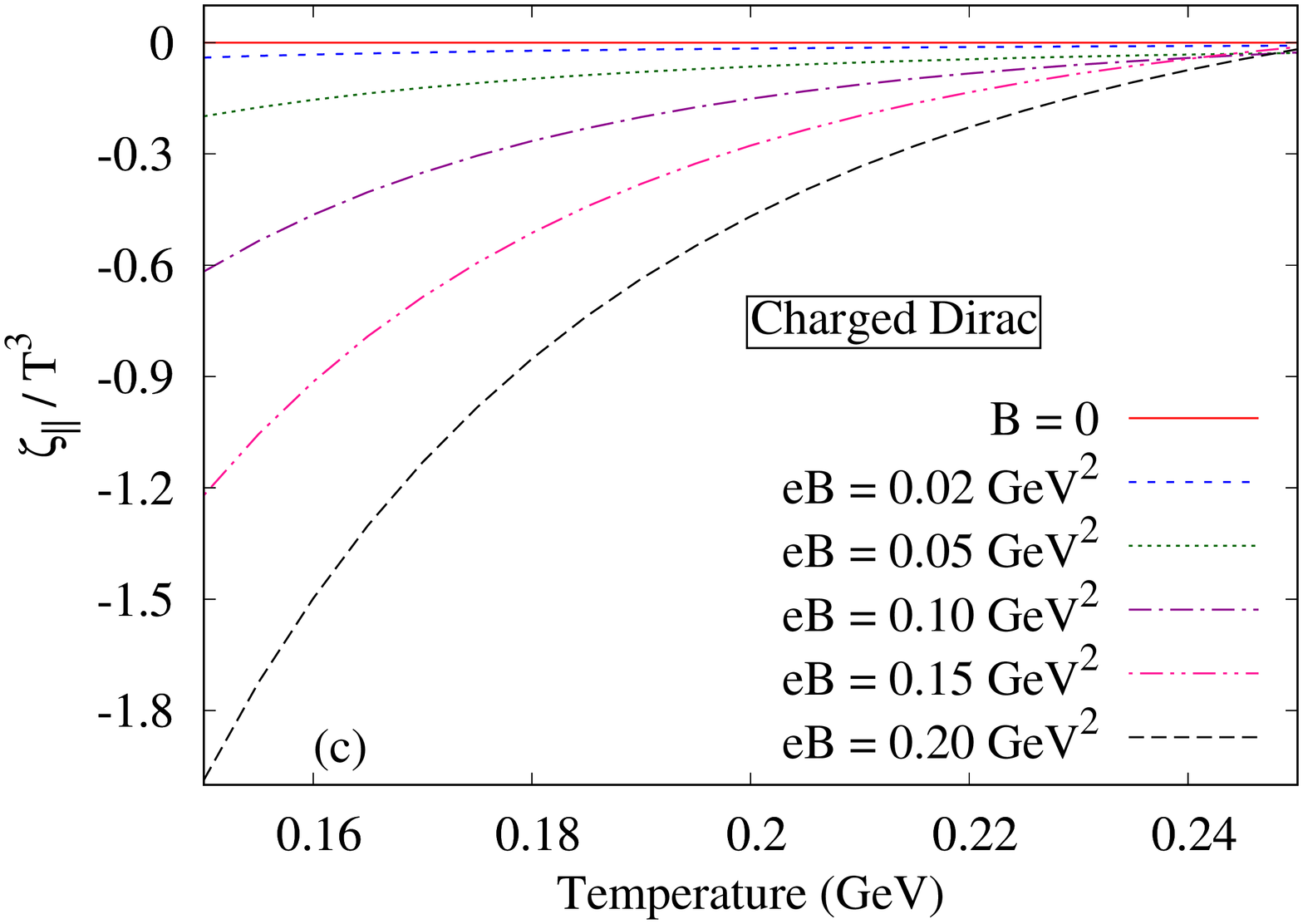}  \includegraphics[angle=-90,scale=0.3]{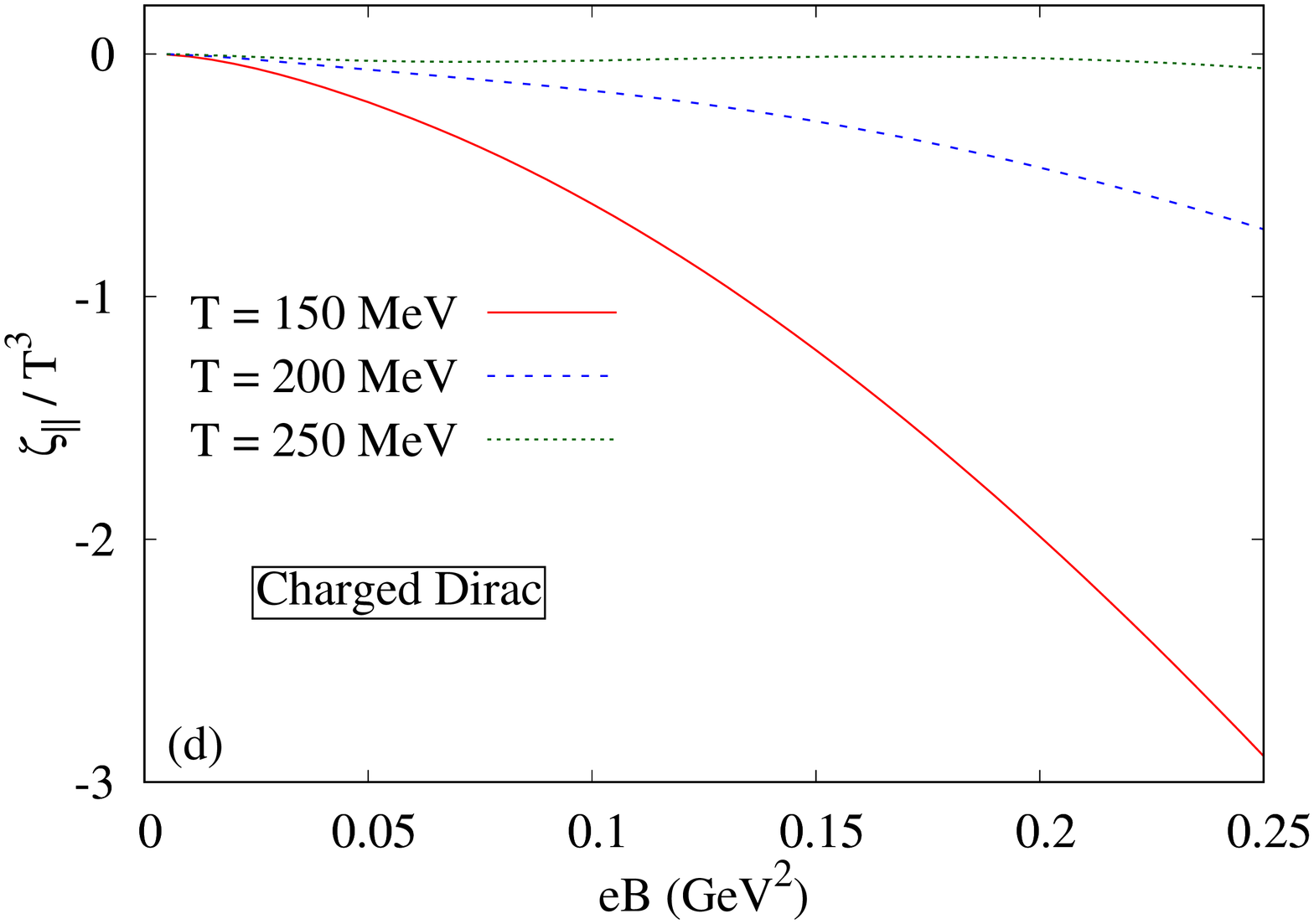}
	\end{center}
	\caption{(Color Online) The variation of $\zeta_\parallel/T^3$ as a function of (a) $T$ and (b) $eB$ for system of massless charged scalar Bosons (spin-0) with relaxation time $\tau_c=1/\Gamma=1$ fm. The variation of $\zeta_\parallel/T^3$ as a function of (c) $T$ and (d) $eB$ for system of massless charged Dirac Fermions (spin-$\frac{1}{2}$) with relaxation time $\tau_c=1/\Gamma=1$ fm.}
	\label{fig:zeta_par}
\end{figure}

Fig.~\ref{fig:eta0_TB} shows the temperature and magnetic field dependence of $\eta_0$ for the two different systems consisting of spin-$0$ scalar Bosons and spin-$\frac{1}{2}$ Dirac Fermions. Sub-figs.~\ref{fig:eta0_TB}(a) and (c) depicts the variation of the dimensionless quantity $\eta_0/T^3$ as a function of temperature for different values of magnetic field whereas Sub-figs.~\ref{fig:eta0_TB}(b) and (d) shows the variation of $\eta_0/T^3$ as a function of magnetic field for different values of temperature. To understand the change in the values of viscosity components due to the magnetic field, we have first estimated viscosities at $B=0$. 
Using Eqs.~(\ref{eta_B0S}) and (\ref{eta_B0D}), the shear viscosity $\eta$ of scalar and Dirac fluids at $B=0$ are estimated and they are plotted in Fig.~\ref{fig:eta0_TB} with solid-red lines. With respect to the $B=0$ curves, we can get a comparative measurement on the values of $\eta_0$ component. For the scalar fluid, $\eta_0$ decreases with $B$ and increases with $T$. One can identify the opposite roles of $T$ and $B$ on transport coefficients, which is also noticed in RTA of kinetic theory approaches~\cite{JD_s,Denicol_s}. Physically, we can also comprehend that temperature is the measurement of randomness, while the magnetic field aligns the system. So their thermodynamical roles on the system are expected to be quite opposite in nature.

A detail quantitative understanding for $T$, $B$ dependence of $\eta_0$ seems to be very cumbersome task (as the analytical expressions are very complicated), but we can try to relate it with its RTA expression~\cite{JD_s,Denicol_s},
\be
\eta_0 = \eta A_0~~~\text{with}~~~A_0=\frac{1}{1+4\Big(\frac{\tau_c}{\tau_B}\Big)^2}
\ee
where the expression of $\eta$ will be same as given in Eqs.~\eqref{eta_B0S} and \eqref{eta_B0D}, $\tau_c=1/\Gamma$ is basically the relaxation time or inverse of the thermal width, and $\tau_B=\frac{\om_{\rm av}}{eB}$ is the inverse of synchrotron frequency. Now in the quantum mechanical picture, the energy difference between the two Landau levels $\Delta\om=(\om_{kl}-\om_{kn})$ might be associated to the synchrotron frequency, i.e. we can grossly write $\Delta\om\approx 1/\tau_B$.
Using this connection in Eq.~(\ref{eq.vtil.B.1}), one can identify the term of effective relaxation time:
\be
\frac{\Gamma}{(\om_{kl}-\om_{kn})^2+\Gamma^2}\approx \tau_c\frac{1}{1+\Big(\frac{\tau_c}{\tau_B}\Big)^2}\approx\tau_c A_0~.
\ee
In massless limit, $\tau_B\approx\frac{3T}{eB}$~\cite{JD_s}, so the dominant $B$ dependent anisotropic factor
\be
A_0(T,B)\approx \Big[1+\Big(\frac{2\tau_c eB}{3T}\Big)^2\Big]^{-1}
\label{A0_m0}
\ee
will mainly control the $T$, $B$ dependence of $\eta_0$. One can find that, $A_0$ decreases with $B$ and increases with $T$, which is mostly
reflected on $\eta_0(T,B)$. Apart from the anisotropic factor $A_0$, $\eta_0$ contains the additional $B$ dependence via the quantized energy relation in the other part of the integrand of Eq.~(\ref{eq.vtil.B.1}). Other part of integrand mainly contains $\tilde{N}_{ln}(\om)$ as well as the thermal distribution function $f(\om)$, which is not much influential for scalar fluid. Therefore, $B$ and $T$ dependence of $\eta_0$ almost follow same trend as observed in $A_0(T,B)$

For the Dirac fluid, similar trend for $\eta_0$ is observed at the lower values of magnetic field and high temperature regions. However, we notice non-monotonic behaviour of $\eta_0$ at higher values of external magnetic field. This is probably due to the non-trivial spin structure contained in $\tilde{N}_{ln}(\om)$ for Dirac case and is mainly responsible for this non-monotonic behaviour.


Since the $\tilde{N}_{ln}$'s corresponding to the other viscous coefficients also carry the magneto-thermmodynamical quantities $\theta= \FB{\frac{\del P}{\del \varepsilon}}_B$ and $ \phi = -B\FB{\frac{\del M}{\del \varepsilon}}_B $, their temperature and magnetic field dependence are separately plotted in Figs~\ref{fig:theta} and \ref{fig:phi}. Their detail calculations are provided in Appendix~\ref{app.thermodynamics}. Sub-figs.~\ref{fig:theta}(a) and (c) depicts the variation of the $\theta$ as a function of temperature whereas Sub-figs.~\ref{fig:theta}(b) and (d) shows the variation of $\theta$ as a function of magnetic field.

We first note that, at $B=0$, $\theta$ is nothing but the squared speed of sound ($c_s^2$) of the medium which is $\frac{1}{3}$ in the mass-less case as clearly shown in Sub-figs.~\ref{fig:theta}(a) and (c) by solid-red horizontal lines. However, at $B\ne0$, the quantity $ \theta = \FB{\frac{\del P}{\del \varepsilon}}_B$  is not equal to the speed of sound of the medium since the speed of sound is most generally defined as $c_s^2 = \FB{\frac{\del P}{\del \varepsilon}}_s \ne \theta$ where $s$ is the entropy of the medium~\cite{Sarwar:2015irq}. At non-zero magnetic field, we find $\theta$ to increase (decrease) monotonically with the increase in temperature for the scalar (Dirac) fluid.  Whereas for the scalar (Dirac) fluid, $\theta$ decreases (increases) with the increases in magnetic field. Thus, we observe completely opposite behaviours for the scalar and Dirac fluid in the $T$ and $B$ dependence of $\theta$. Though, in all the cases, at high temperature, $\theta$ asymptotically approach the corresponding $B=0$ curves.

In Sub-figs.~\ref{fig:phi}(a) and (c), we have shown the variation of the $\phi$ as a function of temperature whereas in Sub-figs.~\ref{fig:phi}(b) and (d), the variation of $\phi$ as a function of magnetic field has been shown. We first note that, at vanishing magnetic field, $\phi$ becomes zero as it is related to the magnetization of the system. We also notice that, $\phi$ decreases (increases) monotonically with the increase in temperature for the scalar (Dirac) fluid; whereas it increases (decreases) with the increases in magnetic field. Thus, similar to $\theta$, here also we observe completely opposite roles of temperature and magnetic field. Interestingly, we get $\phi>0$ for scalar and $\phi<0$ for Dirac systems. Alike $\theta$, in all the cases, at high temperature, $\phi$ also asymptotically approach the corresponding $B=0$ curves.


Next, let us come to the corresponding temperature and magnetic field profiles for the other shear viscosity components $\eta_1$ and $\eta_2$. They are shown respectively in Figs.~\ref{fig:eta1_TB} and \ref{fig:eta2_TB}. Unlike $\eta_0$, these components are purely magnetically induced components as they are completely disappeared at $B=0$, while $\eta_0$ at $B=0$ becomes exactly equal to the $\eta$. In RTA of kinetic theory framework~\cite{JD_s,Denicol_s}, one can find a proportional relation $\eta_{1,2}\propto (\tau_c/\tau_B)^2$, which is missing for $\eta_0$. Therefore, at $B\to 0$ or $\tau_B\to \infty$, one gets $\eta_{1,2}\to 0$ in RTA. Whereas, in the Kubo expressions, this imposition is taken care by the rich structure contained in $\tilde{N}_{ln}$'s of Eq.~\eqref{eq.vtil.B.1} when the continuum limits ($B\to 0$) are considered.

Sub-figs.~\ref{fig:eta1_TB}(a) and (c) depicts the variation of the dimensionless quantity $\eta_1/T^3$ as a function of temperature for different values of magnetic field whereas Sub-figs.~\ref{fig:eta1_TB}(b) and (d) shows the variation of $\eta_1/T^3$ as a function of magnetic field for different values of temperature. For the scalar (Dirac) case, $\eta_1/T^3$ increases (decrease) monotonically with the increase in temperature and it decreases (increases) with the increase in magnetic field; though at high magnetic field region slight non-monotonicity is observed. We also notice that, $\eta_1$ is negative for the scalar fluid and it might not be correct to consider the absolute values of viscosity component; rather the positive or negative sign should be considered as the direction of magnetically induced shear flow. The completely opposite behaviour of the temperature and magnetic field dependence of $\eta_1/T^3$ of the scalar and Dirac fluid can be attributed to several factors; for example the different spin structures contained in $\tilde{N}_{ln}(\om)$'s as well as the opposite thermo-magnetic behaviour of $\theta$ and $\phi$. In all the cases, at high temperature, $\eta_1$ is seen to approach zero which is due to the increase in random thermal motions in the system trying to destroy the magnetic orientations.

Next, Sub-figs.~\ref{fig:eta2_TB}(a) and (c) shows the variation of the dimensionless quantity $\eta_2/T^3$ as a function of temperature whereas Sub-figs.~\ref{fig:eta2_TB}(b) and (d) shows the variation of $\eta_2/T^3$ as a function of magnetic field. We observe that, $\eta_2/T^3$ increases monotonically with the increase in temperature, whereas a non-monotonic behaviour is noticed in its magnetic field dependence. Alike $\eta_1$, $\eta_{2}$ is also varying from positive to negative values at different $T$, $B$ ranges, which should again be considered as the direction of the magnetically induced shear flows.


Since the other shear viscosity components $\eta_3$ and $\eta_4$ are coming zero in QFT calculations because of the anti-symmetric structure, therefore, we have not plotted them. Same vanishing values are realized in the ADS/CFT direction~\cite{ADS_s1,ADS_s2}. However, in the RTA of kinetic theory based calculations~\cite{JD_s,Denicol_s}, one might expect their non-zero values at non-zero chemical potential of the fluid, where particle and anti-particle density becomes different. These two coefficients in kinetic theory framework are realized as the Hall viscosities, which of course are vanished at zero chemical potential i.e. when particle and anti-particle densities are the same; but they should be finite at non-zero values of chemical potential. For the Kubo or the field theoretical expressions, it seems that there is no possibility of getting non-zero Hall viscosity even at finite chemical potential as their vanishing contributions are coming from the anti-symmetric nature of vertex factors $\tilde{N}_{ln}(\om)$. We have not found any discussion on it in any earlier references (off course based on our searching) and we are unable to resolve this discrepancy, which should get a kind attention to the community.


Now, let us turn our attention to the bulk viscosity coefficients $\zeta_{\perp}$ and $\zeta_\parallel$ whose temperature and magnetic field dependence are shown in Figs~\ref{fig:zeta_perp} and \ref{fig:zeta_par} respectively. Before going to finite $B$ cases, let us recapitulate our earlier knowledge of bulk viscosity at $B=0$. The expressions of $\zeta$ at $B=0$ for the scalar and Dirac system are given in Eqs.~\eqref{zeta_B0S} and \eqref{zeta_B0D} respectively from which we see that $\zeta \propto (3\theta-1)$ in the mass-less limit ($m=0$).
%
%
%
%
As already discussed, in the mass-less limit, $\theta=\frac{1}{3}$ for $B=0$ and therefore one gets zero bulk viscosity although the shear viscosity is not necessary to be zero. At the high temperature, QCD behaves like a mass-less and conformal type theory, where we can get $\zeta=0$. However, in the low and intermediate values of temperature, QCD will exhibit its non-conformal nature, which can be measured through the non-vanishing profile of $\zeta(T)$. If we take guidance from the lattice QCD (LQCD) calculations~\cite{SSharma} (see references therein), then a clear deviation of $\theta$ from $\frac{1}{3}$ will be noticed and a non-zero interaction measure $T^\mu_{~\mu}=\FB{\varepsilon-3P}$ of QCD thermodynamics will also be observed in the low and intermediate temperature domains. We can understand that, the trace of the ideal part of the EMT is interaction measure of QCD thermodynamics; similarly the dissipation part of the EMT is basically linked with the bulk viscosity. Non-zero values of both expose the non-conformal nature of QCD system~\cite{KS_MPLA,Tuchin_b}.

Here, we will not go through any particular system like QCD, rather will consider general relativistic scalar and Dirac fluids. And in the numerical point of view, we have focused on mass-less limits, from where we can get an idea of extreme relativistic boundaries of the different quantities. The bulk viscosity, which is zero in the mass-less limit of $B=0$ case, might not remain same at finite $B$ case; one immediate reason is that, $\theta$ deviates from $1/3$ when we switch on the magnetic field, therefore a non-zero bulk viscosity is expected even in mass-less limit. Even if one puts $\theta=\frac{1}{3}$ in Eq.~\eqref{eq.vtil.B.1} by hand for massless system, the bulk viscosity components will still not be vanished due to the presence of other terms in $\tilde{N}_{ln}(\om)$. So, it means that an additional non-conformal picture is growing at finite magnetic field case, which is entering not only through the  $\theta(T,B)$ but also some other $B$ dependent terms in $\tilde{N}_{ln}(\om)$. Hence, we can say that massless relativistic matter in presence of magnetic field can have non-zero bulk viscosity, which corresponds to its non-conformal nature (irrespective system like QED or QCD plasma). The phenomenon might be little mild at high $T$ and low $B$ domain but quite prominent at low $T$ and high $B$ zone, which is known to be quantum zone. This might be considered as a pure quantum field theoretical phenomenon due to field quantization picture.

In Sub-figs.~\ref{fig:zeta_perp}(a) and (c), we depict the variation of the dimensionless quantity $\zeta_\perp/T^3$ as a function of temperature for different values of magnetic field whereas in Sub-figs.~\ref{fig:zeta_perp}(b) and (d), we show the variation of $\zeta_\perp/T^3$ as a function of magnetic field for different values of temperature. For the scalar fluid, $\zeta_\perp/T^3$ weakly depends on the temperature and the dependence is quite non-monotonic. On the other hand, for the Dirac fluid, $\zeta_\perp/T^3$ increases monotonically with the increase in temperature. Whereas with the increase in magnetic field, $\zeta_\perp/T^3$ increases (decreases) for the scalar (Dirac) cases; though a slight non-monotonic behaviour is seen at high temperature region. Alike $\eta_1$ and $\eta_2$, we can see a positive and negative directional shifts of the values of $\zeta_\perp$ from zero corresponding to the direction of magnetically induced bulk flows.

Finally, Sub-figs.~\ref{fig:zeta_par}(a) and (c) depicts the corresponding variation of the dimensionless quantity $\zeta_\parallel/T^3$ as a function of temperature whereas Sub-figs.~\ref{fig:zeta_par}(b) and (d) shows the variation of $\zeta_\parallel/T^3$ as a function of magnetic field. Alike $\zeta_\perp$, for the scalar fluid, $\zeta_\parallel/T^3$ depends weakly on the temperature and the dependence is non-monotonic in nature. For the Dirac fluid, $\zeta_\perp/T^3$ increases monotonically with the increase in temperature. With the increase in magnetic field, $\zeta_\parallel/T^3$ is seen to increase (decrease) for the scalar (Dirac) cases. The positive and negative directional shifts of the values of $\zeta_\parallel$ from zero again correspond to the direction of magnetically induced bulk flows.

Comparing Figs.~\ref{fig:zeta_perp} and \ref{fig:zeta_par}, one can see the difference between parallel and perpendicular components of bulk viscosity i.e. $\zeta_{\parallel}\neq\zeta_{\perp}$ and the difference completely disappeared at $B=0$, where we can get the relation $\zeta_{\parallel}=\zeta_{\perp}=\zeta=0$. This fact reflect the transition from isotropic dissipation at $B=0$ to anisotropic dissipation at $B\neq 0$. Though a detail analysis of differences between scalar and Dirac system for $\zeta_{\perp}/T^3$ and $\zeta_\parallel/T^3$ are not quite easy task as their analytic expressions are quite complicated, but main resources are hidden in $\tilde{N}_{ln}(\om)$ and the thermal distribution functions. Also, the opposite thermo-magnetic behaviours of $\theta$ and $\phi$ might have a role for getting different temperature and magnetic field dependence of $\zeta_{\perp}$ and $\zeta_\parallel$ of the scalar and Dirac fluids.


\section{SUMMARY \& CONCLUSION}\label{sec.summary}
In summary, we have performed a detailed calculation of the one-loop Kubo expressions of the five shear viscosity components and the two bulk viscosity components in the presence of an arbitrary background magnetic field $B$, where a general form of the thermo-magnetic propagators are used based on the real time formalism of finite temperature field theory and the Schwinger proper time formalism. For this, we have taken two different systems: (i) system of charged scalar Bosons (spin-0) and (ii) system of charged Dirac Fermions (spin-$\frac{1}{2}$), and have calculated the corresponding thermo-magnetic spectral functions of the energy-momentum tensors (EMTs) which is the imaginary part of the Fourier transform of the local EMT-EMT two-point correlator. Then viscous coefficients are estimated from these thermo-magnetic spectral functions using the Kubo relations in the covariant tensor basis of Ref.~\cite{XGH1}.

In the absence of magnetic field, it is quite standard that the one-loop Kubo expression~\cite{Ghosh:2014yea,G_etaN,Nicola,Lang,Jeon} of shear and bulk viscosity are exactly identical to the expression from the RTA of kinetic theory approach~\cite{Gavin,Kapusta}, when we identify the inverse relation ($\Gamma=1/\tau_c$) between thermal width $\Gamma$, introduced in the propagators of Kubo method and the relaxation time $\tau_c$ in RTA method. However, in the presence of magnetic field, this exact equality between the Kubo and RTA expressions has not been found as explored in present work. In the recent time, Refs.~\cite{Asutosh_s,JD_s,JD_QP_s,Arpan_s,Denicol_s,G_HRGB,Tuchin_s,NJLB_s} have provided the RTA based kinetic theory expression of the relativistic matter at finite magnetic field; of which Refs~\cite{Asutosh_s,JD_s,JD_QP_s,Arpan_s,Denicol_s,G_HRGB} consider general value of $B$ and Refs.~\cite{Tuchin_s,NJLB_s} make strong magnetic field approximation. In these calculations, five shear viscosity components and two bulk viscosity components become different in magnitudes due to their $B$ dependent anisotropic factor, made by the two time scales: relaxation time $\tau_c$ and the inverse of synchrotron frequency $\tau_B\propto 1/B$, where $B$ actually enters. If we compare those RTA expressions of viscosity components with the corresponding Kubo expressions, obtained in present work, then one can find a
 rich magnetic field dependent structure, which probably reflect a quantum field theoretical effects on viscosity expressions. One may think that, a straight forward extension of Landau quantization of the RTA based kinetic theory expression might be equal to our Kubo expressions, but this is not the case as the present work reveals; we have identified a rich vertex structure and an alternative entry of $\tau_B$ as an inverse of the difference between Landau quantized energies. Just like we realize the time period ($\tau$) of radiation as the transition between the two energy levels in hydrogen atom problem ($\tau^{-1} = \nu=\Delta E$), here we can think of $\tau_B$ as the characteristic time scale for the transition between the two Landau levels and thus $\tau_B$ may be considered as the inverse of the energy difference of two Landau levels $\delta\om_{ln}$.

Although the qualitative magnetic field dependent trends of RTA and Kubo expressions of viscosity components are quite similar, which can also be checked by recovering their isotropic picture, when one imposes the $B\to 0$ limit. Among the five components of shear viscosity $\eta_n$ ($n=0,1,2,3,4$), $\eta_{1,2,3,4}$ components are completely originated due to the magnetic field and therefore, they are disappeared in the $B\to 0$ limit, whereas $\eta_0$ only survives and merges with the isotropic value $\eta$, present at $B=0$ picture. For the bulk viscosity components, isotropic value $\zeta$ at $B=0$ is split into two components along the parallel ($\zeta_{\parallel}$) and the perpendicular ($\zeta_\perp$) direction with respect to the direction of the magnetic field, whose merging to the isotropic value is verified by imposing the numerical $B\to 0$ limit. The qualitative magnetic field dependent structure of the RTA and Kubo expressions are more or less same but we believe that the Kubo expressions proposed here are carrying a rich quantum field theoretical structure, which is probably missing in existing RTA expressions.

\section*{ACKNOWLEDGEMENTS} 
Authors are thankful to Dr. Arghya Mukherjee and Sarthak Satapathy for carefully reading this article and for providing useful comments and suggestions. Snigdha Ghosh is funded by the Department of Higher Education, Government of West Bengal.


\appendix

\section{CALCULATION OF THE EMT CORRELATORS} \label{app.emt.corr}
In this appendix, we will sketch the calculation of the two-point correlation functions $\ensembleaverage{\mcTc T^\munu(x)T^\alphabeta(y)}_{11}$ for both the complex scalar and charged Dirac fields. 
Using Eq.~\eqref{eq.emt.scalar}, we have
\begin{eqnarray}
\ensembleaverage{\mcTc T_\text{Scalar}^\munu(x)T_\text{Scalar}^\alphabeta(y)}_{11} = 
\ensembleaverage{\mcTc \FB{ \del^\mu\phi^\dagger(x)\del^\nu\phi(x)-\frac{1}{2}g^\munu\scrL_\text{Scalar}(x)}
\FB{ \del^\alpha\phi^\dagger(y)\del^\beta\phi(y)-\frac{1}{2}g^\alphabeta\scrL_\text{Scalar}(y)}}_{11} \nn \\ 
+ (\mu \leftrightarrow \nu) +  (\alpha \leftrightarrow \beta) 
+ (\mu \leftrightarrow \nu, \alpha \leftrightarrow \beta). \label{eq.corr.scalar.1}
\end{eqnarray}
Substituting $\scrL_\text{Scalar}$ from Eq.~\eqref{eq.lag.scalar} into Eq.~\eqref{eq.corr.scalar.1} and applying the Wick's theorem~\cite{Peskin:1995ev}, we arrive at
\begin{eqnarray}
\ensembleaverage{\mcTc T_\text{Scalar}^\munu(x)T_\text{Scalar}^\alphabeta(y)}_{11} = 
\wick
{
\ensembleaverage{\mcTc \del^\mu \c2\phi^\dagger(x)\del^\nu \c1\phi(x) \del^\alpha \c1\phi^\dagger(y)\del^\beta\c2 \phi(y) }_{11} 
-\frac{1}{2}g^\munu \ensembleaverage{\mcTc \del^\sigma \c2\phi^\dagger(x)\del_\sigma \c1\phi(x) \del^\alpha  \c1\phi^\dagger(y)\del^\beta\c2 \phi(y) }_{11} } \nn \\
\wick
{
+\frac{1}{2}g^\munu m^2 \ensembleaverage{\mcTc \c2\phi^\dagger(x) \c1\phi(x) \del^\alpha  \c1\phi^\dagger(y)\del^\beta\c2 \phi(y) }_{11} 
-\frac{1}{2}g^\alphabeta \ensembleaverage{\mcTc \del^\mu \c2\phi^\dagger(x)\del^\nu \c1\phi(x) \del^\sigma  \c1\phi^\dagger(y)\del_\sigma\c2 \phi(y) }_{11} } \nn \\
\wick
{
	+\frac{1}{2}g^\alphabeta m^2 \ensembleaverage{\mcTc \del^\mu\c2\phi^\dagger(x) \del^\nu\c1\phi(x)   \c1\phi^\dagger(y)\c2 \phi(y) }_{11} 
	+ \frac{1}{4}g^\munu g^\alphabeta \bigg\{ \ensembleaverage{\mcTc \del^\sigma \c2\phi^\dagger(x)\del_\sigma \c1\phi(x) \del^\rho \c1\phi^\dagger(y)\del_\rho\c2 \phi(y) }_{11}} \nn \\
\wick
{
- m^2 \ensembleaverage{\mcTc \del^\sigma \c2\phi^\dagger(x)\del_\sigma \c1\phi(x)   \c1\phi^\dagger(y)\c2 \phi(y) }_{11} 
- m^2\ensembleaverage{\mcTc \c2\phi^\dagger(x) \c1\phi(x) \del^\sigma  \c1\phi^\dagger(y)\del_\sigma\c2 \phi(y) }_{11} 
} \nn \\ 
\wick
{
+ m^4\ensembleaverage{\mcTc \c2\phi^\dagger(x) \c1\phi(x)   \c1\phi^\dagger(y) \c2 \phi(y) }_{11} 	
}
	\bigg \} + (\mu \leftrightarrow \nu) +  (\alpha \leftrightarrow \beta) 
	+ (\mu \leftrightarrow \nu, \alpha \leftrightarrow \beta).
\end{eqnarray}
Simplification of the above equation yields,
\begin{eqnarray}
\ensembleaverage{\mcTc T_\text{Scalar}^\munu(x)T_\text{Scalar}^\alphabeta(y)}_{11} &=& 
\del^\nu_x\del^\alpha_yD_{11}(x,y)\del^\mu_x\del^\beta_yD_{11}(y,x) -\frac{1}{2}g^\munu \Big\{
\del_\sigma^x\del^\alpha_yD_{11}(x,y)\del^\sigma_x\del^\beta_yD_{11}(y,x) \nn \\ && \hspace{-2cm}
- m^2 \del^\alpha_yD_{11}(x,y)\del^\beta_yD_{11}(y,x) \Big\} 
 -\frac{1}{2}g^\alphabeta \Big\{
\del^\nu_x\del^\sigma_yD_{11}(x,y)\del^\mu_x\del_\sigma^yD_{11}(y,x) - m^2 \del^\nu_xD_{11}(x,y)\del^\mu_xD_{11}(y,x) \Big\} \nn \\ && \hspace{-2cm}
 + \frac{1}{4}g^\munu g^\alphabeta \Big\{
\del_\sigma^x\del^\rho_yD_{11}(x,y)\del^\sigma_x\del_\rho^yD_{11}(y,x) 
- m^2 \del_\sigma^xD_{11}(x,y)\del^\sigma_xD_{11}(y,x) 
- m^2 \del_\sigma^yD_{11}(x,y)\del^\sigma_yD_{11}(y,x) \nn \\ && \hspace{-1cm}
 + m^4 D_{11}(x,y) D_{11}(y,x) \Big\}  + (\mu \leftrightarrow \nu) +  (\alpha \leftrightarrow \beta) 
+ (\mu \leftrightarrow \nu, \alpha \leftrightarrow \beta)
\label{eq.corr.scalar.2}
\end{eqnarray}
where, $\del^\mu_x \equiv \frac{\del}{\del x_\mu}$, $\del^\mu_y \equiv \frac{\del}{\del y_\mu}$ etc. and $D_{11}(x,y) = \wick{\ensembleaverage{\mcTc \c\phi(x) \c\phi^\dagger(y) }_{11}}$ is the 11-component of the real time free thermal scalar propagator in coordinate space. Since, $D_{11}(x,y)=D_{11}(x-y)$ is translationally invariant, it can be Fourier transformed as 
\begin{eqnarray}
D_{11}(x,y) = D_{11}(x-y) = \int\!\!\!\frac{d^4p}{(2\pi)^4}e^{-ip\cdot(x-y)}\FB{-iD_{11}(p;m)}
\label{eq.scalar.prop1}
\end{eqnarray}
where $D_{11}(p;m)$ is the 11-component of the momentum space real time free thermal scalar propagator whose explicit form reads~\cite{Mallik:2016anp,Bellac:2011kqa}
\begin{eqnarray}
D_{11}(p;m) = \TB{\frac{-1}{p^2-m^2+i\epsilon} + \xi(p\cdot u)2\pi i\delta(p^2-m^2)}
\label{eq.D11}
\end{eqnarray}
in which $u^\mu$ is the four-velocity of the medium, $\xi(x)=\Theta(x)f(x)+\Theta(-x)f(-x)$ and 
$f(x)=\TB{e^{x/T}-1}^{-1}$ is the Bose-Einstein distribution function at temperature $T$. In the Local Rest Frame (LRF) of the medium, $u^\mu_\text{LRF}\equiv(1,\vec{0})$.

Substitution of Eq.~\eqref{eq.scalar.prop1} into Eq.~\eqref{eq.corr.scalar.2} yields after some simplifications
\begin{eqnarray}
\ensembleaverage{\mcTc T_\text{Scalar}^\munu(x)T_\text{Scalar}^\alphabeta(y)}_{11} &=& 
-\int\!\!\!\!\int\!\!\frac{d^4p}{(2\pi)^4}\frac{d^4k}{(2\pi)^4} e^{-i(x-y)\cdot(p-k)} D_{11}(p;m)D_{11}(k;m)
\mcN_\text{Scalar}^\mnab(k,p)
\label{eq.corr.scalar.3}
\end{eqnarray}
where, 
\begin{eqnarray}
\mcN_\text{Scalar}^\mnab(k,p) = k^\mu p^\nu p^\alpha k^\beta -\frac{1}{2}\FB{p\cdot k -m^2}\FB{g^\munu p^\alpha k^\beta + g^\alphabeta k^\mu p^\nu } +\frac{1}{4}\FB{p\cdot k -m^2}^2 g^\munu g^\alphabeta
\nn \\ + (\mu \leftrightarrow \nu) +  (\alpha \leftrightarrow \beta) 
+ (\mu \leftrightarrow \nu, \alpha \leftrightarrow \beta).
\label{eq.N.scalar}
\end{eqnarray}

In the calculation of viscous coefficients, we actually need the quantity $\mcN_\text{Scalar}^\mnab(k,k)$ which is easily followed from Eq.~\eqref{eq.N.scalar} as
\begin{eqnarray}
\mcN_\text{Scalar}^\mnab(k,k) &=& 4k^\mu k^\nu k^\alpha k^\beta -2\fb{k^2 -m^2}\fb{g^\munu k^\alpha k^\beta + g^\alphabeta k^\mu k^\nu } +\fb{k^2 -m^2}^2 g^\munu g^\alphabeta.
\label{eq.Nkk.scalar}
\end{eqnarray}


The calculation of EMT correlator for the Dirac field is done in similar ways. Using Eq.~\eqref{eq.emt.dirac}, we have
\begin{eqnarray}
	\ensembleaverage{\mcTc T_\text{Dirac}^\munu(x)T_\text{Dirac}^\alphabeta(y)}_{11} &=& 
	\ensembleaverage{\mcTc \FB{ \frac{i}{4}\FB{\psibar(x)\gamma^\mu\del^\nu\psi(x)-\del^\nu\psibar(x)\gamma^\mu\psi(x)} - \frac{1}{2}g^\munu \scrL_\text{Dirac}(x)} \right. \nn \\ 
		&& \hspace{-2.5cm} \left. 
		\FB{ \frac{i}{4}\FB{\psibar(y)\gamma^\mu\del^\nu\psi(y)-\del^\nu\psibar(y)\gamma^\mu\psi(y)} - \frac{1}{2}g^\munu \scrL_\text{Dirac}(y)}}_{11}  
	+ (\mu \leftrightarrow \nu) +  (\alpha \leftrightarrow \beta) 
	+ (\mu \leftrightarrow \nu, \alpha \leftrightarrow \beta). \label{eq.corr.dirac.1}
\end{eqnarray}
Substituting $\scrL_\text{Dirac}$ from Eq.~\eqref{eq.lag.dirac} into Eq.~\eqref{eq.corr.dirac.1} and applying the Wick's theorem~\cite{Peskin:1995ev}, we arrive at
\begin{eqnarray}
\ensembleaverage{\mcTc T_\text{Dirac}^\munu(x)T_\text{Dirac}^\alphabeta(y)}_{11} = 
-\frac{1}{16}\bigg\{
\wick[offset=1.2em]{\ensembleaverage{\mcTc \c2\psibar(x)\gamma^\mu\del^\nu\c1\psi(x)  \c1\psibar(y)\gamma^\alpha \del^\beta \c2\psi(y)}_{11}} 
- \wick[offset=1.2em]{\ensembleaverage{\mcTc  \c2\psibar(x)\gamma^\mu \del^\nu\c1\psi(x) \del^\beta\c1\psibar(y)\gamma^\alpha\c2\psi(y)}_{11}} \nn \\
- \wick[offset=1.2em]{\ensembleaverage{\mcTc  \del^\nu\c2\psibar(x)\gamma^\mu\c1\psi(x) \c1\psibar(y)\gamma^\alpha \del^\beta\c2\psi(y)}_{11}} 
+ \wick[offset=1.2em]{\ensembleaverage{\mcTc  \del^\nu\c2\psibar(x)\gamma^\mu\c1\psi(x) \del^\beta\c1\psibar(y)\gamma^\alpha \c2\psi(y)}_{11}} \bigg\} \nn \\
+ \frac{1}{16}g^\munu \bigg\{ 
\wick[offset=1.2em]{\ensembleaverage{\mcTc \c2\psibar(x)\gamma^\sigma\del_\sigma\c1\psi(x) \c1\psibar(y)\gamma^\alpha \del^\beta\c2\psi(y)}_{11}} 
- \wick[offset=1.2em]{ \ensembleaverage{\mcTc  \c2\psibar(x)\gamma^\sigma\del_\sigma\c1\psi(x) \del^\beta\c1\psibar(y)\gamma^\alpha \c2\psi(y)}_{11}} \nn \\
- \wick[offset=1.2em]{ \ensembleaverage{\mcTc  \del_\sigma\c2\psibar(x)\gamma^\sigma\c1\psi(x) \c1\psibar(y)\gamma^\alpha \del^\beta \c2\psi(y)}_{11}} 
+ \wick[offset=1.2em]{ \ensembleaverage{\mcTc  \del_\sigma\c2\psibar(x)\gamma^\sigma\c1\psi(x) \del^\beta\c1\psibar(y)\gamma^\alpha \c2\psi(y)}_{11}} \bigg\} \nn \\
 + \frac{i}{8} g^\munu m \bigg\{\wick[offset=1.2em]{ \ensembleaverage{\mcTc \c2\psibar(x)\c1\psi(x) \c1\psibar(y)\gamma^\alpha\del^\beta \c2\psi(y)}_{11}}
- \wick[offset=1.2em]{ \ensembleaverage{\mcTc \c2\psibar(x)\c1\psi(x) \del^\beta\c1\psibar(y)\gamma^\alpha \c2\psi(y)}_{11}}\bigg\} \nn \\
+ \frac{1}{16}g^\alphabeta \bigg\{ 
\wick[offset=1.2em]{\ensembleaverage{\mcTc \c2\psibar(x)\gamma^\mu\del^\nu\c1\psi(x) \c1\psibar(y)\gamma^\sigma \del_\sigma\c2\psi(y)}_{11}} 
- \wick[offset=1.2em]{ \ensembleaverage{\mcTc  \del^\nu\c2\psibar(x)\gamma^\mu\c1\psi(x) \c1\psibar(y)\gamma^\sigma\del_\sigma \c2\psi(y)}_{11}} \nn \\
- \wick[offset=1.2em]{ \ensembleaverage{\mcTc  \c2\psibar(x)\gamma^\mu\del^\nu\c1\psi(x) \del_\sigma\c1\psibar(y)\gamma^\sigma \c2\psi(y)}_{11}} 
+ \wick[offset=1.2em]{ \ensembleaverage{\mcTc  \del^\nu\c2\psibar(x)\gamma^\mu\c1\psi(x) \del_\sigma\c1\psibar(y)\gamma^\sigma \c2\psi(y)}_{11}}\bigg\} \nn \\
+ \frac{i}{8}g^\alphabeta m \bigg\{\wick[offset=1.2em]{ \ensembleaverage{\mcTc \c2\psibar(x)\gamma^\mu\del^\nu\c1\psi(x) \c1\psibar(y) \c2\psi(y)}_{11}}
- \wick[offset=1.2em]{ \ensembleaverage{\mcTc \del^\nu\c2\psibar(x)\gamma^\mu\c1\psi(x) \c1\psibar(y) \c2\psi(y)}_{11}}\bigg\} \nn \\
- \frac{1}{16}g^\munu g^\alphabeta \bigg\{ 
\wick[offset=1.2em]{\ensembleaverage{\mcTc \c2\psibar(x)\gamma^\sigma\del_\sigma\c1\psi(x) \c1\psibar(y)\gamma^\rho \del_\rho\c2\psi(y)}_{11}} 
- \wick[offset=1.2em]{ \ensembleaverage{\mcTc  \del_\sigma\c2\psibar(x)\gamma^\sigma\c1\psi(x) \c1\psibar(y)\gamma^\rho\del_\rho \c2\psi(y)}_{11}} \nn \\
- \wick[offset=1.2em]{ \ensembleaverage{\mcTc  \c2\psibar(x)\gamma^\sigma\del_\sigma\c1\psi(x) \del_\rho\c1\psibar(y)\gamma^\rho \c2\psi(y)}_{11}} 
+ \wick[offset=1.2em]{ \ensembleaverage{\mcTc  \del_\sigma\c2\psibar(x)\gamma^\sigma\c1\psi(x) \del_\rho\c1\psibar(y)\gamma^\rho \c2\psi(y)}_{11}} \bigg\} \nn \\
- \frac{i}{8}g^\munu g^\alphabeta m \bigg\{ 
\wick[offset=1.2em]{\ensembleaverage{\mcTc \c2\psibar(x)\gamma^\sigma\del_\sigma\c1\psi(x) \c1\psibar(y)\c2\psi(y)}_{11}} 
- \wick[offset=1.2em]{ \ensembleaverage{\mcTc  \del_\sigma\c2\psibar(x)\gamma^\sigma\c1\psi(x) \c1\psibar(y) \c2\psi(y)}_{11}} \nn \\
- \wick[offset=1.2em]{ \ensembleaverage{\mcTc  \c2\psibar(x)\c1\psi(x) \del_\rho\c1\psibar(y)\gamma^\rho \c2\psi(y)}_{11}} 
+ \wick[offset=1.2em]{ \ensembleaverage{\mcTc  \c2\psibar(x)\c1\psi(x) \c1\psibar(y)\gamma^\rho\del_\rho \c2\psi(y)}_{11}} \bigg\} \nn \\
+ \frac{1}{4}g^\munu g^\alphabeta m^2 \wick[offset=1.2em]{ \ensembleaverage{\mcTc  \c2\psibar(x)\c1\psi(x) \c1\psibar(y) \c2\psi(y)}_{11}} 
	+ (\mu \leftrightarrow \nu) +  (\alpha \leftrightarrow \beta) 
+ (\mu \leftrightarrow \nu, \alpha \leftrightarrow \beta).
\end{eqnarray}
Simplification of the above equation yields,
\begin{eqnarray}
\ensembleaverage{\mcTc T_\text{Dirac}^\munu(x)T_\text{Dirac}^\alphabeta(y)}_{11} &=&
\frac{1}{16}\Tr\Big\{ 
  \gamma^\mu\del^\nu_xS_{11}(x,y)\gamma^\alpha\del^\beta_yS_{11}(y,x)
- \gamma^\mu\del^\nu_x\del^\beta_yS_{11}(x,y)\gamma^\alpha S_{11}(y,x) \nn \\ && \hspace{-2.0cm}
- \gamma^\mu S_{11}(x,y)\gamma^\alpha \del^\nu_x\del^\beta_y S_{11}(y,x) 
+ \gamma^\mu \del^\beta_y S_{11}(x,y)\gamma^\alpha \del^\nu_x S_{11}(y,x)\Big\}
- \frac{1}{16}g^\munu \Tr \Big\{ 
  \gamma^\sigma\del_\sigma^xS_{11}(x,y)\gamma^\alpha\del^\beta_yS_{11}(y,x) \nn \\ && \hspace{-2.0cm}
- \gamma^\sigma\del_\sigma^x \del^\beta_y S_{11}(x,y)\gamma^\alpha S_{11}(y,x) 
- \gamma^\sigma S_{11}(x,y)\gamma^\alpha\del_\sigma^x\del^\beta_yS_{11}(y,x) 
+ \gamma^\sigma \del^\beta_y S_{11}(x,y)\gamma^\alpha\del_\sigma^xS_{11}(y,x) \Big\} \nn \\ && \hspace{-2.0cm}
- \frac{i}{8}g^\munu m \Tr \Big\{ S_{11}(x,y)\gamma^\alpha\del^\beta_yS_{11}(y,x) 
- \del^\beta_yS_{11}(x,y)\gamma^\alpha S_{11}(y,x)\Big\} 
- \frac{1}{16}g^\alphabeta \Tr \Big\{ 
  \gamma^\mu\del^\nu_xS_{11}(x,y)\gamma^\sigma\del_\sigma^yS_{11}(y,x) \nn \\ && \hspace{-2.0cm}
- \gamma^\mu S_{11}(x,y)\gamma^\sigma \del^\nu_x \del_\sigma^yS_{11}(y,x) 
- \gamma^\mu\del^\nu_x \del_\sigma^yS_{11}(x,y)\gamma^\sigma S_{11}(y,x) 
+ \gamma^\mu \del_\sigma^y S_{11}(x,y)\gamma^\sigma \del^\nu_x S_{11}(y,x) \Big\} \nn \\ && \hspace{-2.0cm}
- \frac{i}{8}g^\alphabeta m \Tr \Big\{ \gamma^\mu \del^\nu_x S_{11}(x,y) S_{11}(y,x) 
- \gamma^\mu S_{11}(x,y) \del^\nu_x S_{11}(y,x)\Big\} 
+ \frac{1}{16}g^\munu g^\alphabeta \Tr\Big\{ 
  \gamma^\sigma\del_\sigma^xS_{11}(x,y)\gamma^\rho\del_\rho^yS_{11}(y,x) \nn \\ && \hspace{-2.0cm}
- \gamma^\sigma\del_\sigma^x \del_\rho^y S_{11}(x,y)\gamma^\rho S_{11}(y,x) 
- \gamma^\sigma S_{11}(x,y)\gamma^\rho \del_\sigma^x \del_\rho^yS_{11}(y,x) 
+ \gamma^\sigma \del_\rho^y S_{11}(x,y)\gamma^\rho \del_\sigma^x S_{11}(y,x)\Big\} \nn \\ && \hspace{-2.00cm}
+ \frac{i}{8}g^\munu g^\alphabeta m \Tr\Big\{ 
\gamma^\sigma\del_\sigma^xS_{11}(x,y)S_{11}(y,x) 
- \gamma^\sigma S_{11}(x,y) \del_\sigma^x S_{11}(y,x) 
-  \del_\rho^y S_{11}(x,y)\gamma^\rho S_{11}(y,x) \nn \\ && \hspace{-2.5cm}
+ S_{11}(x,y)\gamma^\rho \del_\rho^y S_{11}(y,x)\Big\} 
- \frac{1}{4}g^\munu g^\alphabeta m^2 \Tr\Big\{S_{11}(x,y)S_{11}(y,x)\Big\}
+ (\mu \leftrightarrow \nu) +  (\alpha \leftrightarrow \beta) 
+ (\mu \leftrightarrow \nu, \alpha \leftrightarrow \beta)
\label{eq.corr.dirac.2}
\end{eqnarray}
where, $S_{11}(x,y) = \wick[offset=1.2em]{\ensembleaverage{\mcTc \c\psi(x) \c\psibar(y) }_{11}}$ is the 11-component of the real time free thermal Dirac propagator in coordinate space. It may be noted that, the above expression is valid even if the field $\psi$ is a multiplet; in which case the traces in the above equation have to be taken over all the spaces belonging to the multiplet in addition to the Dirac space. Similar to the scalar propagator, $S_{11}(x,y)=S_{11}(x-y)$ is translationally invariant and it can be Fourier transformed as 
\begin{eqnarray}
S_{11}(x,y) = S_{11}(x-y) = \int\!\!\!\frac{d^4p}{(2\pi)^4}e^{-ip\cdot(x-y)}\FB{-iS_{11}(p)}
\label{eq.dirac.prop1}
\end{eqnarray}
where $S_{11}(p)$ is the 11-component of the momentum space real time free thermal Dirac propagator whose explicit form reads~\cite{Mallik:2016anp,Bellac:2011kqa}
\begin{eqnarray}
S_{11}(p) = \FB{\cancel{p}+m}\TB{\frac{-1}{p^2-m^2+i\epsilon} - \xitil(p\cdot u)2\pi i\delta(p^2-m^2)}
\label{eq.S11.1}
\end{eqnarray}
in which $\xitil(x)=\Theta(x)\ftil(x)+\Theta(-x)\ftil(-x)$ and 
$\ftil(x)=\TB{e^{x/T}+1}^{-1}$ is the Fermi-Dirac distribution function at temperature $T$.

Substitution of Eq.~\eqref{eq.dirac.prop1} into Eq.~\eqref{eq.corr.dirac.2} yields after some simplifications
\begin{eqnarray}
\ensembleaverage{\mcTc T_\text{Dirac}^\munu(x)T_\text{Dirac}^\alphabeta(y)}_{11} &=&
\int\!\!\!\!\int\!\!\frac{d^4p}{(2\pi)^4}\frac{d^4k}{(2\pi)^4} e^{-i(x-y)\cdot(p-k)} \Big\{
\frac{1}{16}\FB{p^\nu k^\beta + k^\nu p^\beta + p^\nu p^\beta + k^\nu k^\beta }
\Tr\FB{\gamma^\mu S_{11}(p)\gamma^\alpha S_{11}(k)} \nn \\ && \hspace{-2.5cm}
-\frac{1}{16}g^\munu 
\FB{p_\sigma k^\beta + k_\sigma p^\beta + p_\sigma p^\beta + k_\sigma k^\beta }
\Tr\FB{\gamma^\sigma S_{11}(p)\gamma^\alpha S_{11}(k)}
+ \frac{1}{8} g^\munu m \FB{p^\beta+k^\beta}\Tr\FB{S_{11}(p)\gamma^\alpha S_{11}(k)} \nn \\ && \hspace{-2.5cm}
-\frac{1}{16}g^\alphabeta 
\FB{p^\nu k_\sigma +k^\nu p_\sigma + p^\nu p_\sigma + k^\nu k_\sigma }
\Tr\FB{\gamma^\mu S_{11}(p)\gamma^\sigma S_{11}(k)}
+ \frac{1}{8} g^\alphabeta m \FB{p^\nu+k^\nu}\Tr\FB{\gamma^\mu S_{11}(p) S_{11}(k)} \nn \\ && \hspace{-2.5cm}
+ \frac{1}{16} g^\munu g^\alphabeta 
\FB{p_\sigma k_\rho + k_\sigma p_\rho + p_\sigma p_\rho + k_\sigma k_\rho}
\Tr\FB{\gamma^\sigma S_{11}(p)\gamma^\rho S_{11}(k)} \nn \\ && \hspace{-2.5cm}
- \frac{1}{8} g^\munu g^\alphabeta m \FB{p_\sigma+k_\sigma} 
\big\{\Tr\FB{ S_{11}(p)\gamma^\sigma S_{11}(k)} + \Tr\FB{ \gamma^\sigma S_{11}(p) S_{11}(k)}\big\}
\nn \\ && \hspace{-2.5cm}
+ \frac{1}{4} g^\munu g^\alphabeta m^2 \Tr\FB{ S_{11}(p)S_{11}(k)}\Big\} 
+ (\mu \leftrightarrow \nu) +  (\alpha \leftrightarrow \beta) 
+ (\mu \leftrightarrow \nu, \alpha \leftrightarrow \beta).
\label{eq.corr.dirac.3}
\end{eqnarray}
Substituting $S_{11}(p)$ from Eq.~\eqref{eq.S11.1} into Eq.~\eqref{eq.corr.dirac.3} followed by evaluating the traces over the Dirac matrices and simplifying, we arrive at
\begin{eqnarray}
\ensembleaverage{\mcTc T_\text{Dirac}^\munu(x)T_\text{Dirac}^\alphabeta(y)}_{11} &=& 
-\int\!\!\!\!\int\!\!\frac{d^4p}{(2\pi)^4}\frac{d^4k}{(2\pi)^4} e^{-i(x-y)\cdot(p-k)} \Dtil_{11}(p;m)\Dtil_{11}(k;m)
\mcN_\text{Dirac}^\mnab(k,p)
\label{eq.corr.dirac.4}
\end{eqnarray}
where, 
\begin{eqnarray}
\Dtil_{11}(p;m) = \TB{\frac{-1}{p^2-m^2+i\epsilon} - \xitil(p\cdot u)2\pi i\delta(p^2-m^2)}
\label{eq.D11til}
\end{eqnarray}
and,
\begin{eqnarray}
\mcN_\text{Dirac}^\mnab(k,p) &=& \frac{1}{4} \Big[
-\fb{k^\mu p^\alpha + p^\mu k^\alpha }\fb{k^\beta+p^\beta}\fb{k^\nu+p^\nu}
+g^{\mu\alpha}\fb{k^\beta+p^\beta}\fb{k^\nu+p^\nu}(k\cdot p-m^2) \nn \\
&& + g^\munu \fb{k^\beta+p^\beta} \{ (p^2-m^2)k^\alpha + (k^2-m^2)p^\alpha \}
+  g^\alphabeta \fb{k^\nu+p^\nu} \{ (p^2-m^2)k^\mu + (k^2-m^2)p^\mu \} \nn \\
&& - g^\munu g^\alphabeta \{ 2(k^2-m^2)(p^2-m^2) + (k\cdot p-m^2)(k^2-m^2+p^2-m^2) \}
\Big] \nn \\
&& + (\mu \leftrightarrow \nu) +  (\alpha \leftrightarrow \beta) 
+ (\mu \leftrightarrow \nu, \alpha \leftrightarrow \beta).
\label{eq.N.dirac}
\end{eqnarray}

In the calculation of viscous coefficients, we actually need the quantity $\mcN_\text{Dirac}^\mnab(k,k)$ which is easily followed from Eq.~\eqref{eq.N.dirac} as
\begin{eqnarray}
\mcN_\text{Dirac}^\mnab(k,k) &=& 
-8 k^\mu k^\nu k^\alpha k^\beta 
+  (k^2-m^2) \big\{ g^{\mu\alpha} k^\nu k^\beta + g^{\nu\alpha} k^\mu k^\beta + g^{\mu\beta} k^\nu k^\alpha + g^{\nu\beta} k^\mu k^\alpha \nn \\ && 
+  4 g^\munu k^\alpha k^\beta 
+  4 g^\alphabeta k^\mu k^\nu  \big\} 
- 4(k^2-m^2)^2 g^\munu g^\alphabeta.
\label{eq.Nkk.dirac}
\end{eqnarray}


\section{CALCULATION OF THE EMT CORRELATORS IN PRESENCE OF EXTERNAL MAGNETIC FIELD} \label{app.emt.corr.b}
In this appendix, we will sketch the calculation of the two-point correlation functions $\ensembleaverage{\mcTc T^\munu(x)T^\alphabeta(y)}^B_{11}$ in presence of constant external magnetic field for both the complex scalar and charged Dirac fields. 
Proceeding along the similar lines as done in Appendix~\ref{app.emt.corr}, Eq.~\eqref{eq.corr.scalar.2} for the scalar field now modifies in presence of magnetic field as
\begin{eqnarray}
\ensembleaverage{\mcTc T_\text{Scalar}^\munu(x)T_\text{Scalar}^\alphabeta(y)}^B_{11} &=&
D^\nu_xD^{*\alpha}_yD^B_{11}(x,y)D^{*\mu}_xD^\beta_yD^B_{11}(y,x) -\frac{1}{2}g^\munu \Big\{
D_\sigma^x D^{*\alpha}_yD^B_{11}(x,y)D^{*\sigma}_xD^\beta_yD^B_{11}(y,x) \nn \\ && \hspace{-2.5cm}
- m^2 D^{*\alpha}_yD^B_{11}(x,y)D^\beta_yD^B_{11}(y,x) \Big\} 
 -\frac{1}{2}g^\alphabeta \Big\{
D^\nu_xD^{*\sigma}_yD^B_{11}(x,y)D^{*\mu}_xD_\sigma^yD^B_{11}(y,x) 
- m^2 D^\nu_xD^B_{11}(x,y)D^{*\mu}_xD^B_{11}(y,x) \Big\} \nn \\ && \hspace{-2.5cm}
 + \frac{1}{4}g^\munu g^\alphabeta \Big\{
D_\sigma^xD^{*\rho}_yD^B_{11}(x,y)D^{*\sigma}_xD_\rho^yD^B_{11}(y,x) 
- m^2 D_\sigma^xD^B_{11}(x,y)D^{*\sigma}_xD^B_{11}(y,x) 
- m^2 \del_\sigma^yD^B_{11}(x,y)D^{*\sigma}_yD^B_{11}(y,x) \nn \\ && \hspace{-1cm}
 + m^4 D^B_{11}(x,y) D^B_{11}(y,x) \Big\}  + (\mu \leftrightarrow \nu) +  (\alpha \leftrightarrow \beta) 
+ (\mu \leftrightarrow \nu, \alpha \leftrightarrow \beta)
\label{eq.corr.scalar.B.1}
\end{eqnarray}
where, 
$D^\mu_x \equiv \del^\mu_x + ieA_\text{ext}^\mu(x)$, 
$D^{*\mu}_x \equiv \del^\mu_x - ieA_\text{ext}^\mu(x)$, 
 $D^\mu_y \equiv \del^\mu_y + ieA_\text{ext}^\mu(y)$, 
 $D^{*\mu}_y \equiv \del^\mu_y - ieA_\text{ext}^\mu(y)$ etc.  
 and $D^B_{11}(x,y) = \wick{\ensembleaverage{\mcTc \c\phi(x) \c\phi^\dagger(y) }^B_{11}}$ is the 11-component of the real time free thermo-magnetic scalar propagator in coordinate space. 
 In contrast to the zero magnetic field case, the thermo-magnetic propagator is no longer 
 translationally invariant $D^B_{11}(x,y)=\Phi(x,y)D^B_{11}(x-y)$ and it contains the gauge dependent phase factor $\Phi(x,y)$ which explicitly breaks the translational invariance. 
 However, the translationally invariant piece $D^B_{11}(x-y)$ can be Fourier transformed and the 
 thermo-magnetic propagator can be written as
 \begin{eqnarray}
 D^B_{11}(x-y) = \int\!\!\!\frac{d^4p}{(2\pi)^4}e^{-ip\cdot(x-y)}\FB{-iD^B_{11}(p)}
 \label{eq.scalar.prop.B1}
 \end{eqnarray}
 where $D^B_{11}(p)$ is the 11-component of the momentum space real time free thermo-magnetic scalar propagator whose explicit form reads~\cite{Ayala:2004dx}
 \begin{eqnarray}
 D^B_{11}(p) = \sum_{l=0}^{\infty}2(-1)^le^{-\alpha_p}L_l(2\alpha_p)D_{11}(\ppll,m_l)
 \label{eq.D11.B}
 \end{eqnarray}
 in which $l$ is the Landau level index, $\alpha_p=-\frac{\pper^2}{eB}\ge0$, $m_l=\sqrt{m^2+(2l+1-2s)eB}=\sqrt{m^2+(2l+1)eB}$ is the ``Landau level dependent effective mass'', $s$ is the spin of the particle (for scalar field $s=0$) and $D_{11}$ is defined in Eq.~\eqref{eq.D11}. Due to the external magnetic field in the $\hat{z}$-direction, the decomposition of any four vector $k^\mu$ is done as $k=(\kpll+\kper)$ where $\kpll^\mu=\gpll^\munu k_\nu$ and $\kper^\mu=\gper^\munu k_\nu$; the corresponding decomposition of the metric tensor reads $g^\munu=(\gpll^\munu+\gper^\munu)$ with  $\gpll^\munu=\text{diag}(1,0,0,-1)$ and $\gper^\munu=\text{diag}(0,-1,-1,0)$. Note that, in our convention, $\kper^\mu$ is a space-like vector with $\kper^2=-(k_x^2+k_y^2) <0$.

 Due to the presence of the phase factor $\Phi(x,y)$ in the propagator, it may seem that the quantity $\ensembleaverage{\mcTc T_\text{Scalar}^\munu(x)T_\text{Scalar}^\alphabeta(y)}^B_{11}$ in Eq.~\eqref{eq.corr.scalar.B.1} is not translationally invariant. Fortunately this is not the case. To see this, we first note that, the gauge-dependent phase factor is given by~\cite{Ayala:2004dx} 
 \begin{eqnarray}
\Phi(x,y) = \exp \TB{ie\int_{x}^{y} dx'_\mu A_\text{ext}^\mu(x')}.
\label{eq.phase}
 \end{eqnarray}
 Differentiating the above equations separately with respect to $x$ and $y$ using Leibniz rule yields,
 \begin{eqnarray}
 \del^\mu_x \Phi(x,y) &=& \Phi(x,y) \SB{-ie A_\text{ext}^\mu(x)}, \\
 \del^\mu_y \Phi(x,y) &=& \Phi(x,y) \SB{ie A_\text{ext}^\mu(x)}.
 \end{eqnarray}
 The above two equations can be rewritten as
 \begin{eqnarray}
 D^\mu_x \Phi(x,y)  = D^{*\mu}_y \Phi(x,y) = 0.
 \label{eq.DPhi}
 \end{eqnarray}
Using Eq.~\eqref{eq.DPhi}, it is easy to see that
\begin{eqnarray}
D^\mu_x D^B_{11}(x,y)&=& D^\mu_x \TB{\Phi(x,y)D^B_{11}(x-y)} = \Phi(x,y)\del^\mu_xD^B_{11}(x-y), 
\label{eq.DDx}\\
D^{*\mu}_y D^B_{11}(x,y)&=& D^{*\mu}_y \TB{\Phi(x,y)D^B_{11}(x-y)} = \Phi(x,y)\del^\mu_yD^B_{11}(x-y).
\label{eq.DDy}
\end{eqnarray}
We now use Eqs.~\eqref{eq.DDx} and \eqref{eq.DDy} to simplify Eq.~\eqref{eq.corr.scalar.B.1} and get,
\begin{eqnarray}
\ensembleaverage{\mcTc T_\text{Scalar}^\munu(x)T_\text{Scalar}^\alphabeta(y)}^B_{11} &=& 
\Phi(x,y)\Phi(y,x)\Big[
\del^\nu_x\del^\alpha_yD^B_{11}(x-y)\del^\mu_x\del^\beta_yD^B_{11}(y-x) -\frac{1}{2}g^\munu \Big\{
\del_\sigma^x\del^\alpha_yD^B_{11}(x-y)\del^\sigma_x\del^\beta_yD^B_{11}(y-x) \nn \\ && \hspace{-2.8cm}
- m^2 \del^\alpha_yD^B_{11}(x-y)\del^\beta_yD^B_{11}(y-x) \Big\} 
-\frac{1}{2}g^\alphabeta \Big\{
\del^\nu_x\del^\sigma_yD^B_{11}(x-y)\del^\mu_x\del_\sigma^yD^B_{11}(y-x) - m^2 \del^\nu_xD^B_{11}(x-y)\del^\mu_xD^B_{11}(y-x) \Big\} \nn \\ && \hspace{-3.1cm}
+ \frac{1}{4}g^\munu g^\alphabeta \Big\{
\del_\sigma^x\del^\rho_yD^B_{11}(x-y)\del^\sigma_x\del_\rho^yD^B_{11}(y-x) 
- m^2 \del_\sigma^xD^B_{11}(x-y)\del^\sigma_xD^B_{11}(y-x) 
- m^2 \del_\sigma^yD^B_{11}(x-y)\del^\sigma_yD^B_{11}(y-x) \nn \\ && \hspace{-2cm}
+ m^4 D^B_{11}(x-y) D^B_{11}(y-x) \Big\}\Big]  + (\mu \leftrightarrow \nu) +  (\alpha \leftrightarrow \beta) 
+ (\mu \leftrightarrow \nu, \alpha \leftrightarrow \beta).
\label{eq.corr.scalar.B.2}
\end{eqnarray}
Since the phase factor given in Eq.~\eqref{eq.phase} satisfy the relation $\Phi(x,y)\Phi(y,x)=1$, we notice that, the phase factor in Eq.~\eqref{eq.corr.scalar.B.2} is completely canceled out and 
the quantity $\ensembleaverage{\mcTc T_\text{Scalar}^\munu(x)T_\text{Scalar}^\alphabeta(y)}^B_{11}$ is indeed translationally invariant and gauge independent. 
This type of cancellation of the phase factor for the loops containing particles with equal charges is well known~\cite{Ayala:2004dx,Ghosh:2019fet,Ghosh:2018xhh}. 
If we now compare Eq.~\eqref{eq.corr.scalar.B.2} with Eq.~\eqref{eq.corr.scalar.2}, we notice that the expression of EMT correlator at non-zero magnetic field is identical to the same at zero magnetic field, except the thermal propagator has to be replaced by the translationally invariant piece of the thermo-magnetic propagator.

Let us now substitute Eq.~\eqref{eq.scalar.prop.B1} into Eq.~\eqref{eq.corr.scalar.B.2} and 
we get after some simplifications
\begin{eqnarray}
\ensembleaverage{\mcTc T_\text{Scalar}^\munu(x)T_\text{Scalar}^\alphabeta(y)}^B_{11} &=& 
-\int\!\!\!\!\int\!\!\frac{d^4p}{(2\pi)^4}\frac{d^4k}{(2\pi)^4} e^{-i(x-y)\cdot(p-k)}
\sum_{l=0}^{\infty} \sum_{n=0}^{\infty} D_{11}(\ppll;m_n)D_{11}(\kpll;m_l) 
\mcN_{ln;\text{Scalar}}^\mnab(k,p)
\label{eq.corr.scalar.B.3}
\end{eqnarray}
where, 
\begin{eqnarray}
\mcN_{ln;\text{Scalar}}^\mnab(k,p) = 4(-1)^{l+n}e^{-\alpha_k-\alpha_p}
L_l(2\alpha_k)L_n(2\alpha_p)\mcN_{\text{Scalar}}^\mnab(k,p)
\label{eq.N.scalar.B}
\end{eqnarray}
in which $\mcN_{\text{Scalar}}^\mnab(k,p)$ is defined in Eq.~\eqref{eq.N.scalar}.

In the calculation of viscous coefficients, we actually need the quantity $\mcN_{ln;\text{Scalar}}^\mnab(k,k)$ which is easily followed from Eq.~\eqref{eq.N.scalar.B} as
\begin{eqnarray}
\mcN_{ln;\text{Scalar}}^\mnab(k,k)
&=& 4 \mathcal{A}_{ln}(\kper^2)	\Big\{
4k^\mu k^\nu k^\alpha k^\beta -2\fb{k^2 -m^2}\fb{g^\munu k^\alpha k^\beta + g^\alphabeta k^\mu k^\nu } +\fb{k^2 -m^2}^2 g^\munu g^\alphabeta \Big\} 
\label{eq.Nkk.scalar.B}
\end{eqnarray}
in which
\begin{eqnarray}
\mathcal{A}_{ln}(\kper^2) &=& (-1)^{l+n}e^{-2\alpha_k} L_l(2\alpha_k) L_n(2\alpha_k).
\label{eq.Anl}
\end{eqnarray}


The calculation of the EMT correlator $\ensembleaverage{\mcTc T_\text{Dirac}^\munu(x)T_\text{Dirac}^\alphabeta(y)}^B_{11}$ 
for the Dirac field in presence of external magnetic field can be done in a similar fashion as done for the scalar field. In this case, the zero magnetic field expression of Eq.~\eqref{eq.corr.dirac.3} 
modifies to
\begin{eqnarray}
\ensembleaverage{\mcTc T_\text{Dirac}^\munu(x)T_\text{Dirac}^\alphabeta(y)}^B_{11} &=&
\int\!\!\!\!\int\!\!\frac{d^4p}{(2\pi)^4}\frac{d^4k}{(2\pi)^4} e^{-i(x-y)\cdot(p-k)} \Big\{
\frac{1}{16}\FB{p^\nu k^\beta + k^\nu p^\beta + p^\nu p^\beta + k^\nu k^\beta }
\Tr\FB{\gamma^\mu S^B_{11}(p)\gamma^\alpha S^B_{11}(k)} \nn \\ && \hspace{-2.5cm}
-\frac{1}{16}g^\munu 
\FB{p_\sigma k^\beta + k_\sigma p^\beta + p_\sigma p^\beta + k_\sigma k^\beta }
\Tr\FB{\gamma^\sigma S^B_{11}(p)\gamma^\alpha S^B_{11}(k)}
+ \frac{1}{8} g^\munu m \FB{p^\beta+k^\beta}\Tr\FB{S^B_{11}(p)\gamma^\alpha S^B_{11}(k)} \nn \\ && \hspace{-2.5cm}
-\frac{1}{16}g^\alphabeta 
\FB{p^\nu k_\sigma +k^\nu p_\sigma + p^\nu p_\sigma + k^\nu k_\sigma }
\Tr\FB{\gamma^\mu S^B_{11}(p)\gamma^\sigma S^B_{11}(k)}
+ \frac{1}{8} g^\alphabeta m \FB{p^\nu+k^\nu}\Tr\FB{\gamma^\mu S^B_{11}(p) S^B_{11}(k)} \nn \\ && \hspace{-2.5cm}
+ \frac{1}{16} g^\munu g^\alphabeta 
\FB{p_\sigma k_\rho + k_\sigma p_\rho + p_\sigma p_\rho + k_\sigma k_\rho}
\Tr\FB{\gamma^\sigma S^B_{11}(p)\gamma^\rho S^B_{11}(k)} \nn \\ && \hspace{-2.5cm}
- \frac{1}{8} g^\munu g^\alphabeta m \FB{p_\sigma+k_\sigma} 
\big\{\Tr\FB{ S^B_{11}(p)\gamma^\sigma S^B_{11}(k)} + \Tr\FB{ \gamma^\sigma S^B_{11}(p) S^B_{11}(k)}\big\}
\nn \\  && \hspace{-2.5cm}
+ \frac{1}{4} g^\munu g^\alphabeta m^2 \Tr\FB{ S^B_{11}(p)S^B_{11}(k)}\Big\} 
+ (\mu \leftrightarrow \nu) +  (\alpha \leftrightarrow \beta) 
+ (\mu \leftrightarrow \nu, \alpha \leftrightarrow \beta) 
\label{eq.corr.dirac.B.1}
\end{eqnarray}
where, 
$S^B_{11}(p)$ is the 11-component of the momentum space real time free thermo-magnetic Dirac propagator whose explicit form is given by~\cite{Ayala:2003pv,Schwinger:1951nm}
\begin{eqnarray}
S^B_{11}(p) = \sum_{l=0}^{\infty}(-1)^le^{-\alpha_p} \scrD_l(p)\Dtil_{11}(\ppll,m_l)
\label{eq.S11.B}
\end{eqnarray}
in which $m_l=\sqrt{m^2+(2l+1-2s)eB} = \sqrt{m^2+2leB}$ (for Dirac field, $s=1/2$), $\Dtil_{11}$ is defined in Eq.~\eqref{eq.D11til} and $\scrD_l(p)$ is 
\begin{eqnarray}
\scrD_l(p) = \FB{\cancel{p}_\parallel+m}\TB{\FB{\mathds{1}+i\gamma^1\gamma^2}L_l(2\alpha_p) 
	- \FB{\mathds{1}-i\gamma^1\gamma^2}L_{l-1}(2\alpha_p)} - 4\cancel{p}_\perp L^1_{l-1}(2\alpha_p)
\label{eq.Dl}
\end{eqnarray}
with the convention $L_{-1}(z) = L_{-1}^1(z) = 0$.

We now substitute Eq.~\eqref{eq.S11.B} into Eq.~\eqref{eq.corr.dirac.B.1} and get after some simplifications
\begin{eqnarray}
\ensembleaverage{\mcTc T_\text{Dirac}^\munu(x)T_\text{Dirac}^\alphabeta(y)}^B_{11} &=& 
-\int\!\!\!\!\int\!\!\frac{d^4p}{(2\pi)^4}\frac{d^4k}{(2\pi)^4} e^{-i(x-y)\cdot(p-k)}
\sum_{l=0}^{\infty} \sum_{n=0}^{\infty} \Dtil_{11}(\ppll;m_n)\Dtil_{11}(\kpll;m_l) 
\mcN_{ln;\text{Dirac}}^\mnab(k,p)
\label{eq.corr.dirac.B.2}
\end{eqnarray}
where, 
\begin{eqnarray}
\mcN_{ln;\text{Dirac}}^\mnab(k,p) = -(-1)^{l+n}e^{-\alpha_k-\alpha_p}
\Big\{
\frac{1}{16}\FB{p^\nu k^\beta + k^\nu p^\beta + p^\nu p^\beta + k^\nu k^\beta }
\Tr\FB{\gamma^\mu \scrD_n(p) \gamma^\alpha \scrD_l(l)} \nn \\
-\frac{1}{16}g^\munu 
\FB{p_\sigma k^\beta + k_\sigma p^\beta + p_\sigma p^\beta + k_\sigma k^\beta }
\Tr\FB{\gamma^\sigma  \scrD_n(p)\gamma^\alpha \scrD_l(l)}
+ \frac{1}{8} g^\munu m \FB{p^\beta+k^\beta}\Tr\FB{\scrD_n(p)\gamma^\alpha \scrD_l(l)} \nn \\
-\frac{1}{16}g^\alphabeta 
\FB{p^\nu k_\sigma +k^\nu p_\sigma + p^\nu p_\sigma + k^\nu k_\sigma }
\Tr\FB{\gamma^\mu \scrD_n(p)\gamma^\sigma \scrD_l(l)}
+ \frac{1}{8} g^\alphabeta m \FB{p^\nu+k^\nu}\Tr\FB{\gamma^\mu \scrD_n(p) \scrD_l(l)} \nn \\
+ \frac{1}{16} g^\munu g^\alphabeta 
\FB{p_\sigma k_\rho + k_\sigma p_\rho + p_\sigma p_\rho + k_\sigma k_\rho}
\Tr\FB{\gamma^\sigma \scrD_n(p)\gamma^\rho \scrD_l(l)} \nn \\
- \frac{1}{8} g^\munu g^\alphabeta m \FB{p_\sigma+k_\sigma} 
\big\{\Tr\FB{ \scrD_n(p)\gamma^\sigma \scrD_l(l)} + \Tr\FB{ \gamma^\sigma \scrD_n(p) \scrD_l(l)}\big\}
\nn \\ 
+ \frac{1}{4} g^\munu g^\alphabeta m^2 \Tr\FB{ \scrD_n(p) \scrD_l(l) }\Big\} 
+ (\mu \leftrightarrow \nu) +  (\alpha \leftrightarrow \beta) 
+ (\mu \leftrightarrow \nu, \alpha \leftrightarrow \beta).
\label{eq.N.dirac.B}
\end{eqnarray}

In the calculation of viscous coefficients, we actually need the quantity $\mcN_{ln;\text{Dirac}}^\mnab(k,k)$ which is easily followed from Eq.~\eqref{eq.N.dirac.B} as
\begin{eqnarray}
\mcN_{ln;\text{Dirac}}^\mnab(k,k) &=& 
-\frac{1}{4} \Big[
\mcT^{\mu\alpha}_{ln}(k)k^\nu k^\beta 
- g^\munu \Big\{ \mcT^{\sigma\alpha}_{ln}(k)k_\sigma k^\beta 
- m \mcT^{\alpha}_{ln}(k) k^\beta \Big\}
- g^\alphabeta \Big\{ \mcT^{\mu\sigma}_{ln}(k)k_\sigma k^\nu 
- m \mcT^{\mu}_{nl}(k) k^\nu \Big\} \nn \\
&& + g^\munu g^\alphabeta \Big\{ \mcT^{\sigma\rho}_{ln}(k)k_\sigma k_\rho 
-2m \mcT^{\sigma}_{ln}(k)k_\sigma + m^2 \mcT_{ln}(k)\Big\}
\Big] + (\mu \leftrightarrow \nu) +  (\alpha \leftrightarrow \beta) 
+ (\mu \leftrightarrow \nu, \alpha \leftrightarrow \beta)
\label{eq.Nkk.dirac.B.1}
\end{eqnarray}
where, 
\begin{eqnarray}
\mcT^\munu_{ln}(k) &=& (-1)^{l+n}e^{-2\alpha_k} \Tr \TB{ \gamma^\mu \scrD_n(k)\gamma^\nu \scrD_l(k) } \nn \\
&=&  8\Big[ 8(2\kper^\mu\kper^\nu-\kper^2g^\munu)\mathcal{B}_{ln}(\kper^2)
+ \big\{2\kpll^\mu\kpll^\nu-\gpll^\munu(\kpll^2-m^2)\big\}\mathcal{C}_{ln}(\kper^2) \nn \\ &&
+ (\kpll^2-m^2)\gper^\munu \mathcal{D}_{ln}(\kper^2) + 2(\kpll^\mu\kper^\nu+\kper^\mu\kpll^\nu)\mathcal{E}_{ln}(\kper^2)
\Big], \label{eq.T.munu}\\
\mcT^\mu_{ln}(k) &=& (-1)^{l+n}e^{-2\alpha_k} \Tr \TB{ \scrD_n(k)\gamma^\mu \scrD_l(k) }
= 16m \Big[ \kpll^\mu \mathcal{C}_{ln}(\kper^2) + \kper^\mu \mathcal{D}_{ln}(\kper^2)
\Big] \label{eq.T.mu}
, \\
\mcT_{ln}(k) &=& (-1)^{l+n}e^{-2\alpha_k} \Tr \TB{  \scrD_n(k) \scrD_l(k) }
= 8 \Big[ 8 \kper^2 \mathcal{B}_{ln}(\kper^2) + (\kpll^2+m^2)\mathcal{C}_{ln}(\kper^2)\Big]
\label{eq.T}
\end{eqnarray}
in which,
\begin{eqnarray}
\mathcal{B}_{ln}(\kper^2) &=& (-1)^{l+n}e^{-2\alpha_k} L^1_{l-1}(2\alpha_k) L^1_{n-1}(2\alpha_k) ~,
\label{eq.Bnl}\\
\mathcal{C}_{ln}(\kper^2) &=& (-1)^{l+n}e^{-2\alpha_k} 
\SB{L_{l-1}(2\alpha_k) L_{n-1}(2\alpha_k) + L_{l}(2\alpha_k) L_{n}(2\alpha_k)} \label{eq.Cnl}~, \\
\mathcal{D}_{ln}(\kper^2) &=& (-1)^{l+n}e^{-2\alpha_k} 
\SB{L_{l}(2\alpha_k) L_{n-1}(2\alpha_k) + L_{l-1}(2\alpha_k) L_{n}(2\alpha_k)} \label{eq.Dnl}~, \\
\mathcal{E}_{ln}(\kper^2) &=& (-1)^{l+n}e^{-2\alpha_k} 
\SB{L_{l-1}(2\alpha_k) L^1_{n-1}(2\alpha_k) - L_{l}(2\alpha_k) L^1_{n-1}(2\alpha_k) \right. \nn \\ && \left.
	+~ L^1_{l-1}(2\alpha_k) L_{n-1}(2\alpha_k)- L^1_{l-1}(2\alpha_k) L_{n}(2\alpha_k)}. \label{eq.Enl}
\end{eqnarray}

Substituting Eqs.~\eqref{eq.T.munu}-\eqref{eq.T} into Eq.~\eqref{eq.Nkk.dirac.B.1} and simplifying, we finally obtain
\begin{eqnarray}
\mcN_{ln;\text{Scalar}}^\mnab(k,k) &=& 
-16 \mathcal{B}_{ln}(\kper^2) \Big[ 
k^\nu k^\beta (2\kper^\mu\kper^\alpha-\kper^2g^{\mu\alpha}) 
-g^\munu k^\beta \kper^2(\kper^\alpha-\kpll^\alpha) 
- g^\alphabeta k^\nu \kper^2(\kper^\mu-\kpll^\mu) \nn \\ && \hspace{-1.5cm}
+ g^\munu g^\alphabeta \kper^2(\kper^2-\kpll^2+m^2) \Big]
-2 \mathcal{C}_{ln}(\kper^2) \Big[
k^\nu k^\beta \big\{2\kpll^\mu\kpll^\alpha-(\kpll^2-m^2)\gpll^{\mu\alpha}\big\}
- (\kpll^2-m^2) g^\munu k^\beta\kpll^\alpha \nn \\ && \hspace{-1.5cm}
- (\kpll^2-m^2) g^\alphabeta k^\nu \kpll^\mu
+ g^\munu g^\alphabeta (\kpll^2-m^2)^2 \Big]
-2 \mathcal{D}_{ln}(\kper^2) (\kpll^2-m^2) \Big[
k^\nu k^\beta \gper^{\mu\alpha} 
-  g^\munu k^\beta\kper^\alpha  \nn \\ && \hspace{-1.5cm}
-  g^\alphabeta k^\nu \kper^\mu
+ g^\munu g^\alphabeta \kper^2 \Big]
-4 \mathcal{E}_{ln}(\kper^2) \Big[
k^\nu k^\beta (\kpll^\mu\kper^\alpha+\kper^\mu\kpll^\alpha) 
- g^\munu k^\beta \big\{(\kpll^2-m^2)\kper^\alpha + \kper^2\kpll^\alpha \big\}  \nn \\ && \hspace{-1.5cm}
- g^\alphabeta k^\nu \big\{(\kpll^2-m^2)\kper^\mu + \kper^2\kpll^\mu\big\}
+ 2g^\munu g^\alphabeta \kper^2 (\kpll^2-m^2) \Big]
+ (\mu \leftrightarrow \nu) +  (\alpha \leftrightarrow \beta) 
+ (\mu \leftrightarrow \nu, \alpha \leftrightarrow \beta).
\label{eq.Nkk.dirac.B}
\end{eqnarray}

\section{EXPLICIT ANALYTIC EPRESSIONS OF $\mathcal{A}_{ln}^{(j)}$, $\mathcal{B}_{ln}^{(j)}$, $\cdots$,  $\mathcal{E}_{ln}^{(j)}$}\label{app.integral}
Let us first note that, using the orthogonality of the Laguerre polynomials, the following integral identities can be derived:
\begin{eqnarray}
&&\int\!\! \frac{d^2\kper}{(2\pi)^2}e^{-2\alpha_k} L^1_{l-1}(2\alpha_k)L^1_{n-1}(2\alpha_k)\kper^4
= \frac{(eB)^3}{32\pi}nl\FB{2\delta_{l-1}^{n-1}-\delta_{l-1}^{n}-\delta_{l-1}^{n-2}}, \label{eq.iden.i}\\
&&\int\!\! \frac{d^2\kper}{(2\pi)^2}e^{-2\alpha_k} L^1_{l-1}(2\alpha_k)L^1_{n-1}(2\alpha_k)\kper^2
= - \frac{(eB)^2}{16\pi}n\delta_{l-1}^{n-1}, \\
&&\int\!\! \frac{d^2\kper}{(2\pi)^2}e^{-2\alpha_k} L^1_{l-1}(2\alpha_k)L_{n}(2\alpha_k)\kper^2
= - \frac{(eB)^2}{16\pi}l\FB{\delta_{l-1}^{n}-\delta_{l}^{n}}, \\
&&\int\!\! \frac{d^2\kper}{(2\pi)^2}e^{-2\alpha_k} L_{l}(2\alpha_k)L_{n}(2\alpha_k)\kper^4 
= \frac{(eB)^3}{32\pi}\Big\{(2l+1)^2\delta_{l}^{n}-(2l+1)l\delta_{l}^{n+1} 
 - (2l+1)n\delta_{l}^{n-1}-(2n+1)n\delta_{l+1}^{n} \nn \\ && \hspace{6cm}
  + (l+1)^2\delta_{l+1}^{n+1} + (l+1)n\delta_{l+1}^{n-1} 
  - (2n+1)l\delta_{l-1}^{n} + (n+1)l\delta_{l-1}^{n+1} + l^2 \delta_{l-1}^{n-1}\Big\}, \\
&&\int\!\! \frac{d^2\kper}{(2\pi)^2}e^{-2\alpha_k} L_{l}(2\alpha_k)L_{n}(2\alpha_k)\kper^2
= -\frac{(eB)^2}{16\pi}\Big\{(2n+1)\delta_{l}^{n}-(n+1)\delta_{l}^{n+1} 
- n\delta_{l}^{n-1}\Big\}, \\
&&\int\!\! \frac{d^2\kper}{(2\pi)^2}e^{-2\alpha_k} L_{l}(2\alpha_k)L_{n}(2\alpha_k)
= \frac{eB}{8\pi}\delta_{l}^{n}. \label{eq.iden.f}
\end{eqnarray}

Now, using Eqs.~\eqref{eq.iden.i}-\eqref{eq.iden.f}, we perform the $d^2\kper$ integrals of  Eqs.~\eqref{eq.Aln.j}-\eqref{eq.Eln.j} and obtain
\begin{eqnarray}
\mathcal{A}_{ln}^{(0)} &=& \frac{eB}{8\pi}\delta_{l}^{n}, \\
\mathcal{A}_{ln}^{(2)} &=& -\frac{(eB)^2}{16\pi}\Big\{(2n+1)\delta_{l}^{n}+(n+1)\delta_{l}^{n+1} 
+ n\delta_{l}^{n-1}\Big\} \\
\mathcal{A}_{ln}^{(4)} &=& \frac{(eB)^3}{32\pi}\Big\{(2l+1)^2\delta_{l}^{n}+(2l+1)l\delta_{l}^{n+1} 
+ (2l+1)n\delta_{l}^{n-1}+(2n+1)n\delta_{l+1}^{n} \nn \\ && \hspace{1cm}
+ (l+1)^2\delta_{l+1}^{n+1} + (l+1)n\delta_{l+1}^{n-1} 
+ (2n+1)l\delta_{l-1}^{n} + (n+1)l\delta_{l-1}^{n+1} + l^2 \delta_{l-1}^{n-1}\Big\}, \\
\mathcal{B}_{ln}^{(2)} &=& - \frac{(eB)^2}{16\pi}n\delta_{l-1}^{n-1}, \\
\mathcal{B}_{ln}^{(4)} &=& \frac{(eB)^3}{32\pi}nl\FB{2\delta_{l-1}^{n-1}+\delta_{l-1}^{n}+\delta_{l-1}^{n-2}}, \\
\mathcal{C}_{ln}^{(0)} &=& \frac{eB}{8\pi}\FB{\delta_{l}^{n}+\delta_{l-1}^{n-1}}, \\
\mathcal{C}_{ln}^{(2)} &=& -\frac{(eB)^2}{16\pi}\Big\{(2n+1)\delta_{l}^{n}+(n+1)\delta_{l}^{n+1} 
+ n\delta_{l}^{n-1} + (2n-1)\delta_{l-1}^{n-1}+n\delta_{l-1}^{n} 
+ (n-1)\delta_{l-1}^{n-2} \Big\}, \\
\mathcal{D}_{ln}^{(0)} &=& -\frac{eB}{8\pi}\FB{\delta_{l}^{n-1} + \delta_{l-1}^{n}}, \\
\mathcal{D}_{ln}^{(2)} &=& \frac{(eB)^2}{16\pi}\Big\{(2n-1)\delta_{l}^{n-1}+n\delta_{l}^{n} 
+ (n-1)\delta_{l}^{n-2} + (2n+1)\delta_{l-1}^{n}+(n+1)\delta_{l-1}^{n+1} 
+ n\delta_{l-1}^{n-1} \Big\}, \\
\mathcal{E}_{ln}^{(2)} &=& - \frac{(eB)^2}{16\pi}(l+n)\FB{\delta_{l-1}^{n-1}+\delta_{l-1}^{n}+\delta_{l}^{n-1}+\delta_{l}^{n}}.
\end{eqnarray}
It is important to note that, the Kronecker delta function with a negative index is always zero (i.e. $\delta_{-1}^{-1}=0$) which follows from the convention of the Laguerre polynomials $L_{-1}(z)=L_{-1}^1(z) =0$ used in Eq.~\eqref{eq.Dl}.

\section{THERMODYNAMIC QUANTITIES} \label{app.thermodynamics}
In this appendix, we will derive the thermodynamic quantities $\theta = \FB{\dfrac{\del P}{\del \varepsilon}}_B$ and $\phi = -B\FB{\dfrac{\del M}{\del \varepsilon}}_B$ both at zero and non-zero external magnetic field where $P$ is the longitudinal pressure, $\varepsilon$ is the energy density and $M$ is the magnetization.

Let us first consider the zero magnetic field case. The canonical partition function reads~\cite{Kapusta:2006pm}
\begin{eqnarray}
\ln\mathcal{Z} = -V g\! \!\int\!\!\! \frac{d^3k}{(2\pi)^3} a \ln (1-ae^{-\omega_k/T}) \label{eq.lnZ}
\end{eqnarray}
where,  $g$ is the degeneracy factor. For system of charged scalar Bosons, $g=2$ whereas for the system of charged Dirac Fermions $g=4$. 
All the thermodynamic quantities of interest can be derived from the partition function. The pressure is
\begin{eqnarray}
P = \frac{T}{V}\ln\mathcal{Z} = g \! \int\!\!\! \frac{d^3k}{(2\pi)^3} \frac{\vec{k}^2}{3\omega_k} f_{a}(\omega_k) \label{eq.P}
\end{eqnarray}
where, $f_a(x) = \TB{e^{x/T}-a}^{-1}$ is the equilibrium thermal distributin (Bose-Einstein or Fermi-Dirac). 
Similarly, the energy density is given by
\begin{eqnarray}
\varepsilon = T\FB{\frac{\del P}{\del T}} -P = g \int\!\!\! \frac{d^3k}{(2\pi)^3} \omega_k f_{a}(\omega_k). \label{eq.E}
\end{eqnarray}
It is straight forward to obtain
\begin{eqnarray}
\theta = \FB{\dfrac{\del P}{\del \varepsilon}} = \FB{\dfrac{\del P}{\del T}}\bigg/ \FB{\dfrac{\del \varepsilon}{\del T}}
\end{eqnarray}
where, 
\begin{eqnarray}
\FB{\dfrac{\del P}{\del T}} &=& \frac{\varepsilon+P}{T}, \\
\FB{\dfrac{\del \varepsilon}{\del T}} &=&  \frac{1}{T^2} g \int\!\!\! \frac{d^3k}{(2\pi)^3} \omega_k^2 f_{a}(\omega_k) \SB{1+af_{a}(\omega_k)}.
\end{eqnarray}

Let us now switch on the external magnetic field. The canonical partition function for the charged particles now modifies to
\begin{eqnarray}
\ln\mathcal{Z} = -V\frac{eB}{2\pi^2} \! \sum_{l=0}^{\infty} g_{l} \! \int_{0}^{\infty} \!\! \frac{dk_z}{(2\pi)}  a \ln (1-ae^{-\wkl/T}) \label{eq.lnZ.B}
\end{eqnarray}
where the degeneracy $g_{l}$ can now depend on the Landau level index $l$ as well. For the charged scalars $g_l=2$ whereas for the charged Dirac particles 
$g_l=2(2-\delta_{l}^0)$ implying that the Lowest Landau Level (LLL) is spin non-degenerate. 
The longitudinal pressure and the energy density are given by
\begin{eqnarray}
P = \frac{T}{V}\ln \mathcal{Z} = T \frac{eB}{2\pi^2}\!\sum_{l=0}^{\infty} g_{l}a \int_{0}^{\infty} \!\! \! dk_z \ln\SB{1+af_{a}(\wkl)}, \\
\varepsilon = T\FB{\frac{\del P}{\del T}}_B -P = \frac{eB}{2\pi^2}\!\sum_{l=0}^{\infty}g_{l} \int_{0}^{\infty} \!\! \!  dk_z \wkl f_{a}(\wkl)
\end{eqnarray}
whereas the magnetization becomes,
\begin{eqnarray}
M = \FB{\frac{\del P}{\del B}}_T  = \frac{1}{B} \TB{P - \frac{(eB)^2}{4\pi^2}\sum_{l=0}^{\infty} g_{l} (2l+1-2s) \int_{0}^{\infty} \!\! \!  dk_z \frac{1}{\wkl} f_{a}(\wkl)}.
\end{eqnarray}

The calculation of the quantities $\theta$ and $\phi$ at non-zero magnetic field are obtained from 
\begin{eqnarray}
\theta = \FB{\dfrac{\del P}{\del \varepsilon}}_B = \FB{\dfrac{\del P}{\del T}}_B\bigg/ \FB{\dfrac{\del \varepsilon}{\del T}}_B~, \\
\phi = -B\FB{\dfrac{\del M}{\del \varepsilon}}_B = -B\FB{\dfrac{\del M}{\del T}}_B\bigg/ \FB{\dfrac{\del \varepsilon}{\del T}}_B 
\end{eqnarray}
where,
\begin{eqnarray}
\FB{\dfrac{\del P}{\del T}}_B &=& \frac{\varepsilon+P}{T}, \\
\FB{\dfrac{\del \varepsilon}{\del T}}_B &=&  \frac{1}{T^2}\frac{eB}{2\pi^2} \sum_{l=0}^{\infty} g_{l} \int_{0}^{\infty} \!\! \!  dk_z \wkl^2 f_{a}(\wkl)\SB{1+a f_{a}(\wkl)}, \\
B\FB{\dfrac{\del M}{\del T}}_B &=& \FB{\dfrac{\del P}{\del T}}_B - \frac{1}{T^2}\frac{(eB)^2}{4\pi^2}\sum_{l=0}^{\infty}g_{l} (2l+1-2s)\int_{0}^{\infty} \!\! \!  dk_z  f_{a}(\wkl)\SB{1+af_{a}(\wkl)}.
\end{eqnarray}

\bibliographystyle{apsrev4-1}
\bibliography{snigdha}

\begin{thebibliography}{66}%
\makeatletter
\providecommand \@ifxundefined [1]{%
 \@ifx{#1\undefined}
}%
\providecommand \@ifnum [1]{%
 \ifnum #1\expandafter \@firstoftwo
 \else \expandafter \@secondoftwo
 \fi
}%
\providecommand \@ifx [1]{%
 \ifx #1\expandafter \@firstoftwo
 \else \expandafter \@secondoftwo
 \fi
}%
\providecommand \natexlab [1]{#1}%
\providecommand \enquote  [1]{``#1''}%
\providecommand \bibnamefont  [1]{#1}%
\providecommand \bibfnamefont [1]{#1}%
\providecommand \citenamefont [1]{#1}%
\providecommand \href@noop [0]{\@secondoftwo}%
\providecommand \href [0]{\begingroup \@sanitize@url \@href}%
\providecommand \@href[1]{\@@startlink{#1}\@@href}%
\providecommand \@@href[1]{\endgroup#1\@@endlink}%
\providecommand \@sanitize@url [0]{\catcode `\\12\catcode `\$12\catcode
  `\&12\catcode `\#12\catcode `\^12\catcode `\_12\catcode `\%12\relax}%
\providecommand \@@startlink[1]{}%
\providecommand \@@endlink[0]{}%
\providecommand \url  [0]{\begingroup\@sanitize@url \@url }%
\providecommand \@url [1]{\endgroup\@href {#1}{\urlprefix }}%
\providecommand \urlprefix  [0]{URL }%
\providecommand \Eprint [0]{\href }%
\providecommand \doibase [0]{http://dx.doi.org/}%
\providecommand \selectlanguage [0]{\@gobble}%
\providecommand \bibinfo  [0]{\@secondoftwo}%
\providecommand \bibfield  [0]{\@secondoftwo}%
\providecommand \translation [1]{[#1]}%
\providecommand \BibitemOpen [0]{}%
\providecommand \bibitemStop [0]{}%
\providecommand \bibitemNoStop [0]{.\EOS\space}%
\providecommand \EOS [0]{\spacefactor3000\relax}%
\providecommand \BibitemShut  [1]{\csname bibitem#1\endcsname}%
\let\auto@bib@innerbib\@empty
\bibitem [{\citenamefont {Tuchin}(2013)}]{Tuchin}%
  \BibitemOpen
  \bibfield  {author} {\bibinfo {author} {\bibfnamefont {K.}~\bibnamefont
  {Tuchin}},\ }\href {\doibase 10.1155/2013/490495} {\bibfield  {journal}
  {\bibinfo  {journal} {Adv. High Energy Phys.}\ }\textbf {\bibinfo {volume}
  {2013}},\ \bibinfo {pages} {490495} (\bibinfo {year} {2013})},\ \Eprint
  {http://arxiv.org/abs/1301.0099} {arXiv:1301.0099 [hep-ph]} \BibitemShut
  {NoStop}%
\bibitem [{\citenamefont {Kharzeev}\ \emph {et~al.}(2016)\citenamefont
  {Kharzeev}, \citenamefont {Liao}, \citenamefont {Voloshin},\ and\
  \citenamefont {Wang}}]{Rev1}%
  \BibitemOpen
  \bibfield  {author} {\bibinfo {author} {\bibfnamefont {D.}~\bibnamefont
  {Kharzeev}}, \bibinfo {author} {\bibfnamefont {J.}~\bibnamefont {Liao}},
  \bibinfo {author} {\bibfnamefont {S.}~\bibnamefont {Voloshin}}, \ and\
  \bibinfo {author} {\bibfnamefont {G.}~\bibnamefont {Wang}},\ }\href {\doibase
  10.1016/j.ppnp.2016.01.001} {\bibfield  {journal} {\bibinfo  {journal} {Prog.
  Part. Nucl. Phys.}\ }\textbf {\bibinfo {volume} {88}},\ \bibinfo {pages} {1}
  (\bibinfo {year} {2016})},\ \Eprint {http://arxiv.org/abs/1511.04050}
  {arXiv:1511.04050 [hep-ph]} \BibitemShut {NoStop}%
\bibitem [{\citenamefont {Andersen}\ \emph {et~al.}(2016)\citenamefont
  {Andersen}, \citenamefont {Naylor},\ and\ \citenamefont {Tranberg}}]{Rev2}%
  \BibitemOpen
  \bibfield  {author} {\bibinfo {author} {\bibfnamefont {J.~O.}\ \bibnamefont
  {Andersen}}, \bibinfo {author} {\bibfnamefont {W.~R.}\ \bibnamefont
  {Naylor}}, \ and\ \bibinfo {author} {\bibfnamefont {A.}~\bibnamefont
  {Tranberg}},\ }\href {\doibase 10.1103/RevModPhys.88.025001} {\bibfield
  {journal} {\bibinfo  {journal} {Rev. Mod. Phys.}\ }\textbf {\bibinfo {volume}
  {88}},\ \bibinfo {pages} {025001} (\bibinfo {year} {2016})},\ \Eprint
  {http://arxiv.org/abs/1411.7176} {arXiv:1411.7176 [hep-ph]} \BibitemShut
  {NoStop}%
\bibitem [{\citenamefont {Bandyopadhyay}\ and\ \citenamefont
  {Farias}(2020)}]{Rev3}%
  \BibitemOpen
  \bibfield  {author} {\bibinfo {author} {\bibfnamefont {A.}~\bibnamefont
  {Bandyopadhyay}}\ and\ \bibinfo {author} {\bibfnamefont {R.~L.}\ \bibnamefont
  {Farias}},\ }\href@noop {} {\  (\bibinfo {year} {2020})},\ \Eprint
  {http://arxiv.org/abs/2003.11054} {arXiv:2003.11054 [hep-ph]} \BibitemShut
  {NoStop}%
\bibitem [{\citenamefont {Farias}\ \emph {et~al.}(2014)\citenamefont {Farias},
  \citenamefont {Gomes}, \citenamefont {Krein},\ and\ \citenamefont
  {Pinto}}]{Rev4}%
  \BibitemOpen
  \bibfield  {author} {\bibinfo {author} {\bibfnamefont {R.}~\bibnamefont
  {Farias}}, \bibinfo {author} {\bibfnamefont {K.}~\bibnamefont {Gomes}},
  \bibinfo {author} {\bibfnamefont {G.}~\bibnamefont {Krein}}, \ and\ \bibinfo
  {author} {\bibfnamefont {M.}~\bibnamefont {Pinto}},\ }\href {\doibase
  10.1103/PhysRevC.90.025203} {\bibfield  {journal} {\bibinfo  {journal} {Phys.
  Rev. C}\ }\textbf {\bibinfo {volume} {90}},\ \bibinfo {pages} {025203}
  (\bibinfo {year} {2014})},\ \Eprint {http://arxiv.org/abs/1404.3931}
  {arXiv:1404.3931 [hep-ph]} \BibitemShut {NoStop}%
\bibitem [{\citenamefont {Tuchin}(2012)}]{Tuchin_s}%
  \BibitemOpen
  \bibfield  {author} {\bibinfo {author} {\bibfnamefont {K.}~\bibnamefont
  {Tuchin}},\ }\href {\doibase 10.1088/0954-3899/39/2/025010} {\bibfield
  {journal} {\bibinfo  {journal} {J. Phys. G}\ }\textbf {\bibinfo {volume}
  {39}},\ \bibinfo {pages} {025010} (\bibinfo {year} {2012})},\ \Eprint
  {http://arxiv.org/abs/1108.4394} {arXiv:1108.4394 [nucl-th]} \BibitemShut
  {NoStop}%
\bibitem [{\citenamefont {Li}\ and\ \citenamefont {Yee}(2018)}]{Li_s}%
  \BibitemOpen
  \bibfield  {author} {\bibinfo {author} {\bibfnamefont {S.}~\bibnamefont
  {Li}}\ and\ \bibinfo {author} {\bibfnamefont {H.-U.}\ \bibnamefont {Yee}},\
  }\href {\doibase 10.1103/PhysRevD.97.056024} {\bibfield  {journal} {\bibinfo
  {journal} {Phys. Rev. D}\ }\textbf {\bibinfo {volume} {97}},\ \bibinfo
  {pages} {056024} (\bibinfo {year} {2018})},\ \Eprint
  {http://arxiv.org/abs/1707.00795} {arXiv:1707.00795 [hep-ph]} \BibitemShut
  {NoStop}%
\bibitem [{\citenamefont {Mohanty}\ \emph {et~al.}(2019)\citenamefont
  {Mohanty}, \citenamefont {Dash},\ and\ \citenamefont {Roy}}]{Asutosh_s}%
  \BibitemOpen
  \bibfield  {author} {\bibinfo {author} {\bibfnamefont {P.}~\bibnamefont
  {Mohanty}}, \bibinfo {author} {\bibfnamefont {A.}~\bibnamefont {Dash}}, \
  and\ \bibinfo {author} {\bibfnamefont {V.}~\bibnamefont {Roy}},\ }\href
  {\doibase 10.1140/epja/i2019-12705-7} {\bibfield  {journal} {\bibinfo
  {journal} {Eur. Phys. J. A}\ }\textbf {\bibinfo {volume} {55}},\ \bibinfo
  {pages} {35} (\bibinfo {year} {2019})},\ \Eprint
  {http://arxiv.org/abs/1804.01788} {arXiv:1804.01788 [nucl-th]} \BibitemShut
  {NoStop}%
\bibitem [{\citenamefont {Ghosh}\ \emph
  {et~al.}(2019{\natexlab{a}})\citenamefont {Ghosh}, \citenamefont
  {Chatterjee}, \citenamefont {Mohanty}, \citenamefont {Mukharjee},\ and\
  \citenamefont {Mishra}}]{NJLB_s}%
  \BibitemOpen
  \bibfield  {author} {\bibinfo {author} {\bibfnamefont {S.}~\bibnamefont
  {Ghosh}}, \bibinfo {author} {\bibfnamefont {B.}~\bibnamefont {Chatterjee}},
  \bibinfo {author} {\bibfnamefont {P.}~\bibnamefont {Mohanty}}, \bibinfo
  {author} {\bibfnamefont {A.}~\bibnamefont {Mukharjee}}, \ and\ \bibinfo
  {author} {\bibfnamefont {H.}~\bibnamefont {Mishra}},\ }\href {\doibase
  10.1103/PhysRevD.100.034024} {\bibfield  {journal} {\bibinfo  {journal}
  {Phys. Rev. D}\ }\textbf {\bibinfo {volume} {100}},\ \bibinfo {pages}
  {034024} (\bibinfo {year} {2019}{\natexlab{a}})},\ \Eprint
  {http://arxiv.org/abs/1804.00812} {arXiv:1804.00812 [hep-ph]} \BibitemShut
  {NoStop}%
\bibitem [{\citenamefont {Dey}\ \emph {et~al.}(2019{\natexlab{a}})\citenamefont
  {Dey}, \citenamefont {Satapathy}, \citenamefont {Murmu},\ and\ \citenamefont
  {Ghosh}}]{JD_s}%
  \BibitemOpen
  \bibfield  {author} {\bibinfo {author} {\bibfnamefont {J.}~\bibnamefont
  {Dey}}, \bibinfo {author} {\bibfnamefont {S.}~\bibnamefont {Satapathy}},
  \bibinfo {author} {\bibfnamefont {P.}~\bibnamefont {Murmu}}, \ and\ \bibinfo
  {author} {\bibfnamefont {S.}~\bibnamefont {Ghosh}},\ }\href@noop {} {\
  (\bibinfo {year} {2019}{\natexlab{a}})},\ \Eprint
  {http://arxiv.org/abs/1907.11164} {arXiv:1907.11164 [hep-ph]} \BibitemShut
  {NoStop}%
\bibitem [{\citenamefont {Dey}\ \emph {et~al.}(2019{\natexlab{b}})\citenamefont
  {Dey}, \citenamefont {Satapathy}, \citenamefont {Mishra}, \citenamefont
  {Paul},\ and\ \citenamefont {Ghosh}}]{JD_QP_s}%
  \BibitemOpen
  \bibfield  {author} {\bibinfo {author} {\bibfnamefont {J.}~\bibnamefont
  {Dey}}, \bibinfo {author} {\bibfnamefont {S.}~\bibnamefont {Satapathy}},
  \bibinfo {author} {\bibfnamefont {A.}~\bibnamefont {Mishra}}, \bibinfo
  {author} {\bibfnamefont {S.}~\bibnamefont {Paul}}, \ and\ \bibinfo {author}
  {\bibfnamefont {S.}~\bibnamefont {Ghosh}},\ }\href@noop {} {\  (\bibinfo
  {year} {2019}{\natexlab{b}})},\ \Eprint {http://arxiv.org/abs/1908.04335}
  {arXiv:1908.04335 [hep-ph]} \BibitemShut {NoStop}%
\bibitem [{\citenamefont {Das}\ \emph {et~al.}(2019{\natexlab{a}})\citenamefont
  {Das}, \citenamefont {Mishra},\ and\ \citenamefont {Mohapatra}}]{Arpan_s}%
  \BibitemOpen
  \bibfield  {author} {\bibinfo {author} {\bibfnamefont {A.}~\bibnamefont
  {Das}}, \bibinfo {author} {\bibfnamefont {H.}~\bibnamefont {Mishra}}, \ and\
  \bibinfo {author} {\bibfnamefont {R.~K.}\ \bibnamefont {Mohapatra}},\ }\href
  {\doibase 10.1103/PhysRevD.100.114004} {\bibfield  {journal} {\bibinfo
  {journal} {Phys. Rev. D}\ }\textbf {\bibinfo {volume} {100}},\ \bibinfo
  {pages} {114004} (\bibinfo {year} {2019}{\natexlab{a}})},\ \Eprint
  {http://arxiv.org/abs/1909.06202} {arXiv:1909.06202 [hep-ph]} \BibitemShut
  {NoStop}%
\bibitem [{\citenamefont {Kurian}\ \emph {et~al.}(2019)\citenamefont {Kurian},
  \citenamefont {Mitra}, \citenamefont {Ghosh},\ and\ \citenamefont
  {Chandra}}]{Manu_s1}%
  \BibitemOpen
  \bibfield  {author} {\bibinfo {author} {\bibfnamefont {M.}~\bibnamefont
  {Kurian}}, \bibinfo {author} {\bibfnamefont {S.}~\bibnamefont {Mitra}},
  \bibinfo {author} {\bibfnamefont {S.}~\bibnamefont {Ghosh}}, \ and\ \bibinfo
  {author} {\bibfnamefont {V.}~\bibnamefont {Chandra}},\ }\href {\doibase
  10.1140/epjc/s10052-019-6649-z} {\bibfield  {journal} {\bibinfo  {journal}
  {Eur. Phys. J. C}\ }\textbf {\bibinfo {volume} {79}},\ \bibinfo {pages} {134}
  (\bibinfo {year} {2019})},\ \Eprint {http://arxiv.org/abs/1805.07313}
  {arXiv:1805.07313 [nucl-th]} \BibitemShut {NoStop}%
\bibitem [{\citenamefont {Denicol}\ \emph {et~al.}(2018)\citenamefont
  {Denicol}, \citenamefont {Huang}, \citenamefont {Moln\'ar}, \citenamefont
  {Monteiro}, \citenamefont {Niemi}, \citenamefont {Noronha}, \citenamefont
  {Rischke},\ and\ \citenamefont {Wang}}]{Denicol_s}%
  \BibitemOpen
  \bibfield  {author} {\bibinfo {author} {\bibfnamefont {G.~S.}\ \bibnamefont
  {Denicol}}, \bibinfo {author} {\bibfnamefont {X.-G.}\ \bibnamefont {Huang}},
  \bibinfo {author} {\bibfnamefont {E.}~\bibnamefont {Moln\'ar}}, \bibinfo
  {author} {\bibfnamefont {G.~M.}\ \bibnamefont {Monteiro}}, \bibinfo {author}
  {\bibfnamefont {H.}~\bibnamefont {Niemi}}, \bibinfo {author} {\bibfnamefont
  {J.}~\bibnamefont {Noronha}}, \bibinfo {author} {\bibfnamefont {D.~H.}\
  \bibnamefont {Rischke}}, \ and\ \bibinfo {author} {\bibfnamefont
  {Q.}~\bibnamefont {Wang}},\ }\href {\doibase 10.1103/PhysRevD.98.076009}
  {\bibfield  {journal} {\bibinfo  {journal} {Phys. Rev. D}\ }\textbf {\bibinfo
  {volume} {98}},\ \bibinfo {pages} {076009} (\bibinfo {year} {2018})},\
  \Eprint {http://arxiv.org/abs/1804.05210} {arXiv:1804.05210 [nucl-th]}
  \BibitemShut {NoStop}%
\bibitem [{\citenamefont {Chen}\ \emph {et~al.}(2020)\citenamefont {Chen},
  \citenamefont {Greiner}, \citenamefont {Huang},\ and\ \citenamefont
  {Xu}}]{Greiner_s}%
  \BibitemOpen
  \bibfield  {author} {\bibinfo {author} {\bibfnamefont {Z.}~\bibnamefont
  {Chen}}, \bibinfo {author} {\bibfnamefont {C.}~\bibnamefont {Greiner}},
  \bibinfo {author} {\bibfnamefont {A.}~\bibnamefont {Huang}}, \ and\ \bibinfo
  {author} {\bibfnamefont {Z.}~\bibnamefont {Xu}},\ }\href {\doibase
  10.1103/PhysRevD.101.056020} {\bibfield  {journal} {\bibinfo  {journal}
  {Phys. Rev. D}\ }\textbf {\bibinfo {volume} {101}},\ \bibinfo {pages}
  {056020} (\bibinfo {year} {2020})},\ \Eprint
  {http://arxiv.org/abs/1910.13721} {arXiv:1910.13721 [hep-ph]} \BibitemShut
  {NoStop}%
\bibitem [{\citenamefont {Jaiswal}\ \emph {et~al.}(2021)\citenamefont {Jaiswal}
  \emph {et~al.}}]{DQCD}%
  \BibitemOpen
  \bibfield  {author} {\bibinfo {author} {\bibfnamefont {A.}~\bibnamefont
  {Jaiswal}} \emph {et~al.},\ }\href {\doibase 10.1142/S0218301321300010}
  {\bibfield  {journal} {\bibinfo  {journal} {Int. J. Mod. Phys. E}\ }\textbf
  {\bibinfo {volume} {30}},\ \bibinfo {pages} {2130001} (\bibinfo {year}
  {2021})},\ \Eprint {http://arxiv.org/abs/2007.14959} {arXiv:2007.14959
  [hep-ph]} \BibitemShut {NoStop}%
\bibitem [{\citenamefont {Dash}\ \emph {et~al.}(2020)\citenamefont {Dash},
  \citenamefont {Samanta}, \citenamefont {Dey}, \citenamefont {Gangopadhyaya},
  \citenamefont {Ghosh},\ and\ \citenamefont {Roy}}]{G_HRGB}%
  \BibitemOpen
  \bibfield  {author} {\bibinfo {author} {\bibfnamefont {A.}~\bibnamefont
  {Dash}}, \bibinfo {author} {\bibfnamefont {S.}~\bibnamefont {Samanta}},
  \bibinfo {author} {\bibfnamefont {J.}~\bibnamefont {Dey}}, \bibinfo {author}
  {\bibfnamefont {U.}~\bibnamefont {Gangopadhyaya}}, \bibinfo {author}
  {\bibfnamefont {S.}~\bibnamefont {Ghosh}}, \ and\ \bibinfo {author}
  {\bibfnamefont {V.}~\bibnamefont {Roy}},\ }\href {\doibase
  10.1103/PhysRevD.102.016016} {\bibfield  {journal} {\bibinfo  {journal}
  {Phys. Rev. D}\ }\textbf {\bibinfo {volume} {102}},\ \bibinfo {pages}
  {016016} (\bibinfo {year} {2020})},\ \Eprint
  {http://arxiv.org/abs/2002.08781} {arXiv:2002.08781 [nucl-th]} \BibitemShut
  {NoStop}%
\bibitem [{\citenamefont {Hattori}\ \emph
  {et~al.}(2017{\natexlab{a}})\citenamefont {Hattori}, \citenamefont {Huang},
  \citenamefont {Rischke},\ and\ \citenamefont {Satow}}]{Hattori_b}%
  \BibitemOpen
  \bibfield  {author} {\bibinfo {author} {\bibfnamefont {K.}~\bibnamefont
  {Hattori}}, \bibinfo {author} {\bibfnamefont {X.-G.}\ \bibnamefont {Huang}},
  \bibinfo {author} {\bibfnamefont {D.~H.}\ \bibnamefont {Rischke}}, \ and\
  \bibinfo {author} {\bibfnamefont {D.}~\bibnamefont {Satow}},\ }\href
  {\doibase 10.1103/PhysRevD.96.094009} {\bibfield  {journal} {\bibinfo
  {journal} {Phys. Rev. D}\ }\textbf {\bibinfo {volume} {96}},\ \bibinfo
  {pages} {094009} (\bibinfo {year} {2017}{\natexlab{a}})},\ \Eprint
  {http://arxiv.org/abs/1708.00515} {arXiv:1708.00515 [hep-ph]} \BibitemShut
  {NoStop}%
\bibitem [{\citenamefont {Agasian}(2013)}]{Agasian_b1}%
  \BibitemOpen
  \bibfield  {author} {\bibinfo {author} {\bibfnamefont {N.}~\bibnamefont
  {Agasian}},\ }\href {\doibase 10.1134/S1063778813100025} {\bibfield
  {journal} {\bibinfo  {journal} {Phys. Atom. Nucl.}\ }\textbf {\bibinfo
  {volume} {76}},\ \bibinfo {pages} {1382} (\bibinfo {year}
  {2013})}\BibitemShut {NoStop}%
\bibitem [{\citenamefont {Agasian}(2012)}]{Agasian_b2}%
  \BibitemOpen
  \bibfield  {author} {\bibinfo {author} {\bibfnamefont {N.}~\bibnamefont
  {Agasian}},\ }\href {\doibase 10.1134/S0021364012040029} {\bibfield
  {journal} {\bibinfo  {journal} {JETP Lett.}\ }\textbf {\bibinfo {volume}
  {95}},\ \bibinfo {pages} {171} (\bibinfo {year} {2012})},\ \Eprint
  {http://arxiv.org/abs/1109.5849} {arXiv:1109.5849 [hep-ph]} \BibitemShut
  {NoStop}%
\bibitem [{\citenamefont {Kadam}(2015)}]{Kadam_b}%
  \BibitemOpen
  \bibfield  {author} {\bibinfo {author} {\bibfnamefont {G.}~\bibnamefont
  {Kadam}},\ }\href {\doibase 10.1142/S0217732315500315} {\bibfield  {journal}
  {\bibinfo  {journal} {Mod. Phys. Lett. A}\ }\textbf {\bibinfo {volume}
  {30}},\ \bibinfo {pages} {1550031} (\bibinfo {year} {2015})},\ \Eprint
  {http://arxiv.org/abs/1412.5303} {arXiv:1412.5303 [hep-ph]} \BibitemShut
  {NoStop}%
\bibitem [{\citenamefont {Hattori}\ and\ \citenamefont
  {Satow}(2016)}]{Hattori_e1}%
  \BibitemOpen
  \bibfield  {author} {\bibinfo {author} {\bibfnamefont {K.}~\bibnamefont
  {Hattori}}\ and\ \bibinfo {author} {\bibfnamefont {D.}~\bibnamefont
  {Satow}},\ }\href {\doibase 10.1103/PhysRevD.94.114032} {\bibfield  {journal}
  {\bibinfo  {journal} {Phys. Rev. D}\ }\textbf {\bibinfo {volume} {94}},\
  \bibinfo {pages} {114032} (\bibinfo {year} {2016})},\ \Eprint
  {http://arxiv.org/abs/1610.06818} {arXiv:1610.06818 [hep-ph]} \BibitemShut
  {NoStop}%
\bibitem [{\citenamefont {Hattori}\ \emph
  {et~al.}(2017{\natexlab{b}})\citenamefont {Hattori}, \citenamefont {Li},
  \citenamefont {Satow},\ and\ \citenamefont {Yee}}]{Hattori_e2}%
  \BibitemOpen
  \bibfield  {author} {\bibinfo {author} {\bibfnamefont {K.}~\bibnamefont
  {Hattori}}, \bibinfo {author} {\bibfnamefont {S.}~\bibnamefont {Li}},
  \bibinfo {author} {\bibfnamefont {D.}~\bibnamefont {Satow}}, \ and\ \bibinfo
  {author} {\bibfnamefont {H.-U.}\ \bibnamefont {Yee}},\ }\href {\doibase
  10.1103/PhysRevD.95.076008} {\bibfield  {journal} {\bibinfo  {journal} {Phys.
  Rev. D}\ }\textbf {\bibinfo {volume} {95}},\ \bibinfo {pages} {076008}
  (\bibinfo {year} {2017}{\natexlab{b}})},\ \Eprint
  {http://arxiv.org/abs/1610.06839} {arXiv:1610.06839 [hep-ph]} \BibitemShut
  {NoStop}%
\bibitem [{\citenamefont {Kurian}\ and\ \citenamefont
  {Chandra}(2017)}]{Manu_e1}%
  \BibitemOpen
  \bibfield  {author} {\bibinfo {author} {\bibfnamefont {M.}~\bibnamefont
  {Kurian}}\ and\ \bibinfo {author} {\bibfnamefont {V.}~\bibnamefont
  {Chandra}},\ }\href {\doibase 10.1103/PhysRevD.96.114026} {\bibfield
  {journal} {\bibinfo  {journal} {Phys. Rev. D}\ }\textbf {\bibinfo {volume}
  {96}},\ \bibinfo {pages} {114026} (\bibinfo {year} {2017})},\ \Eprint
  {http://arxiv.org/abs/1709.08320} {arXiv:1709.08320 [nucl-th]} \BibitemShut
  {NoStop}%
\bibitem [{\citenamefont {Harutyunyan}\ and\ \citenamefont
  {Sedrakian}(2016)}]{Sedrakian_e1}%
  \BibitemOpen
  \bibfield  {author} {\bibinfo {author} {\bibfnamefont {A.}~\bibnamefont
  {Harutyunyan}}\ and\ \bibinfo {author} {\bibfnamefont {A.}~\bibnamefont
  {Sedrakian}},\ }\href {\doibase 10.1103/PhysRevC.94.025805} {\bibfield
  {journal} {\bibinfo  {journal} {Phys. Rev. C}\ }\textbf {\bibinfo {volume}
  {94}},\ \bibinfo {pages} {025805} (\bibinfo {year} {2016})},\ \Eprint
  {http://arxiv.org/abs/1605.07612} {arXiv:1605.07612 [astro-ph.HE]}
  \BibitemShut {NoStop}%
\bibitem [{\citenamefont {Nam}(2012)}]{Nam_e}%
  \BibitemOpen
  \bibfield  {author} {\bibinfo {author} {\bibfnamefont {S.-i.}\ \bibnamefont
  {Nam}},\ }\href {\doibase 10.1103/PhysRevD.86.033014} {\bibfield  {journal}
  {\bibinfo  {journal} {Phys. Rev. D}\ }\textbf {\bibinfo {volume} {86}},\
  \bibinfo {pages} {033014} (\bibinfo {year} {2012})},\ \Eprint
  {http://arxiv.org/abs/1207.3172} {arXiv:1207.3172 [hep-ph]} \BibitemShut
  {NoStop}%
\bibitem [{\citenamefont {Das}\ \emph {et~al.}(2020)\citenamefont {Das},
  \citenamefont {Mishra},\ and\ \citenamefont {Mohapatra}}]{Arpan_e1}%
  \BibitemOpen
  \bibfield  {author} {\bibinfo {author} {\bibfnamefont {A.}~\bibnamefont
  {Das}}, \bibinfo {author} {\bibfnamefont {H.}~\bibnamefont {Mishra}}, \ and\
  \bibinfo {author} {\bibfnamefont {R.~K.}\ \bibnamefont {Mohapatra}},\ }\href
  {\doibase 10.1103/PhysRevD.101.034027} {\bibfield  {journal} {\bibinfo
  {journal} {Phys. Rev. D}\ }\textbf {\bibinfo {volume} {101}},\ \bibinfo
  {pages} {034027} (\bibinfo {year} {2020})},\ \Eprint
  {http://arxiv.org/abs/1907.05298} {arXiv:1907.05298 [hep-ph]} \BibitemShut
  {NoStop}%
\bibitem [{\citenamefont {Das}\ \emph {et~al.}(2019{\natexlab{b}})\citenamefont
  {Das}, \citenamefont {Mishra},\ and\ \citenamefont {Mohapatra}}]{Arpan_e2}%
  \BibitemOpen
  \bibfield  {author} {\bibinfo {author} {\bibfnamefont {A.}~\bibnamefont
  {Das}}, \bibinfo {author} {\bibfnamefont {H.}~\bibnamefont {Mishra}}, \ and\
  \bibinfo {author} {\bibfnamefont {R.~K.}\ \bibnamefont {Mohapatra}},\ }\href
  {\doibase 10.1103/PhysRevD.99.094031} {\bibfield  {journal} {\bibinfo
  {journal} {Phys. Rev. D}\ }\textbf {\bibinfo {volume} {99}},\ \bibinfo
  {pages} {094031} (\bibinfo {year} {2019}{\natexlab{b}})},\ \Eprint
  {http://arxiv.org/abs/1903.03938} {arXiv:1903.03938 [hep-ph]} \BibitemShut
  {NoStop}%
\bibitem [{\citenamefont {Ghosh}\ \emph {et~al.}(2002)\citenamefont {Ghosh},
  \citenamefont {Ghosh}, \citenamefont {Goswami}, \citenamefont {Chakrabarty},\
  and\ \citenamefont {Goyal}}]{Somnath_e}%
  \BibitemOpen
  \bibfield  {author} {\bibinfo {author} {\bibfnamefont {S.}~\bibnamefont
  {Ghosh}}, \bibinfo {author} {\bibfnamefont {S.}~\bibnamefont {Ghosh}},
  \bibinfo {author} {\bibfnamefont {K.}~\bibnamefont {Goswami}}, \bibinfo
  {author} {\bibfnamefont {S.}~\bibnamefont {Chakrabarty}}, \ and\ \bibinfo
  {author} {\bibfnamefont {A.}~\bibnamefont {Goyal}},\ }\href {\doibase
  10.1142/S0218271802002098} {\bibfield  {journal} {\bibinfo  {journal} {Int.
  J. Mod. Phys. D}\ }\textbf {\bibinfo {volume} {11}},\ \bibinfo {pages} {843}
  (\bibinfo {year} {2002})},\ \Eprint {http://arxiv.org/abs/astro-ph/0106153}
  {arXiv:astro-ph/0106153} \BibitemShut {NoStop}%
\bibitem [{\citenamefont {Ghosh}\ \emph {et~al.}(2020)\citenamefont {Ghosh},
  \citenamefont {Bandyopadhyay}, \citenamefont {Farias}, \citenamefont {Dey},\
  and\ \citenamefont {Krein}}]{G_NJLB_e}%
  \BibitemOpen
  \bibfield  {author} {\bibinfo {author} {\bibfnamefont {S.}~\bibnamefont
  {Ghosh}}, \bibinfo {author} {\bibfnamefont {A.}~\bibnamefont
  {Bandyopadhyay}}, \bibinfo {author} {\bibfnamefont {R.~L.~S.}\ \bibnamefont
  {Farias}}, \bibinfo {author} {\bibfnamefont {J.}~\bibnamefont {Dey}}, \ and\
  \bibinfo {author} {\bibfnamefont {G.~a.}\ \bibnamefont {Krein}},\ }\href
  {\doibase 10.1103/PhysRevD.102.114015} {\bibfield  {journal} {\bibinfo
  {journal} {Phys. Rev. D}\ }\textbf {\bibinfo {volume} {102}},\ \bibinfo
  {pages} {114015} (\bibinfo {year} {2020})},\ \Eprint
  {http://arxiv.org/abs/1911.10005} {arXiv:1911.10005 [hep-ph]} \BibitemShut
  {NoStop}%
\bibitem [{\citenamefont {Thakur}\ and\ \citenamefont
  {Srivastava}(2019)}]{Lata_e}%
  \BibitemOpen
  \bibfield  {author} {\bibinfo {author} {\bibfnamefont {L.}~\bibnamefont
  {Thakur}}\ and\ \bibinfo {author} {\bibfnamefont {P.}~\bibnamefont
  {Srivastava}},\ }\href {\doibase 10.1103/PhysRevD.100.076016} {\bibfield
  {journal} {\bibinfo  {journal} {Phys. Rev. D}\ }\textbf {\bibinfo {volume}
  {100}},\ \bibinfo {pages} {076016} (\bibinfo {year} {2019})},\ \Eprint
  {http://arxiv.org/abs/1910.12087} {arXiv:1910.12087 [hep-ph]} \BibitemShut
  {NoStop}%
\bibitem [{\citenamefont {Samanta}\ \emph {et~al.}(2020)\citenamefont
  {Samanta}, \citenamefont {Dey}, \citenamefont {Satapathy},\ and\
  \citenamefont {Ghosh}}]{G_HRGBQM_e}%
  \BibitemOpen
  \bibfield  {author} {\bibinfo {author} {\bibfnamefont {S.}~\bibnamefont
  {Samanta}}, \bibinfo {author} {\bibfnamefont {J.}~\bibnamefont {Dey}},
  \bibinfo {author} {\bibfnamefont {S.}~\bibnamefont {Satapathy}}, \ and\
  \bibinfo {author} {\bibfnamefont {S.}~\bibnamefont {Ghosh}},\ }\href@noop {}
  {\  (\bibinfo {year} {2020})},\ \Eprint {http://arxiv.org/abs/2002.04434}
  {arXiv:2002.04434 [nucl-th]} \BibitemShut {NoStop}%
\bibitem [{\citenamefont {Rath}\ and\ \citenamefont {Patra}(2020)}]{Patra_e}%
  \BibitemOpen
  \bibfield  {author} {\bibinfo {author} {\bibfnamefont {S.}~\bibnamefont
  {Rath}}\ and\ \bibinfo {author} {\bibfnamefont {B.~K.}\ \bibnamefont
  {Patra}},\ }\href {\doibase 10.1140/epjc/s10052-020-8331-x} {\bibfield
  {journal} {\bibinfo  {journal} {Eur. Phys. J. C}\ }\textbf {\bibinfo {volume}
  {80}},\ \bibinfo {pages} {747} (\bibinfo {year} {2020})},\ \Eprint
  {http://arxiv.org/abs/2005.00997} {arXiv:2005.00997 [hep-ph]} \BibitemShut
  {NoStop}%
\bibitem [{\citenamefont {Kalikotay}\ \emph {et~al.}(2020)\citenamefont
  {Kalikotay}, \citenamefont {Ghosh}, \citenamefont {Chaudhuri}, \citenamefont
  {Roy},\ and\ \citenamefont {Sarkar}}]{SS_PR_e}%
  \BibitemOpen
  \bibfield  {author} {\bibinfo {author} {\bibfnamefont {P.}~\bibnamefont
  {Kalikotay}}, \bibinfo {author} {\bibfnamefont {S.}~\bibnamefont {Ghosh}},
  \bibinfo {author} {\bibfnamefont {N.}~\bibnamefont {Chaudhuri}}, \bibinfo
  {author} {\bibfnamefont {P.}~\bibnamefont {Roy}}, \ and\ \bibinfo {author}
  {\bibfnamefont {S.}~\bibnamefont {Sarkar}},\ }\href {\doibase
  10.1103/PhysRevD.102.076007} {\bibfield  {journal} {\bibinfo  {journal}
  {Phys. Rev. D}\ }\textbf {\bibinfo {volume} {102}},\ \bibinfo {pages}
  {076007} (\bibinfo {year} {2020})},\ \Eprint
  {http://arxiv.org/abs/2009.10493} {arXiv:2009.10493 [hep-ph]} \BibitemShut
  {NoStop}%
\bibitem [{\citenamefont {Lifshitz}\ and\ \citenamefont
  {L.P.}(1981)}]{Landau_10}%
  \BibitemOpen
  \bibfield  {author} {\bibinfo {author} {\bibfnamefont {E.}~\bibnamefont
  {Lifshitz}}\ and\ \bibinfo {author} {\bibfnamefont {P.}~\bibnamefont
  {L.P.}},\ }\href {\doibase 10.1016/C2009-0-25523-1} {\emph {\bibinfo {title}
  {{Physical Kinetics}}}}\ (\bibinfo  {publisher} {Elsevier India},\ \bibinfo
  {year} {1981})\BibitemShut {NoStop}%
\bibitem [{\citenamefont {Huang}\ \emph {et~al.}(2011)\citenamefont {Huang},
  \citenamefont {Sedrakian},\ and\ \citenamefont {Rischke}}]{XGH1}%
  \BibitemOpen
  \bibfield  {author} {\bibinfo {author} {\bibfnamefont {X.-G.}\ \bibnamefont
  {Huang}}, \bibinfo {author} {\bibfnamefont {A.}~\bibnamefont {Sedrakian}}, \
  and\ \bibinfo {author} {\bibfnamefont {D.~H.}\ \bibnamefont {Rischke}},\
  }\href {\doibase 10.1016/j.aop.2011.08.001} {\bibfield  {journal} {\bibinfo
  {journal} {Annals Phys.}\ }\textbf {\bibinfo {volume} {326}},\ \bibinfo
  {pages} {3075} (\bibinfo {year} {2011})},\ \Eprint
  {http://arxiv.org/abs/1108.0602} {arXiv:1108.0602 [astro-ph.HE]} \BibitemShut
  {NoStop}%
\bibitem [{\citenamefont {Kubo}(1957)}]{Kubo}%
  \BibitemOpen
  \bibfield  {author} {\bibinfo {author} {\bibfnamefont {R.}~\bibnamefont
  {Kubo}},\ }\href {\doibase 10.1143/JPSJ.12.570} {\bibfield  {journal}
  {\bibinfo  {journal} {J. Phys. Soc. Jap.}\ }\textbf {\bibinfo {volume}
  {12}},\ \bibinfo {pages} {570} (\bibinfo {year} {1957})}\BibitemShut
  {NoStop}%
\bibitem [{\citenamefont {Ghosh}(2014{\natexlab{a}})}]{Ghosh:2014yea}%
  \BibitemOpen
  \bibfield  {author} {\bibinfo {author} {\bibfnamefont {S.}~\bibnamefont
  {Ghosh}},\ }\href {\doibase 10.1142/S0217751X14500547} {\bibfield  {journal}
  {\bibinfo  {journal} {Int. J. Mod. Phys. A}\ }\textbf {\bibinfo {volume}
  {29}},\ \bibinfo {pages} {1450054} (\bibinfo {year} {2014}{\natexlab{a}})},\
  \Eprint {http://arxiv.org/abs/1404.4788} {arXiv:1404.4788 [nucl-th]}
  \BibitemShut {NoStop}%
\bibitem [{\citenamefont {Ghosh}(2014{\natexlab{b}})}]{G_etaN}%
  \BibitemOpen
  \bibfield  {author} {\bibinfo {author} {\bibfnamefont {S.}~\bibnamefont
  {Ghosh}},\ }\href {\doibase 10.1103/PhysRevC.90.025202} {\bibfield  {journal}
  {\bibinfo  {journal} {Phys. Rev. C}\ }\textbf {\bibinfo {volume} {90}},\
  \bibinfo {pages} {025202} (\bibinfo {year} {2014}{\natexlab{b}})},\ \Eprint
  {http://arxiv.org/abs/1503.06927} {arXiv:1503.06927 [nucl-th]} \BibitemShut
  {NoStop}%
\bibitem [{\citenamefont {Fernandez-Fraile}\ and\ \citenamefont
  {Gomez~Nicola}(2009)}]{Nicola}%
  \BibitemOpen
  \bibfield  {author} {\bibinfo {author} {\bibfnamefont {D.}~\bibnamefont
  {Fernandez-Fraile}}\ and\ \bibinfo {author} {\bibfnamefont {A.}~\bibnamefont
  {Gomez~Nicola}},\ }\href {\doibase 10.1140/epjc/s10052-009-0935-0} {\bibfield
   {journal} {\bibinfo  {journal} {Eur. Phys. J. C}\ }\textbf {\bibinfo
  {volume} {62}},\ \bibinfo {pages} {37} (\bibinfo {year} {2009})},\ \Eprint
  {http://arxiv.org/abs/0902.4829} {arXiv:0902.4829 [hep-ph]} \BibitemShut
  {NoStop}%
\bibitem [{\citenamefont {Lang}\ \emph {et~al.}(2012)\citenamefont {Lang},
  \citenamefont {Kaiser},\ and\ \citenamefont {Weise}}]{Lang}%
  \BibitemOpen
  \bibfield  {author} {\bibinfo {author} {\bibfnamefont {R.}~\bibnamefont
  {Lang}}, \bibinfo {author} {\bibfnamefont {N.}~\bibnamefont {Kaiser}}, \ and\
  \bibinfo {author} {\bibfnamefont {W.}~\bibnamefont {Weise}},\ }\href
  {\doibase 10.1140/epja/i2012-12109-3} {\bibfield  {journal} {\bibinfo
  {journal} {Eur. Phys. J. A}\ }\textbf {\bibinfo {volume} {48}},\ \bibinfo
  {pages} {109} (\bibinfo {year} {2012})},\ \Eprint
  {http://arxiv.org/abs/1205.6648} {arXiv:1205.6648 [hep-ph]} \BibitemShut
  {NoStop}%
\bibitem [{\citenamefont {Jeon}(1995)}]{Jeon}%
  \BibitemOpen
  \bibfield  {author} {\bibinfo {author} {\bibfnamefont {S.}~\bibnamefont
  {Jeon}},\ }\href {\doibase 10.1103/PhysRevD.52.3591} {\bibfield  {journal}
  {\bibinfo  {journal} {Phys. Rev. D}\ }\textbf {\bibinfo {volume} {52}},\
  \bibinfo {pages} {3591} (\bibinfo {year} {1995})},\ \Eprint
  {http://arxiv.org/abs/hep-ph/9409250} {arXiv:hep-ph/9409250} \BibitemShut
  {NoStop}%
\bibitem [{\citenamefont {Gavin}(1985)}]{Gavin}%
  \BibitemOpen
  \bibfield  {author} {\bibinfo {author} {\bibfnamefont {S.}~\bibnamefont
  {Gavin}},\ }\href {\doibase 10.1016/0375-9474(85)90190-3} {\bibfield
  {journal} {\bibinfo  {journal} {Nucl. Phys. A}\ }\textbf {\bibinfo {volume}
  {435}},\ \bibinfo {pages} {826} (\bibinfo {year} {1985})}\BibitemShut
  {NoStop}%
\bibitem [{\citenamefont {Chakraborty}\ and\ \citenamefont
  {Kapusta}(2011)}]{Kapusta}%
  \BibitemOpen
  \bibfield  {author} {\bibinfo {author} {\bibfnamefont {P.}~\bibnamefont
  {Chakraborty}}\ and\ \bibinfo {author} {\bibfnamefont {J.}~\bibnamefont
  {Kapusta}},\ }\href {\doibase 10.1103/PhysRevC.83.014906} {\bibfield
  {journal} {\bibinfo  {journal} {Phys. Rev. C}\ }\textbf {\bibinfo {volume}
  {83}},\ \bibinfo {pages} {014906} (\bibinfo {year} {2011})},\ \Eprint
  {http://arxiv.org/abs/1006.0257} {arXiv:1006.0257 [nucl-th]} \BibitemShut
  {NoStop}%
\bibitem [{\citenamefont {Bellac}(2011)}]{Bellac:2011kqa}%
  \BibitemOpen
  \bibfield  {author} {\bibinfo {author} {\bibfnamefont {M.~L.}\ \bibnamefont
  {Bellac}},\ }\href {\doibase 10.1017/CBO9780511721700} {\emph {\bibinfo
  {title} {{Thermal Field Theory}}}},\ Cambridge Monographs on Mathematical
  Physics\ (\bibinfo  {publisher} {Cambridge University Press},\ \bibinfo
  {year} {2011})\BibitemShut {NoStop}%
\bibitem [{\citenamefont {Mallik}\ and\ \citenamefont
  {Sarkar}(2016)}]{Mallik:2016anp}%
  \BibitemOpen
  \bibfield  {author} {\bibinfo {author} {\bibfnamefont {S.}~\bibnamefont
  {Mallik}}\ and\ \bibinfo {author} {\bibfnamefont {S.}~\bibnamefont
  {Sarkar}},\ }\href {\doibase 10.1017/9781316535585} {\emph {\bibinfo {title}
  {{Hadrons at Finite Temperature}}}}\ (\bibinfo  {publisher} {Cambridge
  University Press},\ \bibinfo {address} {Cambridge},\ \bibinfo {year}
  {2016})\BibitemShut {NoStop}%
\bibitem [{\citenamefont {Greiner}\ and\ \citenamefont
  {Reinhardt}(1996)}]{Greiner:1996zu}%
  \BibitemOpen
  \bibfield  {author} {\bibinfo {author} {\bibfnamefont {W.}~\bibnamefont
  {Greiner}}\ and\ \bibinfo {author} {\bibfnamefont {J.}~\bibnamefont
  {Reinhardt}},\ }\href@noop {} {\emph {\bibinfo {title} {{Field
  quantization}}}}\ (\bibinfo {year} {1996})\BibitemShut {NoStop}%
\bibitem [{\citenamefont {Peskin}\ and\ \citenamefont
  {Schroeder}(1995)}]{Peskin:1995ev}%
  \BibitemOpen
  \bibfield  {author} {\bibinfo {author} {\bibfnamefont {M.~E.}\ \bibnamefont
  {Peskin}}\ and\ \bibinfo {author} {\bibfnamefont {D.~V.}\ \bibnamefont
  {Schroeder}},\ }\href@noop {} {\emph {\bibinfo {title} {{An Introduction to
  quantum field theory}}}}\ (\bibinfo  {publisher} {Addison-Wesley},\ \bibinfo
  {address} {Reading, USA},\ \bibinfo {year} {1995})\BibitemShut {NoStop}%
\bibitem [{\citenamefont {Lahiri}\ and\ \citenamefont
  {Pal}(2005)}]{Lahiri:2005sm}%
  \BibitemOpen
  \bibfield  {author} {\bibinfo {author} {\bibfnamefont {A.}~\bibnamefont
  {Lahiri}}\ and\ \bibinfo {author} {\bibfnamefont {P.~B.}\ \bibnamefont
  {Pal}},\ }\href@noop {} {\emph {\bibinfo {title} {{A First Book of Quantum
  Field Theory}}}}\ (\bibinfo {year} {2005})\BibitemShut {NoStop}%
\bibitem [{\citenamefont {Schwartz}(2013)}]{Schwartz}%
  \BibitemOpen
  \bibfield  {author} {\bibinfo {author} {\bibfnamefont {M.~D.}\ \bibnamefont
  {Schwartz}},\ }\href
  {https://www.cambridge.org/tw/academic/subjects/physics/theoretical-physics-and-mathematical-physics/quantum-field-theory-and-standard-model}
  {\emph {\bibinfo {title} {{Quantum Field Theory and the Standard Model}}}}\
  (\bibinfo  {publisher} {Cambridge University Press},\ \bibinfo {year}
  {2013})\BibitemShut {NoStop}%
\bibitem [{\citenamefont {Ghosh}\ \emph
  {et~al.}(2019{\natexlab{b}})\citenamefont {Ghosh}, \citenamefont {Mukherjee},
  \citenamefont {Roy},\ and\ \citenamefont {Sarkar}}]{Ghosh:2019fet}%
  \BibitemOpen
  \bibfield  {author} {\bibinfo {author} {\bibfnamefont {S.}~\bibnamefont
  {Ghosh}}, \bibinfo {author} {\bibfnamefont {A.}~\bibnamefont {Mukherjee}},
  \bibinfo {author} {\bibfnamefont {P.}~\bibnamefont {Roy}}, \ and\ \bibinfo
  {author} {\bibfnamefont {S.}~\bibnamefont {Sarkar}},\ }\href {\doibase
  10.1103/PhysRevD.99.096004} {\bibfield  {journal} {\bibinfo  {journal} {Phys.
  Rev. D}\ }\textbf {\bibinfo {volume} {99}},\ \bibinfo {pages} {096004}
  (\bibinfo {year} {2019}{\natexlab{b}})},\ \Eprint
  {http://arxiv.org/abs/1901.02290} {arXiv:1901.02290 [hep-ph]} \BibitemShut
  {NoStop}%
\bibitem [{\citenamefont {Ghosh}\ \emph {et~al.}(2017)\citenamefont {Ghosh},
  \citenamefont {Mukherjee}, \citenamefont {Mandal}, \citenamefont {Sarkar},\
  and\ \citenamefont {Roy}}]{Ghosh:2017rjo}%
  \BibitemOpen
  \bibfield  {author} {\bibinfo {author} {\bibfnamefont {S.}~\bibnamefont
  {Ghosh}}, \bibinfo {author} {\bibfnamefont {A.}~\bibnamefont {Mukherjee}},
  \bibinfo {author} {\bibfnamefont {M.}~\bibnamefont {Mandal}}, \bibinfo
  {author} {\bibfnamefont {S.}~\bibnamefont {Sarkar}}, \ and\ \bibinfo {author}
  {\bibfnamefont {P.}~\bibnamefont {Roy}},\ }\href {\doibase
  10.1103/PhysRevD.96.116020} {\bibfield  {journal} {\bibinfo  {journal} {Phys.
  Rev. D}\ }\textbf {\bibinfo {volume} {96}},\ \bibinfo {pages} {116020}
  (\bibinfo {year} {2017})},\ \Eprint {http://arxiv.org/abs/1704.05319}
  {arXiv:1704.05319 [hep-ph]} \BibitemShut {NoStop}%
\bibitem [{\citenamefont {Ghosh}\ and\ \citenamefont
  {Chandra}(2018)}]{Ghosh:2018xhh}%
  \BibitemOpen
  \bibfield  {author} {\bibinfo {author} {\bibfnamefont {S.}~\bibnamefont
  {Ghosh}}\ and\ \bibinfo {author} {\bibfnamefont {V.}~\bibnamefont
  {Chandra}},\ }\href {\doibase 10.1103/PhysRevD.98.076006} {\bibfield
  {journal} {\bibinfo  {journal} {Phys. Rev. D}\ }\textbf {\bibinfo {volume}
  {98}},\ \bibinfo {pages} {076006} (\bibinfo {year} {2018})},\ \Eprint
  {http://arxiv.org/abs/1808.05176} {arXiv:1808.05176 [hep-ph]} \BibitemShut
  {NoStop}%
\bibitem [{\citenamefont {Mamo}\ and\ \citenamefont
  {Yee}(2016)}]{Mamo:2015xkw}%
  \BibitemOpen
  \bibfield  {author} {\bibinfo {author} {\bibfnamefont {K.~A.}\ \bibnamefont
  {Mamo}}\ and\ \bibinfo {author} {\bibfnamefont {H.-U.}\ \bibnamefont {Yee}},\
  }\href {\doibase 10.1103/PhysRevD.93.065053} {\bibfield  {journal} {\bibinfo
  {journal} {Phys. Rev. D}\ }\textbf {\bibinfo {volume} {93}},\ \bibinfo
  {pages} {065053} (\bibinfo {year} {2016})},\ \Eprint
  {http://arxiv.org/abs/1512.01316} {arXiv:1512.01316 [hep-ph]} \BibitemShut
  {NoStop}%
\bibitem [{\citenamefont {Huang}\ \emph {et~al.}(2010)\citenamefont {Huang},
  \citenamefont {Huang}, \citenamefont {Rischke},\ and\ \citenamefont
  {Sedrakian}}]{XGH2}%
  \BibitemOpen
  \bibfield  {author} {\bibinfo {author} {\bibfnamefont {X.-G.}\ \bibnamefont
  {Huang}}, \bibinfo {author} {\bibfnamefont {M.}~\bibnamefont {Huang}},
  \bibinfo {author} {\bibfnamefont {D.~H.}\ \bibnamefont {Rischke}}, \ and\
  \bibinfo {author} {\bibfnamefont {A.}~\bibnamefont {Sedrakian}},\ }\href
  {\doibase 10.1103/PhysRevD.81.045015} {\bibfield  {journal} {\bibinfo
  {journal} {Phys. Rev. D}\ }\textbf {\bibinfo {volume} {81}},\ \bibinfo
  {pages} {045015} (\bibinfo {year} {2010})},\ \Eprint
  {http://arxiv.org/abs/0910.3633} {arXiv:0910.3633 [astro-ph.HE]} \BibitemShut
  {NoStop}%
\bibitem [{\citenamefont {Hongo}\ and\ \citenamefont
  {Hattori}(2021)}]{Hongo:2020qpv}%
  \BibitemOpen
  \bibfield  {author} {\bibinfo {author} {\bibfnamefont {M.}~\bibnamefont
  {Hongo}}\ and\ \bibinfo {author} {\bibfnamefont {K.}~\bibnamefont
  {Hattori}},\ }\href {\doibase 10.1007/JHEP02(2021)011} {\bibfield  {journal}
  {\bibinfo  {journal} {JHEP}\ }\textbf {\bibinfo {volume} {02}},\ \bibinfo
  {pages} {011} (\bibinfo {year} {2021})},\ \Eprint
  {http://arxiv.org/abs/2005.10239} {arXiv:2005.10239 [hep-th]} \BibitemShut
  {NoStop}%
\bibitem [{\citenamefont {Sarwar}\ \emph {et~al.}(2017)\citenamefont {Sarwar},
  \citenamefont {Chatterjee},\ and\ \citenamefont {Alam}}]{Sarwar:2015irq}%
  \BibitemOpen
  \bibfield  {author} {\bibinfo {author} {\bibfnamefont {G.}~\bibnamefont
  {Sarwar}}, \bibinfo {author} {\bibfnamefont {S.}~\bibnamefont {Chatterjee}},
  \ and\ \bibinfo {author} {\bibfnamefont {J.-e.}\ \bibnamefont {Alam}},\
  }\href {\doibase 10.1088/1361-6471/aa61b3} {\bibfield  {journal} {\bibinfo
  {journal} {J. Phys. G}\ }\textbf {\bibinfo {volume} {44}},\ \bibinfo {pages}
  {055101} (\bibinfo {year} {2017})},\ \Eprint
  {http://arxiv.org/abs/1512.06496} {arXiv:1512.06496 [nucl-th]} \BibitemShut
  {NoStop}%
\bibitem [{\citenamefont {Finazzo}\ \emph {et~al.}(2016)\citenamefont
  {Finazzo}, \citenamefont {Critelli}, \citenamefont {Rougemont},\ and\
  \citenamefont {Noronha}}]{ADS_s1}%
  \BibitemOpen
  \bibfield  {author} {\bibinfo {author} {\bibfnamefont {S.~I.}\ \bibnamefont
  {Finazzo}}, \bibinfo {author} {\bibfnamefont {R.}~\bibnamefont {Critelli}},
  \bibinfo {author} {\bibfnamefont {R.}~\bibnamefont {Rougemont}}, \ and\
  \bibinfo {author} {\bibfnamefont {J.}~\bibnamefont {Noronha}},\ }\href
  {\doibase 10.1103/PhysRevD.94.054020} {\bibfield  {journal} {\bibinfo
  {journal} {Phys. Rev. D}\ }\textbf {\bibinfo {volume} {94}},\ \bibinfo
  {pages} {054020} (\bibinfo {year} {2016})},\ \bibinfo {note} {[Erratum:
  Phys.Rev.D 96, 019903 (2017)]},\ \Eprint {http://arxiv.org/abs/1605.06061}
  {arXiv:1605.06061 [hep-ph]} \BibitemShut {NoStop}%
\bibitem [{\citenamefont {Critelli}\ \emph {et~al.}(2014)\citenamefont
  {Critelli}, \citenamefont {Finazzo}, \citenamefont {Zaniboni},\ and\
  \citenamefont {Noronha}}]{ADS_s2}%
  \BibitemOpen
  \bibfield  {author} {\bibinfo {author} {\bibfnamefont {R.}~\bibnamefont
  {Critelli}}, \bibinfo {author} {\bibfnamefont {S.}~\bibnamefont {Finazzo}},
  \bibinfo {author} {\bibfnamefont {M.}~\bibnamefont {Zaniboni}}, \ and\
  \bibinfo {author} {\bibfnamefont {J.}~\bibnamefont {Noronha}},\ }\href
  {\doibase 10.1103/PhysRevD.90.066006} {\bibfield  {journal} {\bibinfo
  {journal} {Phys. Rev. D}\ }\textbf {\bibinfo {volume} {90}},\ \bibinfo
  {pages} {066006} (\bibinfo {year} {2014})},\ \Eprint
  {http://arxiv.org/abs/1406.6019} {arXiv:1406.6019 [hep-th]} \BibitemShut
  {NoStop}%
\bibitem [{\citenamefont {Sharma}(2013)}]{SSharma}%
  \BibitemOpen
  \bibfield  {author} {\bibinfo {author} {\bibfnamefont {S.}~\bibnamefont
  {Sharma}},\ }\href {\doibase 10.1155/2013/452978} {\bibfield  {journal}
  {\bibinfo  {journal} {Adv. High Energy Phys.}\ }\textbf {\bibinfo {volume}
  {2013}},\ \bibinfo {pages} {452978} (\bibinfo {year} {2013})},\ \Eprint
  {http://arxiv.org/abs/1403.2102} {arXiv:1403.2102 [hep-lat]} \BibitemShut
  {NoStop}%
\bibitem [{\citenamefont {Saha}\ \emph {et~al.}(2016)\citenamefont {Saha},
  \citenamefont {Upadhaya},\ and\ \citenamefont {Ghosh}}]{KS_MPLA}%
  \BibitemOpen
  \bibfield  {author} {\bibinfo {author} {\bibfnamefont {K.}~\bibnamefont
  {Saha}}, \bibinfo {author} {\bibfnamefont {S.}~\bibnamefont {Upadhaya}}, \
  and\ \bibinfo {author} {\bibfnamefont {S.}~\bibnamefont {Ghosh}},\ }\href
  {\doibase 10.1142/S0217732317500183} {\bibfield  {journal} {\bibinfo
  {journal} {Mod. Phys. Lett. A}\ }\textbf {\bibinfo {volume} {32}},\ \bibinfo
  {pages} {1750018} (\bibinfo {year} {2016})},\ \Eprint
  {http://arxiv.org/abs/1505.00177} {arXiv:1505.00177 [hep-ph]} \BibitemShut
  {NoStop}%
\bibitem [{\citenamefont {Kharzeev}\ and\ \citenamefont
  {Tuchin}(2008)}]{Tuchin_b}%
  \BibitemOpen
  \bibfield  {author} {\bibinfo {author} {\bibfnamefont {D.}~\bibnamefont
  {Kharzeev}}\ and\ \bibinfo {author} {\bibfnamefont {K.}~\bibnamefont
  {Tuchin}},\ }\href {\doibase 10.1088/1126-6708/2008/09/093} {\bibfield
  {journal} {\bibinfo  {journal} {JHEP}\ }\textbf {\bibinfo {volume} {09}},\
  \bibinfo {pages} {093} (\bibinfo {year} {2008})},\ \Eprint
  {http://arxiv.org/abs/0705.4280} {arXiv:0705.4280 [hep-ph]} \BibitemShut
  {NoStop}%
\bibitem [{\citenamefont {Ayala}\ \emph {et~al.}(2005)\citenamefont {Ayala},
  \citenamefont {Sanchez}, \citenamefont {Piccinelli},\ and\ \citenamefont
  {Sahu}}]{Ayala:2004dx}%
  \BibitemOpen
  \bibfield  {author} {\bibinfo {author} {\bibfnamefont {A.}~\bibnamefont
  {Ayala}}, \bibinfo {author} {\bibfnamefont {A.}~\bibnamefont {Sanchez}},
  \bibinfo {author} {\bibfnamefont {G.}~\bibnamefont {Piccinelli}}, \ and\
  \bibinfo {author} {\bibfnamefont {S.}~\bibnamefont {Sahu}},\ }\href {\doibase
  10.1103/PhysRevD.71.023004} {\bibfield  {journal} {\bibinfo  {journal} {Phys.
  Rev. D}\ }\textbf {\bibinfo {volume} {71}},\ \bibinfo {pages} {023004}
  (\bibinfo {year} {2005})},\ \Eprint {http://arxiv.org/abs/hep-ph/0412135}
  {arXiv:hep-ph/0412135} \BibitemShut {NoStop}%
\bibitem [{\citenamefont {Ayala}\ \emph {et~al.}(2004)\citenamefont {Ayala},
  \citenamefont {Bashir},\ and\ \citenamefont {Sahu}}]{Ayala:2003pv}%
  \BibitemOpen
  \bibfield  {author} {\bibinfo {author} {\bibfnamefont {A.}~\bibnamefont
  {Ayala}}, \bibinfo {author} {\bibfnamefont {A.}~\bibnamefont {Bashir}}, \
  and\ \bibinfo {author} {\bibfnamefont {S.}~\bibnamefont {Sahu}},\ }\href
  {\doibase 10.1103/PhysRevD.69.045008} {\bibfield  {journal} {\bibinfo
  {journal} {Phys. Rev. D}\ }\textbf {\bibinfo {volume} {69}},\ \bibinfo
  {pages} {045008} (\bibinfo {year} {2004})},\ \Eprint
  {http://arxiv.org/abs/hep-ph/0308212} {arXiv:hep-ph/0308212} \BibitemShut
  {NoStop}%
\bibitem [{\citenamefont {Schwinger}(1951)}]{Schwinger:1951nm}%
  \BibitemOpen
  \bibfield  {author} {\bibinfo {author} {\bibfnamefont {J.~S.}\ \bibnamefont
  {Schwinger}},\ }\href {\doibase 10.1103/PhysRev.82.664} {\bibfield  {journal}
  {\bibinfo  {journal} {Phys. Rev.}\ }\textbf {\bibinfo {volume} {82}},\
  \bibinfo {pages} {664} (\bibinfo {year} {1951})}\BibitemShut {NoStop}%
\bibitem [{\citenamefont {Kapusta}\ and\ \citenamefont
  {Gale}(2011)}]{Kapusta:2006pm}%
  \BibitemOpen
  \bibfield  {author} {\bibinfo {author} {\bibfnamefont {J.}~\bibnamefont
  {Kapusta}}\ and\ \bibinfo {author} {\bibfnamefont {C.}~\bibnamefont {Gale}},\
  }\href {\doibase 10.1017/CBO9780511535130} {\emph {\bibinfo {title}
  {{Finite-temperature field theory: Principles and applications}}}},\
  Cambridge Monographs on Mathematical Physics\ (\bibinfo  {publisher}
  {Cambridge University Press},\ \bibinfo {year} {2011})\BibitemShut {NoStop}%
\end{thebibliography}%

\end{document}